\documentclass[preprint]{aastex}
\usepackage{epsfig}
\usepackage{lscape}
\usepackage{graphicx}
\usepackage{latexsym}
\usepackage{color}
\usepackage{rotating}
\usepackage{natbib}

\shorttitle{Merger Simulations}
\shortauthors{Motl et al.}

\newcommand{\beq}{\begin{equation}}
\newcommand{\eeq}{\end{equation}}

\newcommand{\numsims}{23 }

\usepackage{natbib}

\usepackage{color}
\definecolor{orange}{rgb}{1,0.5,0}

\begin{document}

\title{
A Comparison of Grid-Based and SPH Binary Mass-Transfer and Merger Simulations
}

\author{
    Patrick M.\ Motl\altaffilmark{1},
    Juhan Frank\altaffilmark{2},
    Jan Staff\altaffilmark{3},
    Geoffrey C.\ Clayton\altaffilmark{2},
    Christopher L.\ Fryer\altaffilmark{4},
    Wesley Even\altaffilmark{5},
    Steven Diehl\altaffilmark{4,5},    
     and
    Joel E.\ Tohline\altaffilmark{2}}
\altaffiltext{1}{Indiana University Kokomo, School of Sciences, P.O. Box 9003,
Kokomo, IN, 46903-9004, USA; pmotl@iuk.edu}
\altaffiltext{2}{Louisiana State University,
Department of Physics \& Astronomy, 202 Nicholson Hall, Baton
Rouge, LA 70803-4001, USA}
\altaffiltext{3}{Department of Astronomy, University of Florida, Gainesville, FL, 32611, USA}
\altaffiltext{4}{Computational Methods Group
(CCS-2), Los Alamos National Laboratory, P.O. Box 1663, Los
Alamos, NM, 87545, USA}
\altaffiltext{5}{Theoretical
Astrophysics Group (T-6)/Computational Methods Group (CCS-2),
Los Alamos National Laboratory, P.O. Box 1663, Los Alamos, NM
87545, USA}

\begin{abstract}
Currently there is great interest in the outcomes 
and astrophysical implications of mergers of double degenerate 
binaries. In a commonly adopted approximation, the components 
of such binaries are represented by polytropes with an index 
$n=3/2$. We present detailed comparisons of stellar mass-transfer 
and merger simulations of polytropic binaries that have been 
carried out using two very different numerical algorithms --- a 
finite-volume ``grid'' code and a smoothed-particle hydrodynamics 
(SPH) code. 
We find that there is agreement in both the ultimate outcomes of the 
evolutions and the intermediate stages if 
the initial conditions for each code are chosen to match 
as closely as possible. 
We find that even with closely matching initial setups, 
the time it takes to reach a concordant evolution differs between the 
two codes because the initial depth of contact 
cannot be matched exactly. There is a general tendency for 
SPH to yield higher mass transfer rates and faster evolution to the final 
outcome. We also present comparisons of simulations calculated from 
 two different energy equations: in one series 
we assume a polytropic equation of state and in the other series an 
ideal gas equation of state. In the latter series of simulations an 
atmosphere forms around the accretor which can exchange angular 
momentum and cause a more rapid loss of orbital angular momentum. 
In the simulations presented here, the effect of the ideal equation of 
state is to de-stabilize the binary in both SPH and grid simulations, but the 
effect is more pronounced in the grid code.

\end{abstract}


\section{Introduction}
\label{sec:introduction}

Dynamically unstable mass transfer and stellar mergers involve
complex hydrodynamical flows of material through a varying gravitational
potential. Investigating these rapid phases of stellar evolution requires
the solution of Poisson's equation (for the self-gravity of the
binary)
coupled to the equations of hydrodynamics to predict the
outcome of the mass transfer event.  The complexity of these coupled
equations necessitate a numerical solution, of which several possible
computational techniques are available -- each with their own strengths,
weaknesses  and numerical artifacts.

In this paper we investigate mass transfer, tidal disruption and merger events
in
binaries using two very different numerical codes and 
we compare the results obtained. We follow the commonly adopted 
approximation in which the 
two stars are represented by  
$n=3/2$ polytropes, and we limit our scope to cases where the stellar 
components are of comparable size.  
We are thus able to approximate, in some of our simulations, the scenario 
of two non-relativistic white dwarfs (WDs) in a binary when the less massive
component first reaches contact with its Roche lobe.
Even with this limitation such events 
are suspected to connect binary progenitors to a wide range of astrophysically 
interesting objects.  This class of binary interactions and mergers is
currently attracting considerable attention
\citep[see e.g.][and references therein]{Staff12,2012MNRAS...424.2222P,2012MNRAS...422.2417D,2014MNRAS...438...14D,Pakmor13,Pakmor16,2013ApJ...767..164Z}.
These efforts and others, in some cases using newly developed codes and novel
techniques, also highlight the importance of benchmark runs to help establish the
advantages and disadvantages of different codes. To help facilitate this process,
the initial data for the runs presented in this paper are available to other
groups interested in simulating interacting binary star systems through direct numerical
simulations by contacting the corresponding author.

Recent  observations and population synthesis models imply that there are
a significant number of close double degenerate (DD) binaries in the Galaxy
\citep{Han:1998kl,2005A&A...440.1087N,2010ApJ...716..122K,2011MNRAS.413L.101K,2011ApJ...735L..30P,2011ApJ...737L..23B,2012ApJ...749L..11B}.
A majority of binaries, close enough to interact sometime during their evolution,
will end up as DD systems where both stars are WDs
\citep{2005A&A...440.1087N,2012ApJ...749L..11B}. The formation of such
a close binary system will include two mass transfer phases, with at least
one, more likely two, common envelope phases. If the resulting DD system has a short
enough period ($\lesssim$ 0.2 hr) it will enter a late phase of rapid mass-transfer
and may merge in less than a Hubble time due to the loss of energy due to
gravitational radiation 
\citep[e.g.][and references therein]{PosYun06}. 
This may result in a Type Ia supernova (SN)
 explosion if the total mass of the DD system is sufficiently large,
 or in a
 Type .Ia SNe
 explosion, a hot DQ white dwarf, sdO, R Coronae Borealis
 (RCB) or  Hydrogen-Deficient Carbon (HdC) star if the mass is lower than this limit
\citep {IbenTut84,1984ApJ...277..355W,1994ApJ...420..336Y}.
Longer-lived, interacting DD
 systems, such as AM CVn binaries, will produce gravitational radiation that
 should be detected by the evolved Laser Interferometer Space
 Antenna (eLISA), New Gravitational Wave Observatory (NGO) or their descendants
 \citep{2013GWN.....6....4A}.

Most of the simulations of WD mergers in the literature have been conducted using various
versions of Smoothed Particle Hydrodynamics (SPH), with examples of recent contributions from the LANL
group including \citep{2009MNRAS.399L.156R,2010ApJ...724..111R,2010ApJ...725..296F,2012ApJ...746...62R}.
In contrast, the LSU group is one of the few in the world using
finite difference hydrodynamics (FDH) for this purpose 
(\citeauthor{Mot02} \citeyear{Mot02}; 
 \citeauthor{Dso06} \citeyear{Dso06} - henceforth DMTF06; 
 \citeauthor{Mot07} \citeyear{Mot07} - henceforth MFTD07; 
 \citeauthor{Staff12} \citeyear{Staff12}). 
It has been known for a few years that the onset
of mass transfer, the time the binary system lingers before entering the final phase of the merger or tidal 
disruption, and details of this final phase, are sensitive to initial conditions, the binary parameters, the equation of state assumed, the method used to drive the system to contact, and the choice of simulation technique 
(DMTF06, MFTD07, Dan et al. 2011, 2012). 
In particular, it is important to develop an understanding of the differences observed and documented in the literature between simulations conducted using SPH and those using FDH for nearly identical binary parameters.
However, for this comparison to be meaningful, care must be taken to ensure that 
all the above referenced variables are  controlled and chosen to be as close to identical as 
possible (see \S 2). In practice, however, small differences in the initial conditions 
could not be avoided because of the different way the fluid
is represented in each method and
technical limitations in the construction of initial models.
The optimal way to achieve these goals 
is to put together a team with experts drawn from both camps and
to design the comparison simulations to be concordant. Here we present the 
results of such a team effort.

 We have conducted a suite of simulations of
binary models with two independent codes based on very different numerical
techniques.  One code is based on an Eulerian formulation of the partial
differential equations for fluid flow and the second employs the SPH technique which leverages the Lagrangian
approach to computational fluid dynamics (See \S  \ref{sec:numerical}).  To enable a detailed comparison
of these simulations we have taken care to ensure that both codes
(i) begin from the same equilibrium binaries as initial data, (ii) drive
these binaries into contact in the same manner to attain a mass transfer
rate that can be resolved numerically, and (iii) analyze the time series of data
for simulations with both codes using the same diagnostic tools and
visualization techniques.

In \S \ref{sec:numerical} we describe both our grid-based ``Eulerian" code and SPH code
in addition to detailing how we generate our initial data through the Self-Consistent
Field (SCF) technique, formulate a particle distribution to match these grid-based
models for SPH simulations as well as our choices for computing diagnostics
from the simulations to compare.  Then, in \S \ref{sec:results} we step through
\numsims different numerical evolutions performed with our two codes.  We have
used four different initial models that span the range of initial mass ratios where
dynamical mass transfer instability and merger are possible.  In addition to
comparing the two numerical codes head-to-head,  we vary numerical resolution,
depth of contact
and the assumed equation of state to better elucidate the differences between
our two numerical techniques. Finally, in \S \ref{sec:summary} we summarize
our conclusions and highlight the elements that are common to both techniques
for numerical simulations in addition to the distinctions we have identified as
being likely the result of numerical artifacts arising in one formulation or the other.

\section{Numerical Algorithms}
\label{sec:numerical}

From a computational science point of view, our separate groups have
shared the goal of developing computational fluid dynamic (CFD)
algorithms that can be used to accurately simulate binary mass-transfer
events in a variety of astrophysical settings.  We treat both stars as
continuum fluid structures whose Newtonian gravitational fields and
tidal interactions are modeled in a fully self-consistent fashion.  We also
employ numerical schemes that are designed to spatially resolve the
interior structures of both stars as well as faithfully represent the complex,
nonlinear dynamical flows that result from mass exchange between the
two stars.  The SPH
technique serves as the foundation of the numerical code designed
by the ``LANL''  group; the numerical code designed by the ``LSU'' group
--- hereafter, referred to as the ``grid'' code ---
employs an Eulerian grid and uses a finite-volume discretization technique.
The distinguishing attributes of both codes are described in \S\ref{sec:CFD},
below.  Here we briefly summarize key elements that are common to both
codes.

Using explicit techniques, we integrate forward in time
the set of partial differential equations (PDEs) that govern
inviscid, compressible fluid flows.  These include the continuity
equation governing mass conservation, a vector equation of
motion governing conservation of the three components of momentum,
and an energy conservation equation.
Only two physical source terms explicitly appear in the equation of motion:
an acceleration due to gravity and an acceleration arising from
local gradients in the gas pressure.  In the context of the simulations
being conducted here, principally for comparison purposes,
we adopt one of two algebraic equations of state (EOS).   For ``polytropic''
simulations, gas pressure $P$ is derived from the mass density $\rho$
via the relation,
\begin{equation}
   \label{eq:PolytropicEOS}
   P = \kappa_i\rho^\gamma \, ,
\end{equation}
where the adopted adiabatic exponent is $\gamma = \frac{5}{3}$ and the
polytropic constant $\kappa_i$, which specifies the specific entropy of a
fluid element, can have one of two values depending
on whether the fluid element is originally part of the donor star ($i=\mathrm{D}$) or
part of the accretor ($i=\mathrm{A}$).  (Values of $\kappa_\mathrm{A}$ and $\kappa_\mathrm{D}$,
are reported in Table \ref{tab:InitialModels} for the initial models considered
here.)  For ``ideal gas'' simulations, we use the relation,
\begin{equation}
   \label{eq:IdealGasEOS}
   P = (\gamma - 1)\epsilon\rho \, ,
\end{equation}
where $\gamma = \frac{5}{3}$ and
where, at each integration time step, the specific internal energy of the gas,
$\epsilon$, is determined from the PDE that governs energy conservation.
As noted in Table \ref{tab:InitialModels}, this equation of state allows us to simulate
low mass (compared to the Chandrasekhar mass) double white binaries when 
both binary components are on the same adiabat. In other cases, the binary components 
will have the mass -- radius relation appropriate to low-mass, fully convective stars.

Both codes employ mechanisms for handling the formation and movement
of shock discontinuities.
Via the energy equation, for example, some kinetic energy is converted
to heat when the fluid passes through a shock.  Hence, the specific entropy
of a fluid element should also increase
when the fluid passes through a shock.  In practice, this additional heat
will be realized by an increase in $\epsilon$ that is above what would be
expected through adiabatic compression alone.  When the ideal gas EOS is
used, the gas pressure also will reflect this heat gain because, via
Eq.~(\ref{eq:IdealGasEOS}), $P$ is proportional to $\epsilon$.
However, when the polytropic EOS is used, this additional heat is
effectively lost from the system because the specific entropy of each fluid element
is forced to remain constant, independent of whether or not the fluid
has passed through a shock.  By evolving otherwise identical systems,
first, with a polytropic (P) EOS then, second, with an ideal gas (I) EOS,
the set of simulations presented here not only permits us to compare
results from different algorithmic techniques but also to
isolate the effects that shock heating has on binary mass-transfer
evolutions.

Results from CFD simulations can depend sensitively on initial
conditions.  With this in mind, we have made a concerted effort to
start each simulation from a well-defined, equilibrium binary
configuration, and to start each pair of comparison simulations from initial
configurations that are nearly identical.  The initial binary configuration
for each of the \numsims simulations identified in Table \ref{tab:Prescription}
was drawn from one of the equilibrium configurations listed
in Table \ref{tab:InitialModels}.  By design, the models span a
range of initial binary mass ratios, 
$1.323 \geq q_0=M_{\rm D}/M_{\rm A} \geq 0.4085$ that
includes the regime where mass transfer is dynamically unstable
($q_{0} > q_\mathrm{crit}$) to smaller initial mass ratios where mass transfer may
be stable depending on the fate of the angular momentum of the accretion
stream.  For our chosen polytropic exponent of $\gamma = \frac{5}{3}$,
the standard critical mass ratio for dynamical stability is $q_\mathrm{crit} = \frac{2}{3}$
\citep{1982ApJ...254..616R, 1997A&A...327..620S}, but could be significantly lower
for direct impact accretion (e.g. 
$q_\mathrm{crit} = 0.22$ 
\citeauthor{2004MNRAS...350..113M} \citeyear{2004MNRAS...350..113M};
\citeauthor{2007ApJ...655.1010G} \citeyear{2007ApJ...655.1010G}). Table
\ref{tab:InitialModels} includes the radii of each binary component as well as
the accretor's radii in units of the circularization radius, $R_{\mathrm{circ}}$
(where $R_{\mathrm{circ}}$ is calculated from equation 4.20 of 
\citeauthor{Frank02} \citeyear{Frank02}),
showing that all of our initial models will experience direct impact accretion.

The initial models identified in Table \ref{tab:InitialModels} as Q$1.3$, Q$0.5$,
and Q$0.4a$ were selected for this investigation because their polytropic
simulations had already been studied using the grid code (see DMTF06
and MFTD07).  In each of these cases the detailed structure of the initial equilibrium
configuration was derived using an SCF technique as described
in \S\ref{subsec:SCF}, below. In the context of our present investigation,
these three initial model configurations were each
mapped to a particle distribution as described in \S\ref{subsec:SPHinitial}
that was suitable for insertion as an initial model into the SPH code.  We
decided that we should also study the evolution of a binary system
with an initial mass ratio $q_0 = 0.7$ because this was near the
analytically predicted critical value, $q_\mathrm{crit} = 2/3$, and filled
a gap in parameter values that would otherwise exist between models
Q$1.3$ and Q$0.5$.  The standard SCF technique was used to construct
the model identified as Q$0.7a$ in Table \ref{tab:InitialModels} as the
additional initial model for the grid code.  But, in this case, a separate 
relaxation technique
was used to generate the model identified in Table
\ref{tab:InitialModels} as Q$0.7b$ for insertion into the 
SPH code (See \S \ref{subsec:SPHinitial}).
We note that models Q$0.7a$ and Q$0.7b$ are very similar but not identical to
one another; for example $\kappa_\mathrm{A} \neq \kappa_\mathrm{D}$ in Q$0.7a$.
An additional SCF model, Q$0.4b$, at higher numerical
resolution, was also constructed. As detailed in Table \ref{tab:InitialModels}, the polytropic
models Q$0.4a$ and Q$0.4b$ differ slightly in their parameters due to the difference
in numerical resolution used for the SCF code.

As constructed, the donor star did not quite fill its Roche lobe in each initial
model.  After introducing an
initial model into one of the two CFD codes, we ``drove'' the donor into
contact with its Roche lobe by explicitly extracting angular momentum
from the binary system at rate of one percent per orbit for a specified
``driving duration.''  (The precise manner by which angular momentum
was extracted from the binary system in the two separate CFD codes
is described in \S\ref{sec:CFD}.)  Column 5 in Table \ref{tab:Prescription}
lists the driving duration that was chosen for each simulation.

\subsection{Initial Conditions Setup}

\subsubsection{The Self-Consistent Field Method (SCF)}
\label{subsec:SCF}

We use the SCF technique developed by Hachisu
\citep{Hach86, Hachetal86}
to generate our initial models.  This approach allows us to form
semi-detached equilibrium structures for the binary components that are
synchronously rotating assuming a polytropic equation of state.
These equilibrium structures naturally
incorporate the effects of tidal distortion in their self-consistent Roche
geometry as well as rotational flattening of the stars.

In a frame of reference co-rotating with the binary, the fluid equations
reduce to a single scalar equation for each binary component that
equates the sum of the gravitational potential, centrifugal potential and enthalpy
to an integration constant.  Three boundary points are set along the
line of centers where the mass density is forced to vanish and these
boundary points allow us to algebraically solve for the integration
constant for each star and the angular frequency of the rotating
frame.

We iteratively solve the scalar equations starting from spherical Gaussian
density distributions and the Keplerian estimate for the orbital frequency.
Upon solving Poisson's equation (on a cylindrical mesh identical to the
one used in the grid-based CFD code) we have an estimate of the enthalpy field.
For our chosen polytropic equation of state, the enthalpy determines
the next guess at the matter density where we rescale the density to
maintain prescribed maximum density values for each star and an
improved estimate of the centrifugal potential is computed about the current
center of mass of the binary.  Poisson's equation is solved again and
new values for the integration constants and orbital frequency are
computed from the boundary conditions.

The iteration proceeds until the structure changes by an amount smaller
than a given tolerance in each subsequent iteration.  The converged value
of the orbital frequency, $\Omega_0$, is used to initialize the velocity field in our
binary simulations and to define the angular frequency of the grid
code's rotating frame of reference.  
The equilibrium solution is not exact however and the orbit will not be precisely circular
but rather exhibit epicyclic oscillations about the mean binary separation even if artificial
driving is not imposed. For the 
models constructed in this paper, the amplitude of $\frac{\dot{a}}{a}$, where time is measured
in the initial orbital period of the binary, ranges from $\sim 0.02$ to almost $0.1$ at the
end of the driving phase.
We note as well that, throughout this paper,
$P_0 \equiv 2\pi/\Omega_0$, as quantified in Table \ref{tab:InitialModels},
is used to normalize evolutionary times.  In the grid-based simulations, the
converged SCF model is shifted slightly within the grid, via an interpolation
scheme, before it is introduced into the CFD code in order to better ensure that
the center of mass of the binary system initially coincides with the origin of
the cylindrical coordinate system.

\subsubsection{Application to SPH Initial Conditions}
\label{subsec:SPHinitial}

For SPH simulations, the particle distribution has to be chosen
{\it a priori}. Since we want to pay particular attention to
resolving the Roche lobe overflow accretion stream, we need
sufficient resolution (i.e. particles) in the outer layers of
the donor. Since the SCF method yields asymmetric stellar
configurations, we developed a new setup technique for SPH
initial conditions that can handle arbitrary spatial
configurations. This technique is based on weighted Voronoi
tessellation (WVT) and is discussed in greater detail in
\citet{Die12}. For the sake of completeness, we here discuss
the essentials as well as details specific to the SCF setup problem.

The WVT initial condition setup is based on an approach
developed by \citet{Die06} to adaptively bin two-dimensional
astronomical X-ray images 
to a constant signal-to-noise ratio. We extend
this technique to work in arbitrary dimensions, and optimize
the algorithm to work with an underlying, arbitrary probability
distribution instead of images. The algorithm starts out
with an initial guess for the particle distribution, which we
choose to be random positions drawn from the probability
distribution. The WVT setup then iteratively applies artificial
short-range (typically within 2-3 smoothing lengths) forces to
the particles until they settle into a smooth particle
distribution. The strengths of these forces are determined by
the desired input resolution at the particle positions. This
process generally converges rather quickly ($<40$ iterations)
and results in a setup with very low particle-noise, and the
desired particle distribution. For uniform distributions, the
configuration resembles that of gravitational glasses, which
are often employed in cosmological simulations
\citep[e.g.][]{Die12,Wan07}.

\begin{figure*}
\centering 
\scalebox{0.7}{\includegraphics{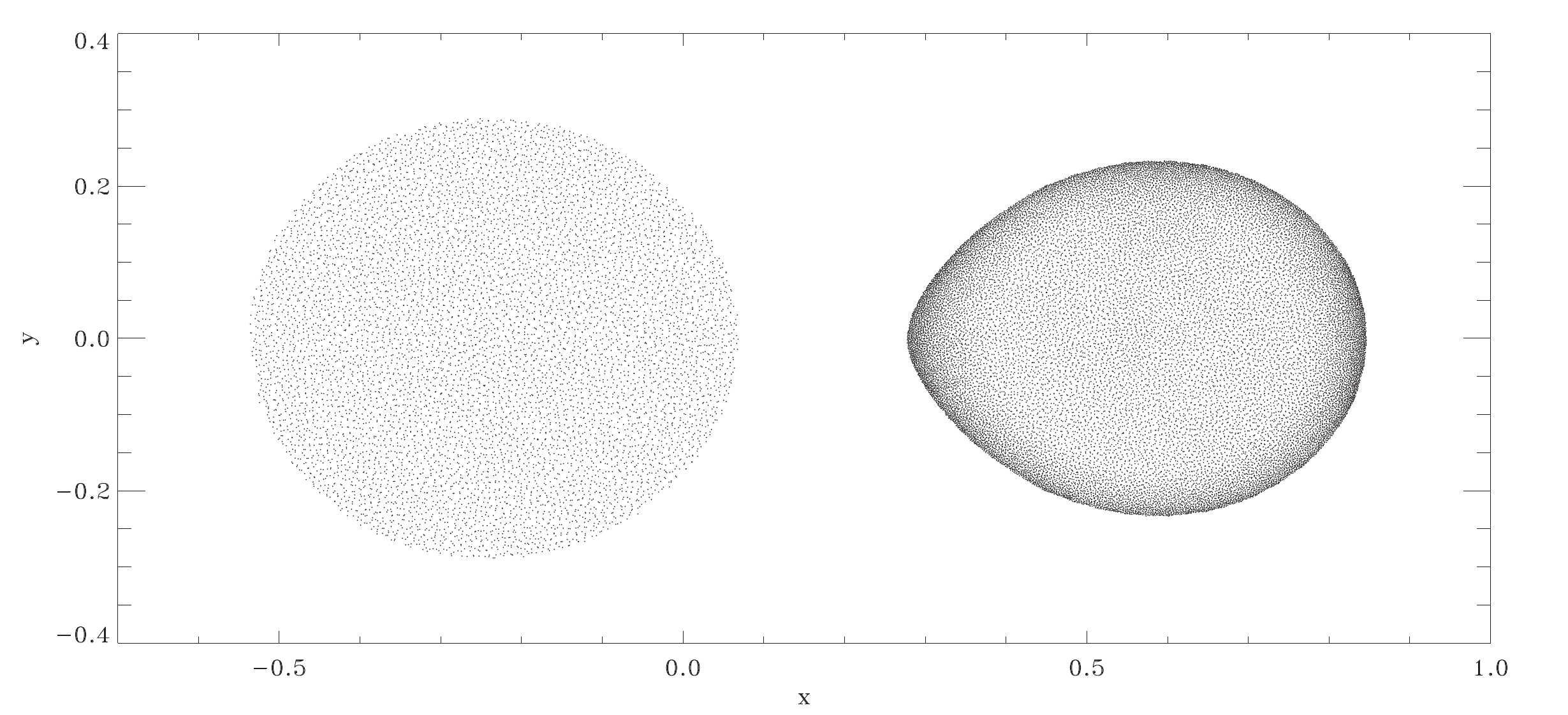}}
\figcaption[DDQ04setup.jpg]{Slice through the meridional plane of
the Q$0.4a$ SPH setup, only particles within one smoothing length
of the plane are shown. Note how the outer layers of the donor
(right side) are much better resolved with a smaller particle
spacing, whereas the accretor (left side) is set up with a
uniform particle density. 
\label{fig:DDQ04setup}}
\end{figure*}

In the context of this investigation, we apply a uniform particle
distribution for the accretor.
In a high-resolution SPH simulation,
each accretor is modeled with a total of 150,000 SPH particles.
In order to properly resolve the accretion stream and beat down
the noise in the
mass transfer rate due to a small number of SPH particles, we
decided to resolve the outer layers of the donor star with much
higher precision. The resolution at the center of the donor star
is chosen to match approximately that of the accretor. However,
the smoothing lengths in the outer layers are approximately $4$
times smaller, putting effectively $\sim 64$ more resolution
elements in these layers, adding up to a total of 850,000 SPH
particles in the donor star and, hence, approximately 1,000,000
particles for each full simulation. Figure \ref{fig:DDQ04setup}
shows a slice through the meridional plane of the WVT SPH setup
for initial model Q$0.4a$, as defined in Table \ref{tab:InitialModels}.
For clarity, only particles within one
smoothing length of the meridional plane are plotted.

\begin{figure}[htb!]
\centering
\scalebox{0.60}{\plotone{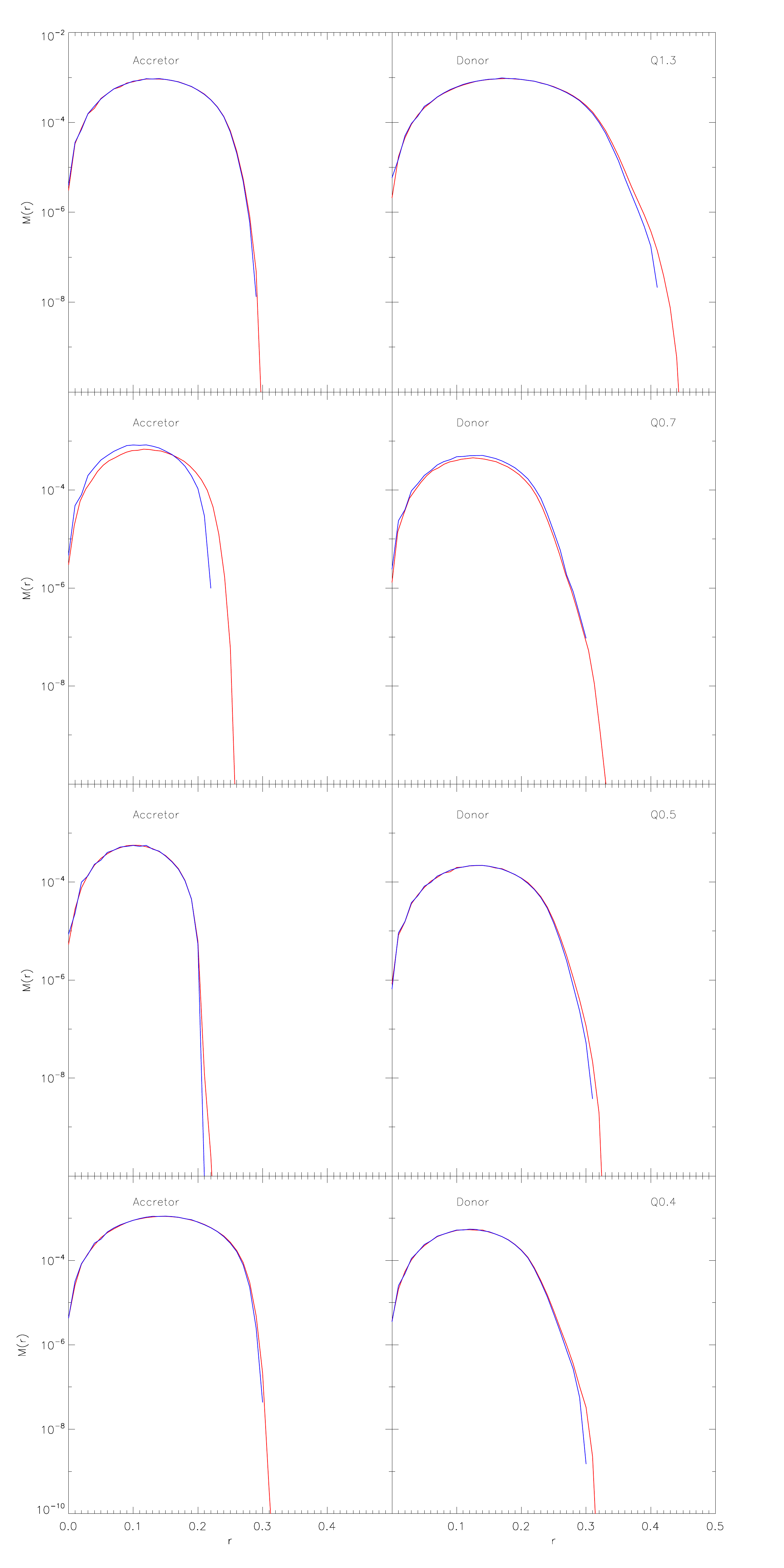}}
\figcaption[MassProfiles]{The mass profile of each stellar component taken from
spherical shells centered on the center of mass of the accreting star
(left column) and donor star (right column) for the four initial binary configurations
Q1.3, Q0.7a and Q0.7b, Q0.5, and Q0.4a.
The SPH initial data is plotted in blue while the 
Eulerian initial data is shown in red. Both codes are thus proceeding 
from very similar initial states and in the Eulerian representation, the donor 
stars (which nearly reach their Roche lobe) are resolved to a somewhat 
lower density level.
\label{fig:InitialMassProfiles}}
\end{figure}

The SPH density interpolation of the particle masses in the WVT
method generally reproduces the SCF grid setup to
better than $1\%$, as shown in Figure
\ref{fig:InitialMassProfiles} where we show mass profiles for the stellar
components in the initial states. Despite the fact that the stars are not 
spherically symmetric, we construct the mass profile as the mass within
spherical shells centered on the center of mass of each stellar component.
Integral quantities for the initial states including the total angular momentum,
orbital angular momentum and the spin angular momentum of each
binary component agree to a level of $1\%$ or better for models
Q$0.4a$, Q$0.5$ and Q$1.3$.
For model Q$0.7b$, a separate, simpler relaxation method was used 
starting with two perfect spheres placed at the expected orbital separation.
Centrifugal and tidal distortions were then introduced adiabatically 
allowing the components to relax to equilibrium.
 
\subsection{Computational Fluid Dynamic Techniques}
\label{sec:CFD}

\subsubsection{Finite-volume ``grid'' code}
\label{subsec:grid}

Our finite-volume ``grid'' code is an explicit code that is second
order accurate in both time and space and has been used extensively to
simulate self-gravitating astrophysical flows 
\citep{Mot02,Dso06,Mot07,Staff12}. 
The code utilizes cylindrical coordinates and evolves equations for
the conservation of mass, radial, vertical, and angular
momentum and an energy equation, as described in more detail below.
With our choice of coordinates and formulation of the fluid equations
we are guaranteed that the total angular momentum of the system will not
change due to advection or due to gravitational or pressure forces.

The hydrodynamics treatment is similar to 
that in the Zeus code \citep{StoNor92}; we use
advection operators based on Van Leer slopes and a similar numerical
viscosity to stabilize the fluid equations in the presence of shocks.  The
fluid equations are formulated for a rotating reference frame (including source
terms for the associated Coriolis and centrifugal forces) and we set
the code's reference frame to rotate at the initial orbital frequency of the
binary to minimize the effects of numerical diffusion through the grid.
Further details of the hydrodynamics algorithm are available in \citet{Mot02} 
and \citet{ Dso06}.

For polytropic simulations our energy equation is cast in terms of an entropy
tracer
\begin{equation}
   \tau = \left( \epsilon \rho \right)^{1 / \gamma}
\end{equation}
where $\epsilon$ is the specific internal energy of the fluid and $\rho$ is the
mass density.  The entropy tracer evolves with an advection equation
similar to the continuity equation for polytropic simulations.  For simulations
with an ideal gas equation of state we evolve the total energy density, $E$,
with
\begin{equation}
 \frac{\partial E}{\partial t}
 + \mathbf{\nabla} \cdot \mathbf{v} E
 =
 - \mathbf{\nabla}\cdot p \mathbf{v}
 - \rho \mathbf{v} \cdot \mathbf{\nabla}  \Phi_{\mathrm{eff}}
\end{equation}
where $\Phi_{\mathrm{eff}}$ includes both the gravitational and centrifugal
potentials.
We compute the internal energy of the fluid by subtracting the kinetic energy
density of the fluid from the total energy density.

The fluid equations evolve densities of conserved quantities like mass density
or momentum density but updating the fluid state requires additional 
quantities that don't obey conservation laws like the velocity of the fluid.
We are thus continually dividing by the mass density of the fluid and therefore no cell in
the grid can have a vanishing mass density. To prevent division by zero, we floor the 
mass density at every timestep and for the runs presented here the floor density is 
$\rho_{\mathrm{min}} = 1 \times 10^{-10}$ compared to a maximum stellar density of $\sim 1$ and a
characteristic density at the stellar edge of approximately $10^{-5}$. The entropy tracer, 
$\tau$, is floored similarly to $\rho$ in polytropic evolutions to remain above a value
\begin{equation}
\tau_{\mathrm{min}} = \left( \frac{\kappa_{\mathrm{min}}}{\gamma - 1} \right)^{\frac{1}{\gamma}}  
                                     \rho_{\mathrm{min}}
\end{equation}
where $\kappa_{\mathrm{min}}$ is the smaller of the polytropic constants in the binary system.
For simulations with an ideal gas equation of state, the total energy density $E$ is floored to remain
above the smallest internal energy density, $\tau_{\mathrm{min}}^{\gamma}$.

Poisson's equation is solved by performing a Fourier analysis along
the periodic azimuthal direction and an alternating-direction implicit
technique is used for the radial and vertical directions.  The boundary
value of the gravitational potential is computed via a numerical integration
of the mass density with cylindrical Greens functions based on half-integer
degree Legendre polynomials of the second kind \citep{Cohl99, Cohl01}.
We employ outflow boundary conditions at the outer edge of the computational
domain and over the course of a merger simulation, up to $\sim 0.3\%$ of a
system's total mass can be lost from the active grid.

As was alluded to above, each binary simulation was begun with a phase of
artificial ``driving'' that causes the orbit to slowly decay and forces the donor star into
contact with its Roche lobe.  This establishes a mass transfer
stream that can be well resolved by the code, given the chosen numerical resolution.
In the grid code the phase of driving was accomplished by removing a fraction of the
specific angular momentum of each fluid element each time step of the simulation
at a rate such that one percent of
the initial angular momentum is removed per initial orbital period, $P_{0}$.

A typical numerical simulation with the grid code (at a numerical resolution of
162 radial zones by 98 vertical zones by 256 azimuthal zones for a total of 
approximately 4 million cells as listed in Table \ref{tab:Prescription})
required about 24 days of
run time on 64 cores of the Louisiana Optical Network Initiative (LONI) 
machine QueenBee, assuming a polytropic
equation of state.  With the ideal gas equation of state, the sound speed is
higher which further reduces the Courant-Friedrichs-Lewy (CFL)
limited time step and increases the typical
run time to approximately 150 days.  The higher sound speed for ideal 
gas equation of state simulations results from a hot, low density envelope 
of shock heated material that
forms around the accreting star. Note that, since we are using a
cylindrical coordinate grid, the CFL limiting constraint on the time step is
often reached near the coordinate axis and is thus significantly smaller
than would be the case with a Cartesian coordinate grid.

\subsubsection{Smooth Particle Hydrodynamics (SPH) code}
\label{subsec:SPH}

The SPH code used in this work
is called SNSPH, and is described in much greater detail in
\citet{Fry06}. Here we will give a global overview of the code,
focusing on the aspects most important to this project.

SNSPH is a tree-based, massively parallel SPH code that has
undergone rigorous testing and has been applied to a large
variety of astrophysical problems
\citep[e.g.][]{Fry02,You06,Die08}. The gravitational forces are
evaluated in an order $N\log N$ gravitational tree, where
long-range forces are replaced by monopole terms. Our multipole
acceptance criterion is described in \citet{War95} and is
designed to provide a direct bound on the error budget
associated with the monopole approximation.

There are several flavors of SNSPH, and each integrates the
hydrodynamics equations in a slightly different manner.
For this investigation we have chosen the version that integrates
the specific internal energy in order to facilitate a direct determination
of temperatures from the SPH simulations. For our SNSPH calculations, 
we use the simple viscosity prescription utilizing both bulk ($\alpha$)  
and von Neumann-Richtmyer ($\beta$) viscosity coefficients
\citep{Ben90b,Mon92,Mon05,Fry06}. 
We do not evolve these constants, and most of our calculations used 
standard values:  $\alpha=0.5$, $\beta=2.0$.  However, we did run a 
test calculation using $\alpha=0.125$ and $\beta=0.5$ and our answers 
did not change noticeably.

For this work, we have included several improvements of the
SNSPH code. First, the traditional n-body Plummer softening of
the gravitational acceleration of the SPH particles has been
replaced with an SPH-kernel based smoothing \citep[see][for
example]{Ben90b,Nel06}, where the SPH kernel is interpreted as
a three-dimensional density distribution, and the gravity is
calculated self-consistently according to this distribution.
The main advantage of this spatially adaptive, but
computationally slightly more intensive prescription is that
the gravitational acceleration between two particles vanishes
as they come very close to each other. This in turn reduces the
frequency of an SPH-specific numerical artifact, known as
``particle-pairing'' or the ``pairing instability''
\citep[e.g.][]{Tho92,Her94,Nel06}, which can drastically reduce
the effective resolution of an SPH simulation. In a sense, the
binary simulations conducted here with a polytropic equation
of state present a worst-case scenario for the
particle-pairing problem. In the Q0.4P simulations (see the last
row of
Table \ref{tab:SCFparameters}), for example,
the donor material that streams onto the surface of the accretor
has a lower specific entropy than the original accretor material, meaning 
that the donor material is convectively unstable and some 
high-resolution donor particles sink deep into the accretor.
In the process, the already small, low-density donor particles get
even further compressed and clump to each other, forcing the
particles to become extremely close.

To further reduce particle-pairing and to avoid
code-specific numerical problems in the tree setup, we also
introduce a repulsive pressure wall within the innermost $2\%$
of the particle smoothing length \citep[see][for a similar
approach]{Tho92}. Due to the small volume affected by this
modified artificial pressure force, this choice does not affect
the overall performance and conservational properties of the
code.

As with the grid-based evolutions, the SPH components were driven into
contact by removing angular momentum from the binary at a prescribed rate
for a specified duration. This was accomplished in the SPH simulations by
reducing the velocity of the SPH particles through a user-specified time
derivative of the angular velocity.

The 1 million particle SPH simulations in this paper generally
ran on 128 processors for about one week clock time on the AMD
Opteron cluster {\it Coyote} located at Los Alamos National
Laboratory.

\subsection{Diagnostics}

Our analysis of each binary simulation has been carried out primarily
using two diagnostic techniques:  (1) Throughout each simulation, a
variety of global binary parameters were evaluated by performing
integrals over the computational volume as detailed in Appendix
\ref{app:Diagnostics}, and then curves
were plotted to show the variation with time of these parameters.
(2) For each simulation, a movie was generated to show the
time-variation of the
matter distribution.  We
determined the degree to which a simulation performed with the
SPH code resembled the simulation performed with the grid code
by comparing the behavior of various parameter curves and/or by
comparing corresponding frames from movies generated with the
two separate codes.

Most of the figures shown in subsequent sections of this paper
display results from either one or both of these diagnostics.  For
example, the curves drawn in Figure
\ref{fig:Q13evolutionA} 
show the time-dependent behavior 
of six binary system parameters
derived from three separate Q1.3P simulations ---
orbital separation, $a$, binary mass ratio, $q$,
log of the donor's mass-transfer rate, $|\dot{M}_\mathrm{D}|$,
orbital angular momentum, $J_\mathrm{orb}$, and spin angular
momentum of the accretor, $J_\mathrm{A}$,
and donor, $J_\mathrm{D}$.
The color contour plots shown in Figure
\ref{fig:Q13evolutionB} 
are movie
frames drawn from two separate points in time during two of these Q1.3P
simulations. 
Since the two numerical codes use different frames of reference, an inertial
reference frame in the case of the SPH simulations and a frame of reference
rotating at the initial frequency of the binary in the grid simulations, there will be
a trivial phase angle offset between the two codes. For the purpose of
aiding comparisons between the two codes, the grid code data are rotated
to match the phase of the SPH simulation for the color contour plots of the 
projected mass density. Throughout the paper, such figures are used to illustrate key
findings from our comparison simulations. 
We also note that, in general, the frames taken 
from our grid code simulations show an outline of the outer edge of the cylindrical grid
at low density levels. This is
a numerical artifact resulting from imperfect outflow boundary conditions applied 
at the outer edge of the computational domain.

Initially we agreed that the group that developed the SPH code and
the group that developed the grid code would perform these
diagnostic analyses independently.  Eventually we realized
that some differences that appeared when comparing
diagnostics from different
simulations arose from subtle differences that appeared in the
volume integrals that were being used by the two different groups
to define various global binary parameters.  There are, for
example, different ways to define the orbital angular momentum
or the spins of the two stars when some of the matter resides in an
envelope that is shared by the two stars.  We ultimately decided,
for diagnostic purposes only, to map results from the SPH simulation
onto a
3D
grid and to use the diagnostic software developed
for the grid code to also analyze results from each SPH simulation.
The diagnostic figures presented throughout the remaining sections
of this paper were derived in this manner.  Appendix \ref{app:Diagnostics} explains how
each volume integral was defined in order to evaluate the various
global binary parameters. 

The initial data for the Q0.7 binary were developed independently by the two
groups resulting in the models Q0.7a and Q0.7b described in Table
\ref{tab:SCFparameters}. We made use of the freedom of the chosen polytropic
equation of state to re-scale the data from the grid evolutions to match the
total mass and initial separation of the SPH model Q0.7b for the purposes of
analyzing the simulation results. Both thee line plots and images of the projected
mass density from the grid code have been rescaled accordingly.

\section{Evolutions}
\label{sec:results}

As is cataloged in Table \ref{tab:Diagnostics},
we are reporting on results from \numsims 
separate binary mass-transfer
simulations. Each of the simulations was followed through at least eight,
usually more than a dozen, and up to 
fifty orbits with each individual
orbit requiring tens of thousands of Courant-limited integration time steps
(for the grid code simulations).
Hence, our critical examination of the behavior of these systems and,
most importantly, the comparison of behavior between different simulations,
has required the analysis of a tremendous amount of simulation data.  In
what follows, our formal discussion of the results is broken down into a series
of subsections that, generally speaking, march through the models in the order
in which they are listed in Table \ref{tab:Prescription}.

Before diving into the details, the reader may find it advantageous to skim
through the figures and/or some of the accompanying movies 
(21 in total, as labeled in column 5 of Table 3)
in order to sample the variety of physical phenomena that arise in these types
of systems.  Examples are:
\begin{itemize}
   \item Binary merger -- see video01 or video02, and Figure
            \ref{fig:Q13evolutionB};
   \item Tidal disruption of the donor -- see video04, video16, or video21,
            and Figure \ref{fig:Q07evolutionB} or Figure
            \ref{fig:Q04idealComparison};
   \item The hot, tenuous common envelope that forms in ``ideal gas,'' as
            opposed to ``polytropic,'' simulations -- compare video07 to video17,
            and see Figure \ref{fig:Q05PolyIdeal};
   \item Nonaxisymmetric modes having four-sided (box-like) and
            three-sided (triangular) shapes excited in an accretion disk -- see
            video07 or video12, and Figure \ref{fig:Q04Presonance};
   \item Binary detachment following a long phase of relatively stable mass
            transfer -- see video07, video09, or video11.
\end{itemize}

\subsection{Polytropic Simulations}

Here we discuss and compare results from simulations that
were carried out using a polytropic equation of state.
The relevant simulations are labeled as Q1.3P, Q0.7P,
Q0.5P, and Q0.4P along with a two-character model identification
suffix (for example, simulation Q1.3P$\_G1$) in the first 13
rows of Table \ref{tab:Prescription}. In each case the structure
of the binary system at time $t=0$ was prescribed by one of
the five, initially uniformly rotating, $n=\frac{3}{2}$ polytropic binary
models identified in Table \ref{tab:SCFparameters}.  At the
start of a simulation, the donor star was driven into contact
with its Roche lobe by artificially draining angular momentum
from the system at a rate of one percent per orbit for the ``driving
duration'' specified in column 5 of Table \ref{tab:Prescription}.
Thereafter the system was evolved in a fully self-consistent fashion
without further external influences. As is documented in column 4
of Table \ref{tab:Prescription}, some SPH simulations were carried
out using 100 thousand ($100k$) particles and others used 1
million ($1M$) particles; most simulations performed with the grid
code used a uniform cylindrical grid containing approximately
four million grid cells but we include one grid simulation of the 
Q0.4P system at a higher resolution of approximately 47 million
($47M$) cells.

To aid in our analysis, we have established
the time of an event common to each merger simulation, the merger time, $t_{\mathrm{merge}}$, as a 
means of synchronizing these simulations.
While the choice of the exact merger time is in some sense arbitrary, we have found empirically
that the first data set where the merged object has only one density concentration
could be identified relatively easily by eye. 
We found this particular choice of merger time could be identified in a repeatable manner for both 
grid and SPH data sets from the animations corresponding to each simulation.
Merger times are recorded in the third column of Table \ref{tab:Diagnostics} for those 
simulations that merge.
With a common event identified in all merger simulations, we can measure time relative 
to $t_{\mathrm{merge}}$ to eliminate trivial phase differences and variations in
instability growth rates in different simulations to compare the highly nonlinear state
of the evolving fluid in a clear way. From $t_{\mathrm{merge}}$ we have established a 
zero point time relative to a particular grid-based simulation as,
\begin{equation}
(t_\mathrm{zpt})_\mathrm{ID} = (t_\mathrm{merge})_{G1}
- (t_\mathrm{merge})_\mathrm{ID} . \label{eq:tzpt}
\end{equation}
Time sequences of data from a particular simulation can be compared then by
plotting the
behavior of various system parameters using
the {\it shifted} time,
\begin{equation}
t_\mathrm{shift} \equiv t + t_\mathrm{zpt} , \label{eq:tshift}
\end{equation}
to label the temporal axis.
For reference, column 4 of Table \ref{tab:Diagnostics} contains $t_{\mathrm{zpt}}$ for
each merger simulation.

\subsubsection{{\rm Q1.3P} Model Simulations}
\label{subsec:Q1.3P}

We carried out three separate
adiabatic simulations of the binary polytrope that had an initial mass ratio
$q_0 = 1.323$.  In each case, in order to initiate mass transfer, the model
was driven at a rate of 1\% per orbit for $2.0 P_0$.  As is indicated by the
note referenced in the last column of Table \ref{tab:Prescription}, some
results from the grid-code simulation Q1.3P\_G1 already have been
presented as simulation Q1.3-D in \S5.1.1 of DMTF06. This model, as shown
in Table \ref{tab:InitialModels}, has an equation of state appropriate for fully
convective, and hence low mass, main sequence stars and in this case it is the 
larger, more massive star that has reached contact with its Roche lobe.

Whether the simulation was conducted using the grid code (model $G1$)
or the SPH code (models $S1$ and $S2$), the Q1.3P simulation was found
to be dynamically unstable to mass transfer.   This was as expected
theoretically because the system had an initial mass ratio greater than
unity and, therefore, also greater than $q_\mathrm{crit}$.  
As Figure 
\ref{fig:Q13evolutionA} 
illustrates,
throughout each model simulation, the binary separation ($a$)
and system mass-ratio ($q$) steadily decreased while the donor
mass-transfer rate ($|\dot{M}_\mathrm{D}|$) steadily increased. As is
shown in the accompanying movies -- video01 from simulation Q1.3P$\_G1$,
video02 from simulation Q1.3P$\_S1$, and video03 from simulation
Q1.3P$\_{S2}$ -- and as is
illustrated by the images shown in 
Figure 
\ref{fig:Q13evolutionB}, 
independent of the chosen numerical
technique the two stars ultimately merged violently in the
sense that the highest density ``core'' region of the donor was
observed to collide and combine with the highest density
``core'' region of the accretor.

The upper right-hand image shown in Figure 
\ref{fig:Q13evolutionB}
is the frame picked from the movie of grid-code simulation
Q1.3P$\_G1$ at the instant the cores of the two stars first merge, that
is, at the instant the donor core is no longer distinguishable from the core
of the accretor.  
Defined in this way, the merger time
for simulation Q1.3P$\_G1$ is  $t_\mathrm{merge}= 12.38 P_0$.
The lower right-hand image in Figure 
\ref{fig:Q13evolutionB}
is the frame picked from  the movie of SPH-code simulation
Q1.3P$\_S1$ that most closely resembles the upper right-hand
image.  Using this frame to define the instant of merger, we deduce
$t_\mathrm{merge}= 7.57 P_0$ for simulation Q1.3P$\_S1$.
In a similar manner we have deduced that
$t_\mathrm{merge}= 8.17 P_0$ for the lower resolution
SPH-code simulation Q1.3P$\_S2$.  

The ``instant of merger'' frames selected from simulations Q1.3P$\_G1$
and Q1.3P$\_S1$ are not just similar qualitatively,
they share a considerable amount of detail.  In both cases the
merged core displays a marked top/bottom asymmetry:  the
bottom edge of the core is flat, revealing the shock front
that was established by the core collision; the top
is elongated largely by donor material, giving the merged core
the shape of an egg.  Toward the bottom, the small, remnant tail of the donor
that has not yet plunged through the shock has the same structure in both
models.  Furthermore, the models display remarkably similar details
where the trailing edges of both spiral
arms branch at sharp angles from the merged core.

The left-hand images in Figure 
\ref{fig:Q13evolutionB} 
present
frames from both movies one quarter of an orbit prior to merger ---
the top frame is drawn from the movie of grid-code simulation
Q1.3P$\_G1$ and the bottom frame is drawn from the movie
of SPH-code simulation Q1.3P$\_S1$.  These two images also
display a high degree of quantitative similarity.  They illustrate
what a careful viewing of video01, video02, and video03 reveals:  the two
different numerical schemes produce very similar
dynamical results, especially as
the simulations each evolve through the time of merger.
Note that the SPH simulations are conducted in an inertial frame whereas 
the grid-code simulations are computed in a frame initially corotating 
with the binary. Consequently, in all the figures showing comparisons of 
equatorial density distributions obtained with both codes, we have rotated the 
density contours obtained with the grid code
to match the orientation of those obtained with the SPH
code.

We cannot ignore the fact that the precise time of
merger was different in the three simulations.  The decidedly
different merger times seem to suggest that different code algorithms or
model resolutions produce different results.  But
a more careful examination reveals that if an appropriate
zero-point time-shift, $t_\mathrm{zpt}$, is added to the
recorded evolutionary time of each simulation, the simulations
show a high degree of agreement.  For this particular set of
models, the merger times have been used to define
$t_\mathrm{zpt}$ as shown in column 4 of Table \ref{tab:Diagnostics}.
In Figure 
\ref{fig:Q13evolutionA} 
the time-evolutionary
behavior of various system parameters has been displayed using
the shifted time that synchronizes merger between the simulations.
 Viewed from this perspective, we conclude again that the
three separate Q1.3P simulations show a remarkable level of
agreement.

Consider first the middle panel on the right-hand-side of
Figure 
\ref{fig:Q13evolutionA}, 
which displays the time-dependent
behavior of the system's orbital angular momentum.
During the first two orbits, $J_\mathrm{orb}$ steadily decreases
at a rate of $1\%$ per orbit due to the imposed driving,
then it remains nearly constant until the merger occurs.
Over the last $\sim 6$ orbits, the (solid blue) curve generated
by the 1M-particle SPH
simulation lies directly on top of the (solid red) curve drawn
from the 4M-cell grid-code simulation. The slight mismatch
exhibited by the (dashed blue) curve drawn from the lower
resolution (100 K) SPH simulation simply
reflects the inability of the lower-resolution model to
accurately represent the system's initial $J_\mathrm{orb}$.
After the zero-point shift in time has been taken into account,
all three curves that describe the
time-evolutionary behavior
of the binary separation (see the top panel on the left-hand
side of Figure 
\ref{fig:Q13evolutionA}) 
also lie on top of
one another over the last $\sim 6 P_{0}$.

All three simulations show that, once mass-transfer commences,
the rate increases by more than two orders of
magnitude in less than 10 orbits; see the middle panel on the
left-hand side of Figure 
\ref{fig:Q13evolutionA}.
It is gratifying to see that, after $t_\mathrm{zpt}$ has been taken
into account, all three $|\dot{M_\mathrm{D}}|$ 
diagnostic curves converge immediately prior to merger.  
However, at earlier times the slope of the
mass-transfer curve produced by the grid-code simulation
does not match the slope of the curves produced by the SPH
simulations.  At the end of the initial phase of driving, the value
of $|\dot{M_\mathrm{D}}|$ in the grid-code simulation is also somewhat lower
than the value of $|\dot{M_\mathrm{D}}|$ in the SPH simulation.
This is the first indication that, at relatively
low mass-transfer rates, the simulations
conducted with our two different numerical techniques generate
results that are quantifiably different from one another.
Figure 
\ref{fig:Q13evolutionA} 
also shows that, while the blue and
red curves illustrating the behavior of $q(t)$ and $J_\mathrm{A}(t)$ overlap
at late times, they do not lie precisely on top of each other at
earlier times.  For both of these binary parameters, this behavior
probably reflects the discrepancy already noted in
mass-transfer rates at early times.

The second indication that simulations conducted with the two
different numerical codes generate results that are quantifiably different from
one another comes from the bottom panel on the right-hand side
of Figure 
\ref{fig:Q13evolutionA}, 
which illustrates the behavior
of the spin of the donor star.  In the grid code,
$J_\mathrm{D}$ slowly but steadily decreases up until the moment
of merger, whereas in the SPH code (at both resolutions)
$J_\mathrm{D}$ initially increases, then decreases for a while, before
increasing dramatically at the moment of merger.  Overall,
the Q1.3P simulation does not appear to be seriously impacted
by this discrepancy in the behavior of $J_\mathrm{D}(t)$.  This is likely
because the donor stores less than $10\%$ of
the system's total angular momentum and, because $q_0  > 1$,
the system is doomed to merge, independent of the precise
treatment of spin-orbit coupling.

\subsubsection{{\rm Q0.7P} Model Simulations}
\label{subsec:Q0.7P}

As is recorded in Table \ref{tab:Prescription}, we carried out three separate
polytropic simulations of binary systems having an initial mass ratio of
$q_0 = 0.7$.  Simulations conducted with the grid code
(models $G1$ and $G2$) were started from the initial model identified in
Table \ref{tab:SCFparameters} as Q0.7a while the simulation conducted with
the SPH code (model $S1$) was started from the initial model
identified as Q0.7b.  These initial models were very similar, but not identical
to one another; for example, the two stars in model Q0.7b were constructed
as $n=3/2$ polytropes having equal polytropic constants whereas the
polytropic constants for the two stars in model Q0.7a were slightly different
from one another.  
For both of the Q0.7 initial models, it is the less massive star that has the larger radius
and hence these systems are closer analogs to low-mass (nonrelativistic) double white dwarf binaries
near the anticipated stability limit for a polytropic equation of state with $n = \frac{3}{2}$ of
$q_{\mathrm{crit}} = \frac{2}{3}$.
Simulations $G1$ and $G2$ differed from one another only
in the duration of the initial phase of driving:  $G2$ was driven for $1.70 P_0$
while $G1$ was driven for $2.28 P_0$.  Model $S1$ was driven for $1.0 P_0$
and was conducted with $100 k$ SPH particles.

As Figure 
\ref{fig:Q07evolutionA} 
illustrates, the time-variation of
key binary parameters exhibited by the various Q0.7P simulations was very similar
and resembled the time-variations seen in simulation
Q1.3P.  For example, throughout each simulation $a$
and $q$ steadily decreased while $|\dot{M}_\mathrm{D}|$ steadily increased.
We conclude that, as with the Q1.3P simulation, the Q0.7P
simulation was dynamically unstable to mass transfer.
It is not entirely appropriate to describe the Q0.7P simulation as a
merger, however.  As is shown in the accompanying movies --
video04 from simulation Q0.7P$\_G1$,
video05 from simulation Q0.7P$\_G2$, 
and video06 from simulation Q0.7P$\_S1$ -- the core of the
donor star does not ultimately plunge into the core of the accretor.  Instead,
the donor star tidally disrupts and its material is dispersed by differential rotation
to form an extended disk surrounding the accretor.

The upper right-hand image shown in Figure 
\ref{fig:Q07evolutionB}
is the frame picked from the movie of grid-code simulation
Q0.7P$\_G1$ (video04) just before tidal disruption of the donor is complete.
Using this frame to define the ``instant of merger,'' we obtain
$t_\mathrm{merge}= 9.70 P_0$ for simulation Q0.7P$\_G1$.
The bottom right-hand image
in Figure 
\ref{fig:Q07evolutionB} 
is the frame picked from the movie of
SPH-code simulation
Q0.7P$\_S1$ (video06) that most closely resembles the upper right-hand
image.  Using this frame to define the ``instant of merger''
we deduce that $t_\mathrm{merge}= 11.51 P_0$ for simulation Q0.7P$\_S1$.
In a similar manner we have deduced that $t_\mathrm{merge}= 21.03 P_0$ for
grid-code simulation Q0.7P$\_G2$.
These merger times are
recorded in the third column of Table \ref{tab:Diagnostics}.
The ``instant of merger'' frames selected from simulations Q0.7P$\_G1$,
Q0.7P$\_G2$, and Q0.7P$\_S1$ share a considerable amount of detail, although
this claim is somewhat weaker than in our earlier comparison of the
model Q1.3P simulations because, in the present case, the SPH simulation
was conducted with only $100 k$ particles.

The left-hand images in Figure 
\ref{fig:Q07evolutionB} 
present
frames from the movies of simulations Q0.7P$\_G1$, Q0.7P$\_G2$, and
Q0.7P$\_S1$ half an orbit
($\Delta t = - 0.5P_0$) prior to $t_\mathrm{merge}$.  These images also
display a high degree of quantitative similarity.  The  pairs
of images in Figure 
\ref{fig:Q07evolutionB} 
illustrate
what a careful viewing of the relevant movies reveals:  the two
different numerical schemes produce very similar
dynamical results, especially as the simulations each evolve
through the phase of tidal disruption of the donor.  Furthermore, a comparison
of the top pair of images with the middle pair of images shows that
the outcome of this particular simulation is relatively insensitive to the
precise amount of driving that is applied in order to initially bring the
surface of the donor into contact with its Roche lobe.

As was seen in the context of the Q1.3P simulations, the time of merger
is different in the three Q0.7P simulations.
It is not surprising that tidal disruption of the donor
occurs earlier in grid-code model $G1$ than in grid-code model $G2$ because,
in $G1$, the ``driving duration" was longer and, hence, the
donor was initially driven into deeper contact with its Roche lobe.
As Figure 
\ref{fig:Q07evolutionA} 
illustrates, all three
Q0.7P simulations show very good agreement when the
time-evolutionary behavior of various system parameters is displayed using the
{\it shifted} time defined by equations (\ref{eq:tzpt}) and (\ref{eq:tshift}).

Again, it is worth pointing out ways in which results from the grid-code simulations
differ from results obtained from the SPH-code simulation.  As Figure
\ref{fig:Q07evolutionA} 
shows,
throughout the simulation the mass-transfer rate increases somewhat more
rapidly in simulation Q0.7P$\_S1$ (solid blue curve)
than in simulations Q0.7P$\_G1$ and Q0.7P$\_G2$ (dashed and solid
red curves, respectively).  This behavior was seen as well in simulations
Q1.3P and, again, probably explains why the $q(t)$ curves do not lie
precisely on top of one another at early times.  The curves
that trace the time-dependent behavior of $J_\mathrm{orb}$ and $J_\mathrm{A}$
in the SPH-code simulation (blue curves) lie slightly below the (red)
curves drawn from the grid-code simulations.  This is almost certainly
because, in this case, the initial models were not identical.  While
the blue curve describing the time-evolutionary behavior of $J_\mathrm{D}$ does
not lie precisely on top of the red curves, the general shape of the curves
derived from the two different numerical techniques appears to be
more similar than was seen in Figure 
\ref{fig:Q13evolutionA} 
for simulation Q1.3P.

We note that the short period ($\sim 1P_0$) oscillations that
are visible in all $a(t)$ curves --- see the upper left-hand
panels of Figures 
\ref{fig:Q13evolutionA} 
and 
\ref{fig:Q07evolutionA} ---
arise because, in practice, none of the binary orbits is perfectly circular.
The amplitude of the resulting ``epicyclic oscillations'' is
determined, in part, by the duration of driving that was initially
imposed on each model simulation.  The absolute phases of the epicyclic
oscillations generally do not --- and should not be expected to ---
match when the $a(t)$ curves are plotted as a function of the {\it shifted} time.

\subsubsection{Model Simulations  {\rm Q0.5P} and {\rm Q0.4P}}
\label{subsec:Q0.5PandQ0.4P}

We have simulated polytropic mass-transfer in three binaries with
initial mass ratios $q_0 < q_\mathrm{crit}$:
Simulations Q0.5P with $q_0 = 0.500$ and polytropic simulations Q0.4a with
$q_0 = 0.4085$ and Q0.4b with a similar $q_{0} = 0.4203$.
Parameters defining the initial properties of
these systems are detailed in Table \ref{tab:SCFparameters}.
As noted in Table \ref{tab:SCFparameters}, model Q0.5 most closely represents a
double white dwarf binary with low mass components as the donor star is the 
less massive component. Models Q0.4a and Q0.4b are similarly semi-detached systems with the
less massive star in contact with its Roche lobe but these particular models have an
accretor with a radius too large for its mass to be on the white dwarf mass -- radius
relation. This exacerbates the role of direct impact accretion in this model.
As column 5 of Table \ref{tab:Prescription} documents, models Q0.5P$\_G1$ and
Q0.5P$\_S1$ were both initially driven into contact at a rate of $1\%$ per
orbit for $2.7P_0$.  The evolutionary behavior of model Q0.5P$\_G1$ is
presented in the movie identified as video07; for comparison, the model Q0.5P$\_S1$
simulation is presented in video08.  (Some results from simulation Q0.5P$\_G1$ already
have been presented as model Q0.5-Da in \S5.2 of DMTF06.)
As Table \ref{tab:Prescription} also details, five separate Q0.4P simulations were
carried out.   Among this group, we begin by discussing models Q0.4P$\_G1$,
Q0.4P$\_{G2}$, 
and Q0.4P$\_S1$, which were
initially driven into contact at a rate of $1\%$ per orbit for $1.6P_0$.
Simulations Q0.4P$\_{G1}$ and Q0.4P$\_{G2}$ differ in their numerical
resolution where Q0.4P$\_{G1}$ was evolved with $4M$ grid cells while
simulation Q0.4P$\_{G2}$ was performed with $47M$ grid cells, partly in
an effort to establish convergence of results with the grid code and to measure
the lowest initial mass transfer rate that can be established in a numerical
simulation of a semi-detached binary.
Movies showing
the evolutionary behavior of these three models are labeled video09, video10,
 and video12, respectively.
The left panel of Figure 
\ref{fig:Q05PQ04P} 
displays the time-evolutionary behavior of
$a$, $q$, $|\dot{M_\mathrm{D}}|$, $J_\mathrm{orb}$, $J_\mathrm{A}$,
and $J_\mathrm{D}$ from simulations Q0.5P$\_G1$ and Q0.5P$\_S1$.
Analogous information is displayed in the right panel of Figure 
\ref{fig:Q05PQ04P} 
from simulations  Q0.4P$\_G1$ (solid curves), Q0.4P$\_{G2}$ (dashed curves)
and Q0.4P$\_S1$.  In all of the Figure 
\ref{fig:Q05PQ04P}
plots, red curves present results from grid-code simulations
and blue curves present results from the $1M$-particle SPH-code simulations.

The Q0.5P and Q0.4P model simulations differ from the simulations
discussed earlier (Q1.3P and Q0.7P) in one especially
significant way: The binary separation, $a$, does not decrease
monotonically with time. Instead, $a$ decreases for only a
brief period of time associated with the initially imposed episode of
driving, then the system reaches some minimum separation,
$a_\mathrm{min}$, and $a$ increases monotonically thereafter.
After $a$ begins to increase, the {\it qualitative} behavior of
each system's subsequent evolution appears to depend on the
depth of contact that was established between the donor and its
surrounding Roche lobe during the phase of driving, as well as the
rate of mass transfer, $|\dot{M_\mathrm{D}}|$, exhibited at the end of
the phase of driving, which correlates with the
depth of contact.  Simulations in which
sufficiently deep contact is established display a
mass-transfer rate that continues to increase with time even as
the system separates; hereafter, these will be referred to as
``Case A'' evolutions. If, on the other hand, the donor established
mass-transfer rate grows for only a short period of time,
reaches a maximum rate $|\dot{M}_\mathrm{max}|$, then levels
off or declines as the system separates; these will be referred
to as ``Case B'' evolutions.

The dashed curves sketched in Figure 
\ref{fig:Q05PQ04Pgeneric}
illustrate in a qualitative fashion how other key parameters of the
Q0.5P and Q0.4P binary systems respond to a steadily increasing
mass-transfer rate in ``Case A'' evolutions; the solid curves
illustrate how these key
parameters respond to the ``Case B'' behavior.
In Case A evolutions ({\it e.g.}, simulation Q0.4P$\_S1$), the
system mass ratio drops at an ever increasing rate and the
accretor steadily spins up, principally extracting angular
momentum from the system's orbit.  After an initial phase
during which the donor also spins up --- perhaps related to the initial, brief
period of imposed driving --- the donor also steadily donates
angular momentum to the accretor.  In Case B evolutions ({\it e.g.}, simulation
Q0.4P$\_G1$), the accretor stops gaining spin angular momentum
and the curves that trace the behavior of $q(t)$,
$J_\mathrm{orb}(t)$, and $J_\mathrm{D}(t)$ each develop an
inflection point, indicating a slowing of the {\it rate} at
which $q$, $J_\mathrm{orb}$, and $J_\mathrm{D}$ are
declining.  In some Case B evolutions, angular momentum is
eventually returned to the donor.

Comparing the two grid-based evolutions Q$0.4\_G1$ and Q$0.4\_G2$, 
we find that the higher-resolution simulation Q$0.4\_G2$ has significantly
shallower slopes in all quantities plotted in Figure \ref{fig:Q05PQ04P}. The 
amplitude of epicyclic motion has been reduced and increasing the numerical
resolution by roughly a factor of 10 has reduced the initial mass transfer rate
by about a factor of two. One may have hoped for second-order spatial convergence
in the mass transfer rate but the edge of the star (where mass transfer is initiated) is 
not in a ``smooth'' region of the flow and only linear convergence, as we have found, is to 
be expected.  This fact offers additional evidence for the computational challenge of
mass transfer evolutions.
The peak mass transfer rate is similarly smaller in
the evolution Q$0.4\_G2$ compared to Q$0.4\_G1$ and in the higher
resolution simulation, the binary is still gently separating after evolving for more
than 50 orbits.

In an effort to better understand how the duration of imposed driving affects
the outcome of low $q_0$ model simulations, we again used both
numerical codes to conduct
polytropic simulations starting from initial model Q0.4 (see
Table \ref{tab:SCFparameters}) but subjected the system
to 1\% per orbit driving for only $1.16P_0$.  These are
identified as simulations Q0.4P$\_G3$ and Q0.4P$\_S2$ in Table
\ref{tab:Prescription}.
(Some results from simulation Q0.4P$\_G3$ already
have been presented as ``baseline simulation Q0.4A'' in \S3 of MTFD07;
note that MTFD07 mistakenly state that the duration of driving in this
model was $1.6P_0$.
A movie showing the first $1.0P_0$ of this ``baseline simulation'' was
published by MTFD07; here video11 carries the same simulation
through $47P_0$).
Key results from these two simulations are
displayed as {\it dashed} curves in Figure 
\ref{fig:Q04Pmerged}
--- the red and blue curves are, respectively, from the grid-code
and SPH-code simulations.  For comparison purposes, the {\it solid}
red and blue curves displayed in Figure 
\ref{fig:Q05PQ04P}b
have been  redrawn in Figure 
\ref{fig:Q04Pmerged}.  
Both of the SPH-code simulations
display characteristics
of a ``Case A'' evolution whereas both of the grid-code simulations
(the dashed and solid red curves) display characteristics
of a ``Case B''  evolution.

Based on the trends seen in Figure 
\ref{fig:Q04Pmerged},
it seems likely that, by subjecting
the initial model to a phase of driving that is even shorter than $1.16P_0$,
the SPH code would produce an evolution with ``Case B'' characteristics.
Alternatively, it seems likely that, by subjecting the initial model to
a phase of driving that is even longer than $1.6P_0$, the
grid code would produce an evolution with ``Case A'' characteristics.
Via a set of grid-code simulations, MTFD07 have previously provided
evidence (see especially their Figure 4) that, as the donor star
is initially driven into deeper and deeper
contact with its Roche lobe, a transition can be made from simulations
that exhibit ``Case B'' characteristics to simulations that display
``Case A'' characteristics.

Our analysis of Q0.4P and Q0.5P model simulations also suggests the
following.   If a simulation of binary mass transfer traverses a ``Case B'' evolutionary
trajectory, the donor may detach from its Roche lobe and, in so doing,
avoid merger or tidal disruption.  Three of the grid-code
simulations actually show the mass-transfer stream detaching from the
accretor late in the simulation:  Video07 shows the binary detaching
in simulation Q0.5P$\_G1$ at $t \approx 34P_0$; as video11 shows,
starting from the same initial state but with a shorter episode of driving,
the stream detaches later, at $46.5P_0$; and video09 shows
the binary detaching in simulation Q0.4P$\_G1$ at $t \approx 40P_0$.
The corresponding SPH simulations (Q0.5P$\_S1$ depicted in video08,
and Q0.4P$\_S1$ depicted in video12) do not confirm this outcome, but
both of these simulations were stopped before completing even twenty
orbits so in both SPH simulations we can, at best, claim that the ultimate fate of
the binaries is undetermined.

\subsubsection{Box- and Triangle-Shaped Resonances}
\label{subsec:Q0.4P}

As has already been mentioned, MFTD07 used the same grid code as we
have used here to analyze the behavior of Q0.4P simulations that were
subjected to a variety of driving rates and driving durations.  In \S4 of their
paper, MFTD07 point out that in the vicinity of the accretor some of the
models developed nonlinear-amplitude ``equatorial distortions with [azimuthal
mode numbers] $6 \geq m \geq 3$.'' The configuration displayed here in the
upper-right-hand corner of Figure 
\ref{fig:Q04Presonance} 
is precisely
the same configuration that was shown in the lower-right-hand corner of
MFTD07's Figure 3. It shows that, at time $t = 15.5P_0$ in MFTD07's
Q$0.4D$ simulation, the disk surrounding the accretor has a triangular shape.
As the image in the upper-left-hand corner of our Figure 
\ref{fig:Q04Presonance}
 shows, at time $t=13.9P_0$ in the same MFTD07 simulation, the disk
surrounding the accretor has a box shape; the time-evolutionary transition between
the box and triangular shapes is shown in the mpeg animation
of simulation Q0.4D that accompanies the MFTD07 publication.

The bottom two images in Figure 
\ref{fig:Q04Presonance} 
have been drawn
from our video12.  They illustrate that analogous box- and triangular-shaped
 distortions developed around the accretor in SPH simulation Q0.4P$\_S1$.
The box-shaped distortion appears at time $t=16.47P_0$ and the
triangular-shaped distortion appears at time $t=17.98P_0$. We were gratified
to see these azimuthal distortions develop spontaneously in simulation
Q0.4P$\_S1$. It demonstrates that our two entirely different numerical
techniques for simulating binary mass-transfer can encounter and follow
with comparable fidelity the development of unexpected nonlinear-amplitude
 structures in the midst of a complex binary interaction.

Box- and triangular-shaped distortions are not apparent in our grid-code
simulation (Q0.4P$\_G1$) that was initially driven at the same rate and
for the same duration as SPH-code simulation Q0.4P$\_S1$. Conversely,
a careful viewing of video07 shows development of box- and triangular-shaped
distortions in the grid-code simulation of model Q0.5P, whereas these
distortions are not apparent in the corresponding SPH simulation (model
Q0.5P$\_S1$; see video08).  We suspect -- as did MFTD07 -- that ``standing
wave'' azimuthal distortions of this type are routinely excited in polytropic
simulations as a result of dynamical interactions between the mass-transfer
stream and the accretion disk.  But the distortions do not grow to nonlinear
amplitude, and therefore are not visible to the eye, unless $|\dot{M}_\mathrm{D}|$
is sufficiently large.  In this light, it is not surprising that our eyes are unable to
detect box- or triangular-shaped distortions in video09 even though they
are present in  video12 because, for model Q0.4P, the mass-transfer
rate is always higher in the ``Case A'' SPH simulation than in the ``Case B''
grid-code simulation (see the Figure 
\ref{fig:Q05PQ04P}b 
plot of the
time-dependent behavior of $|\dot{M}_\mathrm{D}|$).  And it is not surprising
that our eyes are unable to detect nonlinear amplitude, standing-wave
distortions in video08 even though they are present in video07 because,
for model Q0.5P, the mass-transfer rate remains relatively low in the
SPH simulation whereas it grows to moderately high levels in the long, grid-code
simulation (see the Figure \ref{fig:Q05PQ04P} plot of the time-dependent
behavior of $|\dot{M}_\mathrm{D}|$).

\subsection{Ideal Gas Simulations}

Here we discuss and compare results from simulations that
were carried out using an ideal gas  (``I'') equation of state.
The relevant simulations are labeled as Q1.3I, Q0.7I,
Q0.5I, and Q0.4I along with a two-digit model identification suffix
in the last 10 rows of Table \ref{tab:Prescription}.  As with the polytropic
simulations, in each case the structure of the binary system at time
$t=0$ was prescribed by one of the initially uniformly rotating,
$n=3/2$ polytropic binary models identified in Table \ref{tab:SCFparameters}.
Also as was done in the case of the polytropic simulations, at the start
of a simulation, the donor star was driven into contact with its Roche lobe
by artificially draining angular momentum from the system at a rate of one
percent per orbit for the ``driving duration'' specified in column 5 of Table
 \ref{tab:Prescription}. Thereafter the system was evolved in a fully
 self-consistent fashion without further external influences. As is documented in
column 4 of Table \ref{tab:Prescription}, all but one of the SPH simulations
was carried out using 1 million ($1M$) particles; one model (Q0.7I$\_S1$)
used 100 thousand ($100k$) particles.  Every simulation performed with
the grid code used a uniform cylindrical grid containing approximately 4 million grid cells.

In all of the ideal gas simulations, one structural feature develops that
distinguishes each simulation from its polytropic counterpart:  During
the phase of mass transfer, a relatively hot, tenuous ``envelope'' develops
around the accretor.  This happens because the entropy of the material
increases as the donor material passes through the shock front that is
established where the accretion stream impacts the surface of the accretor.
(The shock front also develops in all of the polytropic simulations but, by
design, the entropy of the gas was not allowed to change.)  This low-density
envelope is easy to spot in the movie that accompanies each ideal gas
simulation, especially if it is viewed alongside the movie from its polytropic
counterpart; the lowest density, blue contour region is significantly extended
in each ideal gas simulation. The accretion stream is also noticeably thicker
(hotter) in each ideal gas simulation, relative to its polytropic counterpart.
And in some cases the hot tenuous envelope migrates over to the donor,
establishing a classic ``common envelope'' in which the two stars are
embedded.  The image displayed in the top, right-hand corner of Figure
\ref{fig:Q05PolyIdeal} 
has been extracted at time $t=17.75P_0$ from
video17, which shows the ideal-gas simulation of model Q0.5 conducted
with the grid code -- specifically, simulation Q0.5I$\_G1$. The common
envelope structure is very apparent when compared with a frame (top,
left corner of Figure 
\ref{fig:Q05PolyIdeal}) 
extracted from the middle
($t=14.0P_0$) of video07, which shows the polytropic simulation conducted
with the grid code that started from the same initial model (Q0.5) --
simulation Q0.5P$\_G1$. Similarly, the bottom two panels of Figure
\ref{fig:Q05PolyIdeal} 
 display, for comparison, images from the polytropic
simulation (video08; model Q0.5P$\_S1$) and the ideal gas simulation
 (video18; model Q0.5I$\_S1$) that started from initial model Q0.5 and
 were conducted using the SPH code.

 In ideal gas simulations performed with the grid code, there is a tendency
for a moderately extended ``blue'' atmosphere to develop around the binary components
within the first orbit.  This happens because, in the grid code, the region
surrounding the initial binary system is not actually a vacuum but, rather,
contains a very low density gas at a defined ``floor'' level.  Once a simulation
begins, this low-density gas free-falls down onto the donor, crashes into the
surface supersonically, and heats up as a result.  (The same dynamics occur
in each corresponding polytropic simulation but a hot donor atmosphere does not
result because, again by design, the entropy of the gas was not allowed to
increase as it crashed into the stellar surface.)  By contrast, an extended
atmosphere does not typically develop around the donor during the first
orbit of most SPH simulations because the region surrounding the binary
system initially contains no material.

\subsubsection{{\rm Q1.3I} Model Simulations}
\label{subsec:Q1.3I}

We carried out two separate ideal-gas simulations of binaries that had an
initial mass ratio $q_0 = 1.323$.  In both cases the model
was driven at a rate of 1\% per orbit for $2.0 P_0$ in order to initiate mass transfer.
Whether the simulation was conducted using the grid code (model $G1$)
or the SPH code (model $S1$), the Q1.3I simulation was found
to be dynamically unstable to mass transfer.   As Figure
\ref{fig:Q13idealA} 
illustrates,
throughout both simulations, the binary separation ($a$)
and system mass-ratio ($q$) steadily decreased while the donor
mass-transfer rate ($|\dot{M}_\mathrm{D}|$) steadily increased. As is
shown in the accompanying movies -- video13 from simulation Q1.3I$\_G1$,
and video14 from simulation Q1.3I$\_S1$ --  independent of the chosen numerical
technique the two stars ultimately merged violently in the
sense that the highest density ``core'' region of the donor was
observed to collide and combine with the highest density
``core'' region of the accretor.

There is very little need for us to elaborate further regarding the behavior
and outcome of these simulations because they resemble very closely
our polytropic simulations that started from the same initial models.
The time-dependent behavior of the six system
parameters displayed in Figure 
\ref{fig:Q13idealA} 
is very similar to the
behavior of the same parameters as displayed in Figure 
\ref{fig:Q13evolutionA}
from the corresponding polytropic simulations; and the ``merger'' images
shown in Figure 
\ref{fig:Q13idealB} 
from simulations Q1.3I$\_G1$ (top) and
Q1.3I$\_S1$ (bottom) are nearly indistinguishable from the corresponding merger
images that were drawn from our polytropic simulations and shown earlier
in Figure 
\ref{fig:Q13evolutionB}.  
Upon close examination, merger images
from the ideal gas simulations show a somewhat more extended ``blue''
envelope than in the polytropic case.  But this tenuous atmosphere  did
not significantly influence the overall dynamics of the merger.

It is worth pointing out that, while the merger time was nearly identical
for simulations Q1.3I$\_S1$ and Q1.3P$\_S1$ -- that is, for both the ideal-gas
simulation and the polytropic simulation as produced by the SPH code --
the merger times for simulations Q1.3I$\_G1$ and Q1.3P$\_G1$ differed
by more than $3P_0$.  This is reflected in, for example, a longer period of
time for the mass transfer rate to grow in the grid evolution shown in Figure 
\ref{fig:Q13idealA} 
compared to the polytropic evolution in Figure \ref{fig:Q13evolutionA}.
This delay is likely related to the moderately extended,
tenuous ``blue'' atmosphere that developed in the star's Roche lobes in the grid-code,
but not the SPH, simulation.  A tenuous atmosphere appears early in video13
but is absent in the polytropic simulation shown in video01 and SPH clearly
does not suffer from this artifact in either polytropic (video02) or ideal gas (video14)
evolutions.
With the relatively long driving time required to bring the donor in to contact with its
Roche lobe, the simulation Q$1.3I\_G1$ is bringing a lower density, higher temperature
atmosphere into contact as compared to the polytropic simulation Q$1.3P\_G1$.

\subsubsection{{\rm Q0.7I} Model Simulations}
\label{subsec:Q0.7II}

As part of our comparison, we conducted two simulations of the binary with an initial
mass ratio of $q_{0} = 0.7$  and as with the Q0.7P simulations both numerical
techniques find that the mass transfer grows with time leading to a tidal instability
that disrupts the donor star. As detailed in Table \ref{tab:Prescription}, the SPH simulation
was carried out with a resolution of $100k$  particles and was driven into contact for only
$1.0 P_{0}$ while the grid simulation was conducted with approximately 4 million grid cells
and was driven into a deep contact with a driving period of $2.28 P_{0}$.

The effects of a lower resolution in the SPH simulation Q0.7I$\_S1$ are reflected in
Figure 
\ref{fig:Q07idealA}
 in
both the higher epicyclic amplitude of the orbit and the higher mass transfer rate initially.
As we have seen previously, the higher mass transfer rate dictates a shorter time to
merger for the Q0.7I$\_S1$ compared to the Q0.7I$\_G1$ simulation but once we have
adjusted the time series of data from Q0.7I$\_S1$ with a zero point shift of
$t_{zpt} = 2.33 P_{0}$ both numerical techniques yield a very similar solution for the
tidal instability that destroys the donor star. Images drawn from these simulations during
tidal instability are shown in Figure \ref{fig:Q07idealB}. 

As in the Q$1.3I$ discussion, the SPH code gives nearly identical results regardless of the
assumed equation of state. For the grid code, the results are more similar to the polytropic
evolutions and noticeably, there is no delay in the time to merger and the mass transfer rate
does not stall in the ideal gas simulation. One difference that can be discerned in the grid-based
evolution Q$0.7I\_G1$ is that the spin angular momentum of the donor star remains constant
until the donor is disrupted whereas in the polytropic evolutions Q$0.7P\_G1$ and 
Q$0.7P\_G2$, the donor's spin angular momentum declines gently before rising steeply
as the donor star is disrupted.

\subsubsection{Model Simulations {\rm Q0.5I} and {\rm Q0.4I}}
\label{subsec:Q0.5IandQ0.4I}

To complete our testbed of simulations, we have repeated simulations of the two
low mass ratio models with an ideal gas equation of state yielding the six simulations
Q0.5I$\_G1$, Q0.5I$\_S1$, Q0.4I$\_G1$, Q0.4I$\_G2$, Q0.4I$\_S1$ and Q0.4I$\_S2$.
The simulations of the Q0.5 model offer a good representation of a DD system as both
components in the initial data share the same value of the polytropic constant 
(see Table \ref{tab:SCFparameters}).
As the donor star was
relatively far from contact initially, the system was driven for $2.7 P_{0}$ 
in both the SPH and the grid-code simulations.

The SCF model Q0.4a is somewhat more complicated as the polytropic
constants for the donor and accretor are not equal and the donor star is too large
(compared to what would be the case with a zero-temperature white dwarf 
equation of state).  The relatively
large volume of the donor's Roche lobe that is occupied by the star has important
consequences for the outcome of the ideal gas simulations of this model.  We have run
this model four times, twice with each code where we have varied the initial depth of
contact and hence the initial mass transfer rate.  The simulations Q0.4I$\_G1$ and
Q0.4I$\_S1$ were driven for $1.6 P_{0}$ to establish contact while the simulations
Q0.4I$\_G2$ and Q0.4I$\_S2$ were only driven for $1.16 P_{0}$ to establish a more
shallow initial contact.

Including the ideal gas equation of state introduces a new phenomenon into simulations
of binaries with initial mass ratios below the cutoff for dynamically unstable mass
transfer, $q_{\rm crit}$. Namely, the accreting star will expand to the point that it also
fills its Roche lobe forming a common envelope about both stars with the possibility
that mass is lost from the binary through the outer Lagrange points (see the movies
video17 and video18 as well as the still images in the 
bottom row of
Figure
14). 
The incorporation of
shock heating in the ideal gas equation of state introduces an additional, spurious
effect in grid-based simulations.  In an Eulerian treatment, no grid cell is allowed to
be completely empty and a floor level of the mass density is maintained artificially
even in regions that should clearly be vacant.  At the start of a simulation, this material
will begin to free-fall onto the binary system and shock heat as it crashes into the stars,
heating their outer layer.
As can be seen
in the plot of mass transfer rate for the Q0.5I simulations in the left hand panel of
Figure 
\ref{fig:Q05ideal}, 
the grid simulation Q0.5I$\_G1$ emerges from the driving phase with a mass
transfer rate that is approximately an order of magnitude higher than the SPH simulation
Q0.5I$\_S1$.    The hotter envelope for the donor star in the grid-based simulation means
that the common envelope phase where the separation varies about a constant mean
value is longer than in the SPH simulation Q0.5I$\_S1$.

The quasi-Lagrangian formulation in SPH does not suffer from the ``vacuum'' effects
that arise in the grid code.  In fact, there are indications that for the Q0.5I simulations
that the SPH code is approaching a different solution than the common envelope
followed by tidal instability found with the grid-based simulation Q0.5I$\_G1$.  In the
grid-based simulation a common envelope phase is established where the mass
transfer rate grows with relatively little change in the orbital separation or the
distribution of angular momentum between the orbital motion and the spin of each
component.  This phase is followed by a rapid tidal instability where the orbital angular
momentum plummets while the spin angular momentum of each star grows rapidly.
In the SPH evolution of this model there is still a common envelope phase but the
evolution seems to be approaching the tidal disruption of the donor star, not a tidal
instability.

A similar pattern can be seen in the ideal gas simulations of the Q0.4a initial data.
The right hand panel of Figure 
\ref{fig:Q05ideal} 
shows data for the two simulations
with a greater initial depth of contact (corresponding to a longer driving period of
$1.6 P_{0}$).  The grid simulation Q0.4I$\_G1$ (video19) is very short lived, the mass
transfer rate grows rapidly and tidal instability destroys the donor star within approximately
12 orbits.  In part this is due to the numerical artifact of the vacuum discussed previously
as well as the relatively large fraction of the accretor's Roche lobe that is occupied initially
(meaning that a common envelope forms rapidly and mass exchange and mass loss
effects set in rapidly).  For the same initial driving period of $1.6 P_{0}$ in simulation
Q0.4I$\_S1$ (video21), we again find a common envelope phase followed by tidal
instability.  Note however that without the artificial heating of the envelope, the SPH
simulation is qualitatively similar to the polytropic simulation.  The same is not true
for our grid-based simulations however.

As shown in Figure 
\ref{fig:Q04Imerged}, 
reducing the initial driving period extends
the life of the binary for both codes.  Please note that as described in the appendix, the
mass transfer rate that is plotted is smoothed with a boxcar function with a width of
three initial orbital 
periods meaning that mass transfer rate at the end of driving is not represented
for the Q0.4I$\_G2$ simulation in Figure 
\ref{fig:Q04Imerged}.  
With a shallower
initial depth of contact, the grid-based code now finds tidal instability of the donor
occurring later in the simulation.  With the SPH code however, the mass transfer
rate has turned over -- perhaps indicating that it may avoid destruction of the
donor star and reach detachment with a longer run.

Figure 
14
shows several time slices around the tidal
instability event 
from the simulations Q0.4I$\_S1$ and Q0.4I$\_G1$.  While the
initial mass transfer rates and subsequent time scales are different between the
two runs, there is detailed agreement between the structure of the common
envelope, the tidal disruption of the donor star and the resulting disk of debris
wrapped around the accretor.

\section{Summary and Conclusions}
\label{sec:summary}

For reference, we have summarized our conclusion for the fate of each simulation
(when it could be determined given existing data)  in the seventh column of
Table \ref{tab:Diagnostics}. Possible outcomes range from the donor star detaching from its
Roche lobe, the destruction of the binary through tidal instability (the spin angular momentum of
each component grow rapidly while the orbital angular momentum crashes), and the tidal disruption of
the donor star (where the spin angular momentum of the donor remains relatively constant while
the donor is tidally shredded).  
These outcomes are in some instances modified by the formation of a common
envelope phase which can be identified as relatively steady orbital separation while the 
mass transfer rate grows.

The standard linear analysis of the dynamical stability of mass transfer
in binaries with spherical components, in the 
Roche approximation, and driven by angular momentum losses (AML)
\citep{1982ApJ...254..616R, 1997A&A...327..620S,2004MNRAS...350..113M, 2007ApJ...655.1010G}, reveals two critical values of the mass ratio 
$q=M_{\rm D}/M_{\rm A}$: 
1) if $q > q_a$, the effect of the mass transfer by itself will tend to reduce the binary separation, whereas it will tend to expand the binary otherwise; 
2) if $q<q_\mathrm{crit}$, the mass transfer is stable and will tend toward a 
secular value proportional to the AML rate, whereas the mass transfer will be 
unstable otherwise and grow dynamically for $q > q_{\mathrm{crit}}$. 

In the simplest case of conservative 
mass transfer (no loss of mass from the system) and neglecting the effects of tidal 
distortions and spin-orbit coupling, $q_a=1$, and as already stated earlier, for donors 
obeying a polytropic EOS with $n=3/2$, $q_\mathrm{crit}=2/3<q_a$. 
Therefore, based on these critical mass ratios, the expected evolutionary 
behavior of interacting,
polytropic binary systems can be divided into three broad categories,
depending principally upon whether the system's initial mass ratio $q_0$ falls above
$q_a$, in between $q_\mathrm{crit}$ and $q_a$, or below  $q_\mathrm{crit}$. 
If $q_0>q_a$, both AML and mass transfer contribute to shrinking the binary separation
and driving the donor into deeper contact. Thus we expect the mass-transfer rate to increase monotonically while the separation decreases monotonically, probably 
leading to a merger.
If $q_\mathrm{crit}<q_0<q_a$, AML will shrink the binary initially, but once 
contact is established, the dynamically unstable mass transfer is expected to grow, 
halt the contraction and then rapidly increase the separation. This behavior resembles closely the ``Case A'' evolutions of \S \ref{subsec:Q0.5PandQ0.4P}.
As the mass and density of the donor decrease, we may expect that
the donor material is tidally wrapped around the accretor.
Finally, if $q_0<q_\mathrm{crit}$, after AML shrinks the binary into contact, the 
mass transfer will settle toward the expected secular equilibrium value. 
If, as in our case, $q_\mathrm{crit}<q_a$, the expectation is that the orbit
will settle into a secular expansion, with the donor, and thus the binary, surviving. 
This behavior resembles in part the ``Case B'' evolutions of \S \ref{subsec:Q0.5PandQ0.4P}.

Before attempting to understand our simulations in terms of the above simple 
classification, we must note some important differences between the assumptions
made in the standard treatments and the conditions prevalent in the simulations. 
First, driving by AML is only applied initially for varying lengths of time and then switched off. This results in varying degrees of contact and mass-transfer rates at the point AML ceases. Thereafter, the system's evolution is driven by instabilities and any residual
low-level angular momentum losses or gains due to numerical artifacts or mass loss from the system. Second, 
tidal distortions and spin-orbit coupling occur naturally during the simulation and 
these dominate the simulation especially at late stages. 

The numerical simulations described in this work demonstrate that systems in the first category (relatively large $q_0>q_a$), following the
onset of mass transfer, are dynamically unstable from the outset.  
As the binary separation decreases,
the mass-transfer rate monotonically increases and, in a finite time, the stellar
 components either merge or the donor tidally disrupts.
On the other hand, once
mass-transfer has been initiated in systems that have a relatively small $q_0<q_a$,
the binary separation will generally increase with time.
Our two distinctly
 different CFD algorithms agree on the evolutionary outcome of these
 dynamically unstable systems.  For example, in every one of our Q1.3
simulations, the cores of the two stars merge; whereas, in every Q0.7
simulation, the donor tidally disrupts and material from the donor disperses
to form a rotationally flattened disk around the accretor. As is illustrated by
Figures 
\ref{fig:Q13evolutionB} 
and 
\ref{fig:Q07evolutionB}  
(and their
associated animation sequences), this agreement is more than qualitative:
detailed structures that arise during   merger or tidal disruption events are
present in simulations that are conducted with both codes.

Highly non-linear, transient structures such as the box- and triangle-shaped
resonances in the Q$0.4a$ polytropic evolutions have emerged in simulations
with our two disparate codes as highlighted in Figure \ref{fig:Q04Presonance}. 
These transitory equatorial distortions indicate that both
techniques are yielding solutions remarkably close to one another in the
parameter space of possible numerical solutions despite their different error terms,
numerical artifacts, and fundamental representations of a fluid state. These resonance
features represent a non-trivial verification of our codes.

For a given model, our two different CFD codes generally do not agree on the
precise time of merger (or of tidal disruption), $t_\mathrm{merge}$.  But as is
summarized in Table \ref{tab:Diagnostics}, even the same code will produce
different values of $t_\mathrm{merge}$ when a simulation is repeated at
a different code resolution or with a different driving duration.  These apparent
discrepancies arise because the precise value of $t_\mathrm{merge}$ is
sensitive to the depth of contact that the Roche lobe makes with the donor
star following the initial phase of driving.  All other things being equal, a
relatively brief phase of driving will result in relatively shallow contact, in
which case the binary simulation will begin with a low mass-transfer rate
and it will take the system a relatively long time to merge.  If the phase of
driving lasts longer, the simulation will begin with a higher mass-transfer rate
and, as a result, the time to merger will be shorter. For a given driving duration,
we have found that the time to merger is shorter in an SPH-code simulation
than in a grid-code simulation.  In effect, this implies that for a specified driving
duration the Roche lobe establishes deeper contact with the donor star in the
SPH code than in the grid code. We must emphasize, however, that the ultimate
outcome of these simulations is not sensitive to code resolution or to driving
duration.  Also, as stated earlier, the outcome does not depend on the selected
CFD algorithm:  If the chosen binary system is dynamically unstable toward
merger (or to tidal disruption of the donor), all simulations ultimately converge
to the same solution.

Our collection of simulations demonstrates that the evolutionary outcome
of systems in the third category (relatively small $q_0<q_\mathrm{crit}$) {\it is} sensitive
to the depth of contact that the Roche lobe makes with the donor star
during an initial phase of driving. (See especially the discussion associated
with Figure 
\ref{fig:Q05PQ04Pgeneric}.) 
If, for a given $q_0$, the depth of
contact is sufficiently deep, then the mass-transfer rate will monotonically
increase even as the binary separation is increasing and, ultimately, the
donor star will tidally disrupt.  This is what has been referred to in
\S \ref{subsec:Q0.5PandQ0.4P} as ``Case A'' evolutions.  If, however, the
depth of contact is initially sufficiently shallow, a ``Case B'' evolution will
result, that is, the mass-transfer rate will eventually reach a maximum and,
thereafter, will level off or decrease as the binary separation increases.
For a specified $q_0$ and driving duration --- at least within the  range
of parameters we have explored in this low $q_0$ category, using
a polytropic equation of state  --- simulations
conducted with the SPH code tend to result in ``Case A'' simulations whereas
simulations conducted with the grid code tend to result in ``Case B''
simulations.   As a consequence, we find that  the donor star tends to tidally
disrupt in simulations conducted with the SPH code whereas the donor
tends to survive in simulations conducted with the grid code, at least for
the length of time we've been able to follow the simulations. A given system
can therefore appear to be either unstable or stable, depending on the
CFD algorithm that has been selected to perform the simulation. This is not
a particularly desirable result.  But it is understandable and it is consistent
with the behavior deduced from our analysis of relatively large $q_0$
simulations. For a specified driving duration the Roche lobe typically
establishes deeper contact with the donor star in the SPH code than in
the grid code.  In turn, deeper contact will tend to result in a ``Case A''
rather than a ``Case B'' evolution.

When we compared polytropic and ideal gas simulations of the same initial model 
we noted the formation of a low-density hot envelope and the greater thickness of 
the stream both as a result of the higher temperature of the gas in the ideal gas case. 
However, while the higher temperature of the envelope resulting from the accretion shock appears both in the SPH and grid simulations and is likely a true physical effect, the dissipation of energy due to ``vacuum" material settling on the binary during the first orbit
is only present in grid simulations and should be considered a spurious effect. The spurious heating generates early on a hotter, thicker atmosphere, which allows a wider stream and enhances the mass transfer. 
This tends to speed up the simulation, sets it on a more unstable pathway and 
makes it difficult for the system to return to stability when the donor is near contact initially. 
This has little or no effect on the ultimate fate of the Q1.3I and Q0.7I models, 
but it does affect the simulation of the models Q0.5I and Q0.4I.
In particular, the higher initial mass transfer rate in the grid simulations for the Q0.4I model
results in a faster tidal disruption for the grid simulations than in the SPH simulations.
This is in contrast to most of the other cases presented here, where the mass transfer
 rate in SPH simulations
 is higher yielding also a faster final outcome.

An obvious difficulty in the interpretation of our low $q_0$ simulations 
arises in connection with the uncertainty in the effective value of 
$q_\mathrm{crit}$ because of the noted differences between the 
assumptions made in analytic treatments and the ``reality" of the simulations. 
Since direct impact occurs in our simulations, but also disk-like flows arise 
around the accretor, we may guess that the effective value is somewhere 
in the range $0.22<q_\mathrm{crit}<2/3$. However, tidal effects are obvious
in the images and movies presented here, and our simulations suggest
that the ultimate fate of the low $q_0$ systems may well depend on the 
depth of contact achieved in the initial driving into contact.
The levels of mass transfer in the simulations exceed the likely levels in real systems by many orders of magnitude. This may not be important for systems that are expected to be dynamically unstable, but caution should be exercised when interpreting results for marginally unstable systems.

In summary, we conclude that aside from understandable shifts in the times 
taken by the same models evolved with SPH and grid hydrodynamics to reach the 
same evolutionary stages, and the split into Case A/B evolutions for low $q_0$ systems, 
the agreement between the results produced by our codes is remarkably close
in the final outcomes and during intermediate evolutionary stages.

\acknowledgments We acknowledge valuable interactions that
we have had with Dominic Marcello, Kundan Kadam, and Zachary Byerly.
In addition, we would like to thank the anonymous referee for their very careful and
thoughtful
review of our manuscript.
This work has
been supported, in part, by grants AST-0708551, 
NSF CREATIV grant AST-1240655, and
DGE-0504507 from the U.S. National Science Foundation and,
in part, by grant NNX07AG84G from NASA's ATP program.  This
research also has been made possible by grants of
high-performance computing time on the TeraGrid
(MCA98N043), at LSU, and across LONI (Louisiana Optical
Network Initiative). This material is based upon work supported by the National Science Foundation under Grant No. ACI-0338618l, OCI-0451237, OCI-0535258, and OCI-0504075.
This research was supported in part by the Indiana METACyt Initiative. The Indiana METACyt Initiative of Indiana University is supported in part by Lilly Endowment, Inc.
This work was supported in part by Shared University Research grants from IBM, Inc., to Indiana University.
This material is based upon work supported by the National Science Foundation under Grant No.\ CNS-0521433.

\newpage
\appendix
\section{Diagnostics}\label{app:Diagnostics}

\begin{figure*}
\centering
\figurenum{3a}
\scalebox{0.8}{\plotone{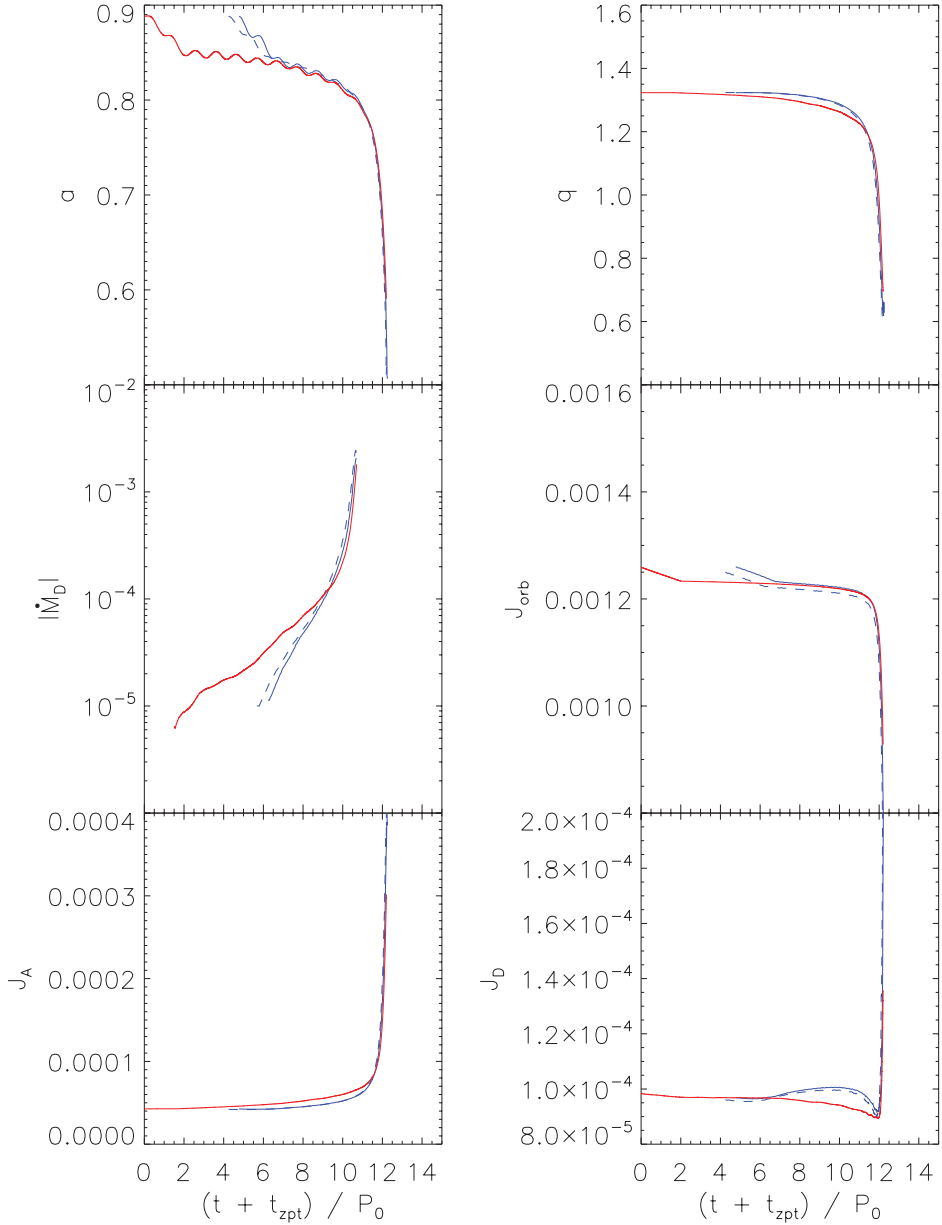}}
\figcaption[Q13Lines]{Simulation Q1.3P. Curves show the time-dependent 
behavior of six system parameters --- orbital separation, $a$, binary mass 
ratio, $q$, log of the donor mass-transfer rate, $|\dot{M}_\mathrm{D}|$, 
orbital angular momentum, $J_\mathrm{orb}$, and spin angular momentum 
of the accretor, $J_\mathrm{A}$, and donor, $J_\mathrm{D}$ --- as derived 
from models Q1.3P$\_G1$ (solid red curves), Q1.3P$\_S1$ (solid blue curves), 
and Q1.3P$\_S2$ (dashed blue curves). For purposes of presentation, each 
curve has been terminated at a time shortly after the system mass ratio has 
dropped below $q=0.7$ and curves from both SPH-code simulations include 
a zero-point shift in time, $t_\mathrm{zpt}$, as detailed in column 4 of Table \ref{tab:Diagnostics}; units used along the vertical axis for each plot are 
defined in Appendix \ref{app:Diagnostics}.} 
\label{fig:Q13evolutionA}
\end{figure*}

\begin{figure*}
\centering
\figurenum{3b}
\scalebox{0.8}{\plotone{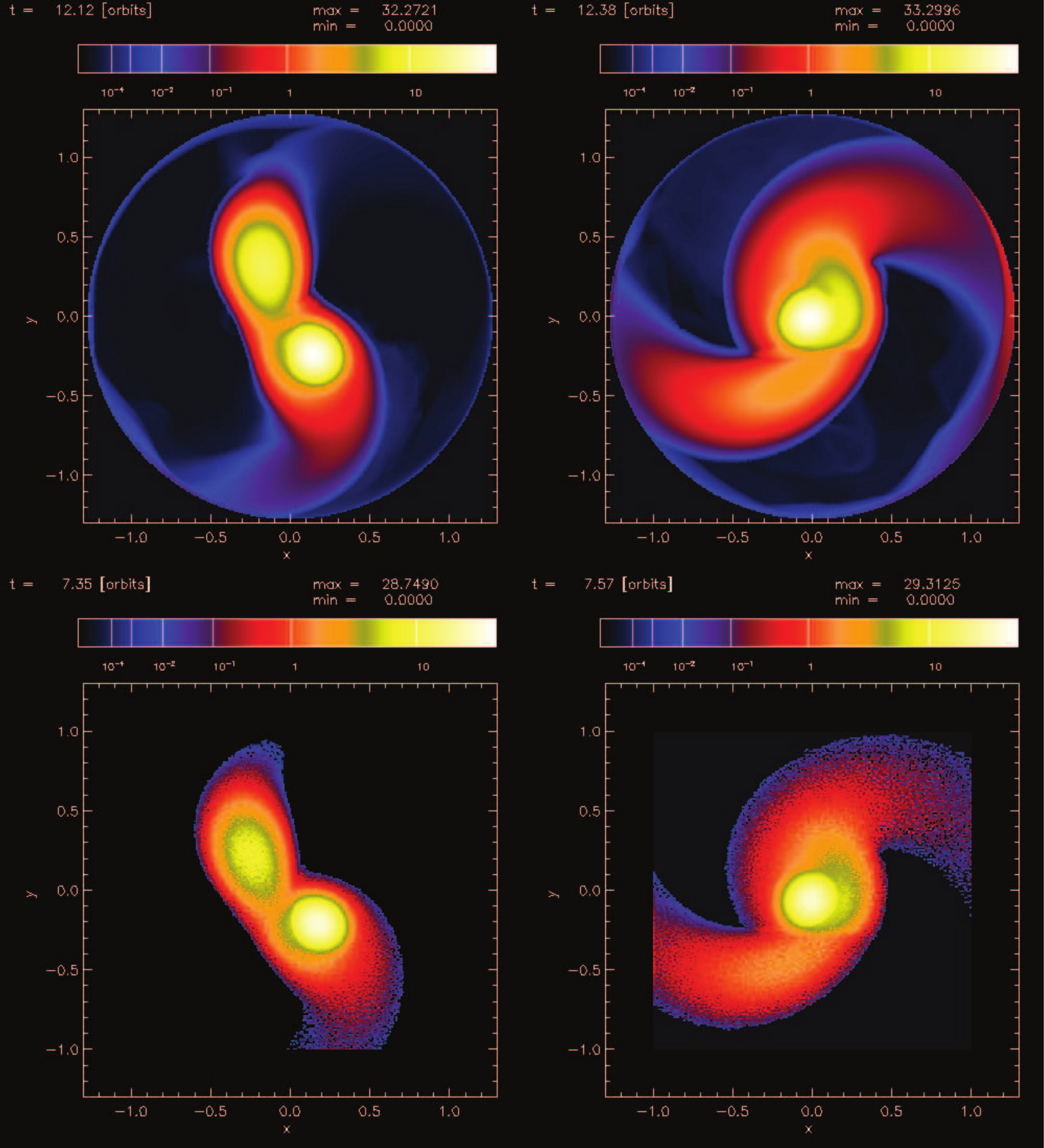}}
\caption[Q13Contours]{Simulation Q1.3P. Images show contours of 
equatorial-plane column densities from model Q1.3P$\_G1$ (top) 
and model Q1.3P$\_S1$ (bottom) at two different points in time as 
the binary merger is occurring --- specifically, {\it left-most images}: 
$t_{G1} = 12.12P_0$ and $t_{S1} = 7.35P_0$; {\it right-most images}: 
$t_{G1} = 12.38P_0$ and $t_{S1} = 7.57P_0$. The top two images 
have been extracted from video01 and the bottom two have been
extracted from video02.
} 
\label{fig:Q13evolutionB}
\end{figure*}

In an effort to quantitatively compare results from our various simulations
of binary mass-transfer events, we have tracked the time-evolutionary
behavior of a set of six key binary system parameters:  Orbital separation,
$a$, binary mass ratio, $q$, mass-transfer rate of the donor, $\dot{M}_\mathrm{D}$,
orbital angular momentum, $J_\mathrm{orb}$, and spin angular momentum
of the accretor, $J_\mathrm{A}$, and donor, $J_\mathrm{D}$.  For each set of model simulations
--- for example, the simulations labeled Q1.3P that are discussed in
\S \ref{subsec:Q1.3P} and illustrated in Figure 
\ref{fig:Q13evolutionA} 
--- the time-evolutionary behavior of each parameter has been illustrated
by showing  curves of the variation of each parameter versus time.
The units that have been used in each plot are as follows: $q$ is
naturally dimensionless in both codes; $a$, $J_{\mathrm{orb}}$, $J_\mathrm{D}$,
and $J_\mathrm{A}$ are in code units (see Table \ref{tab:SCFparameters}); time
is normalized to the binary system's original orbital period, $P_0$;
and $\dot{M}_\mathrm{D}$ has been normalized by a mass transfer rate
corresponding to the initial mass of the donor in the time span of the
initial orbital period, that is,
\begin{equation}
\dot{M}_\mathrm{norm} \equiv \frac{M_\mathrm{D}(t=0)}{P_0} \, .
\end{equation}
For each model, the value of this normalization quantity also can be
determined from the information provided in Table \ref{tab:InitialModels}.

The simulation data from both numerical techniques were analyzed with
the same tools in an effort to ensure that results from both numerical
schemes were viewed through the same quantities. We chose to use
diagnostic tools that were initially developed to analyze data sets from the
cylindrical coordinate-based grid-code simulations.  In order to enable analysis
of the SPH simulations with this same set of diagnostic tools, data sets from
the SPH simulations for mass, velocity, etc. were mapped onto a Cartesian
grid (using a nearest grid point algorithm).
Time series data for the six key variables were then computed for each
simulation as follows.

The maximum density location for each star is identified and the line of
centers of the binary is assumed to run through the two density maxima.
A plane is then constructed parametrically that is perpendicular to the line
of centers and divides the computational domain into two regions (donor
and accretor) near the inner Lagrange point to enable identification of
fluid elements that belong to the donor or accretor. The mass of each
component and the binary separation are then computed from simple
volume integrals of the density and density-weighted position of each
star.   The volume integrals extend over the entire grid region ascribed as
donor or accretor based on the dividing plane.
The momentum and mass within a given grid cell are combined to give the three
components of velocity for the material in each cell.
The orbital angular momentum is then calculated from the center of mass
position and velocity of the two stellar components
while the spin angular momentum of
each component is computed by referencing velocities to the relevant
center of mass.  The mass transfer rate is computed as the centered
finite difference of the mass of the donor star with respect to time and
is quite noisy.  The mass transfer rates shown in the figures are smoothed
using a sliding box car average with a width equal to three times the
initial orbital period, $P_{0}$.  Hence the first data point for the mass  transfer rate
shown is at $1.5 P_{0}$ after the start of the simulation.

\bibliographystyle{apj}
\bibliography{ms}


\newpage
\begin{deluxetable}{llllllllllclccc}
   \tablewidth{0pt}
   \tabletypesize{\scriptsize}
   \tablecaption{Initial (Polytropic) Binary System Parameters
   \label{tab:InitialModels}
   \label{tab:SCFparameters}}
   \tablehead{ \colhead{Model} &
                      \colhead{q} &
                      \colhead{$\kappa_\mathrm{D}$} &
                      \colhead{$\rho_{0,\mathrm{D}}$} &
                      \colhead{$R_{\mathrm{D}}$} &
                      \colhead{$\kappa_\mathrm{A}$} &
                      \colhead{$\rho_{0,\mathrm{A}}$} &
                      \colhead{$R_{\mathrm{A}}$} &
                      \colhead{$\frac{R_{\mathrm{A}}}{R_{\mathrm{circ}}}$} &
                      \colhead{$M_{\mathrm{tot}}$} &
                      \colhead{$a_{\mathrm{sep}}$} &
                      \colhead{$P_{0} = \frac{2 \pi}{\Omega_{0}}$} &
                      \colhead{Analog \tablenotemark{a}} &
                      \colhead{Notes \tablenotemark{b}} }
   \startdata
      Q1.3 & $1.323$  & $0.0372$ & $0.600$  & $0.3508$ & $0.0264$  & $1.000$ &
                  $0.2674$ & $2.602$ & $3.090\times 10^{-2}$ & $0.8882$ & $29.74$ & MS & (1) \\
      Q0.7a & $0.700$  & $0.02512$ & $0.608$ & $0.2865$ & $0.0273$ & $1.000$ &
                    $0.2694$ & $2.302$ & $2.371\times 10^{-2}$ & $0.8394$ & $31.20$ & WD & $\cdots$ \\
   Q0.7b & $0.69985$  & $0.02$ & $0.6$ & $0.2383$ & $0.02$ & $1.3$ &
                 $0.2703$ & $2.584$ & $1.700\times 10^{-2}$ & $0.75$ & $31.70$ & WD & $\cdots$ \\
   Q0.5  & $0.500$  & $0.016$   & $0.235$ & $0.2670$ & $0.016$   & $1.000$ &
                $0.2052$ & $1.496$ & $9.216\times 10^{-3}$ & $0.8764$ & $53.52$ & WD & (2) \\
   Q0.4a  & $0.4085$ & $0.01904$ & $0.710$  & $0.2898$ & $0.03119$ & $1.000$ &
                $0.2437$ & $1.768$ & $2.399\times 10^{-2}$ & $0.8169$ & $29.75$ & WD & (3) \\
   Q0.4b & $0.4203$ & $0.01989$ & $0.740$ & $0.2958$ & $0.03222$ & $1.000$ &
               $0.2479$ & $1.811$ & $2.550 \times 10^{-2}$ & $0.8207$ & $29.05$ & WD &  $\cdots$ \\
   \enddata
   \tablenotetext{a}{Stellar analog of polytropic binary where MS means fully-convective, low mass, main sequence stars and WD means a white dwarf where both components are low in mass compared to the Chandrasekhar limit.}
   \tablenotetext{b}{Information drawn from: (1) Table 4 of DMTF06; (2) Table
5 of DMTF06; (3) Table 1 of MFTD07.}
\end{deluxetable}
\clearpage

\newpage
\begin{deluxetable}{lclccc}
   \tablewidth{0pt}
   \tablecolumns{6}
   \tablecaption{Simulation Prescriptions.
                         \label{tab:Prescription}}
   \tablehead{
                      \colhead{Simulation\tablenotemark{a}} &
                      \colhead{Model} &
                      \colhead{Code} &
                      \colhead{Resolution\tablenotemark{b}} &
                      \colhead{Driving} &
                      \colhead{Notes\tablenotemark{c}} \\
                      \colhead{} &
                      \colhead{ID} &
                      \colhead{ } &
                      \colhead{ } &
                      \colhead{Duration} &
                      \colhead{ } \\
                      \colhead{(1)} &
                      \colhead{(2)} &
                      \colhead{(3)} &
                      \colhead{(4)} &
                      \colhead{(5)} &
                      \colhead{(6)} }
   \startdata
   Q1.3P       & $G1$ & Grid   & $4$ M      & $2.0 P_0$ & (1) \\
                    & $S1$  & SPH   & $1$ M      & $2.0 P_0$ &\\
                    & $S2$  & SPH   & $100$ k   & $2.0 P_0$ &\\
   Q0.7P       & $G1$ & Grid   & $4$ M    & $2.28 P_0$&   \\
                    & $G2$ & Grid   & $4$ M          & $1.70 P_0$ &  \\
                    & $S1$ & SPH    & $100$ k    & $1.0 P_0$ &\\
   Q0.5P       & $G1$ & Grid   & $4$ M    & $2.7 P_0$ & (2) \\
                    & $S1$ & SPH    & $1$ M         & $2.7 P_0$ &\\
   Q0.4P       & $G1$ & Grid   & $4$ M    & $1.6 P_0$ & \\
                    & $G2$ & Grid   & $47$ M           & $1.6 P_0$ & \\
                    & $G3$ & Grid   & $4$ M           & $1.16 P_0$ & (3)\\
                    & $S1$ & SPH    & $1$ M         & $1.6 P_0$ &\\
                    & $S2$ & SPH    & $1$ M         & $1.16 P_0$ &  \\
    Q1.3I       & $G1$ & Grid   & $4$ M     & $2.0 P_0$ &  \\
                    & $S1$ & SPH    & $1$ M         & $2.0 P_0$ &\\
    Q0.7I       & $G1$ & Grid   & $4$ M            & $2.28 P_0$&   \\
                    & $S1$ & SPH    & $100$ k      & $1.0 P_0$ &  \\
    Q0.5I       & $G1$ & Grid   & $4$ M       & $2.7 P_0$ &  \\
                    & $S1$ & SPH    & $1$ M      & $2.7 P_0$ &  \\
    Q0.4I       & $G1$ & Grid   & $4$ M       & $1.6 P_0$ &  \\
                    & $G2$ & Grid   & $4$ M       & $1.16 P_{0}$ & \\
                    & $S1$ & SPH    & $1$ M      & $1.6 P_0$ &\\
                    & $S2$ & SPH    & $1$ M      & $1.16 P_{0}$ & \\
   \enddata
   \tablenotetext{a}{\small{Simulations labelled with ``P'' were evolved with a polytropic equation of state while simulations labelled with ``I'' utilized an ideal gas equation of state.}}
   
   \tablenotetext{b}{\small For SPH simulations ``resolution'' means
number of particles.  For grid simulations, ``resolution'' means
number of uniform cylindrical grid cells; $(N_R,N_\theta,N_Z)
= (162,256,98)$, hence, $4.064$M cells.}

\tablenotetext{c}{\small Identical to model simulation (1) Q1.3-D,
first discussed in \S5.1.1 of DMTF06; (2) Q0.5-Da, first discussed in \S5.2 of
DMTF06; (3) Q0.4A, first discussed in \S3 of MTFD07 -- note: MFTD07
mistakenly state that driving was for $1.6P_0$.}
\end{deluxetable}
\clearpage

\newpage
\begin{deluxetable}{lcrrccl}
   \tablewidth{0pt}
   \tablecolumns{7}
   \tablecaption{Simulation Diagnostics.
                         \label{tab:Diagnostics}}
   \tablehead{
                      \colhead{Simulation} &
                      \colhead{Model} &
                      \colhead{$t_{\mathrm{merge}}$} &
                      \colhead{$t_{\mathrm{zpt}}$} &
                      \colhead{Movie} &
                      \colhead{Outcome\tablenotemark{a}} &
                      \colhead{Elaboration} \\
                      \colhead{ } &
                      \colhead{ID} &
                      \colhead{$P_{0}$} &
                      \colhead{$P_{0}$} &
                      \colhead{ID} & & \\
                      \colhead{(1)} &
                      \colhead{(2)} &
                      \colhead{(3)} &
                      \colhead{(4)} &
                      \colhead{(5)} &
                      \colhead{(6)} &
                      \colhead{(7)} \\  }
   \startdata
   Q1.3P & $G1$ & $12.38$ & $0.00$ & \includegraphics[scale=0.03]{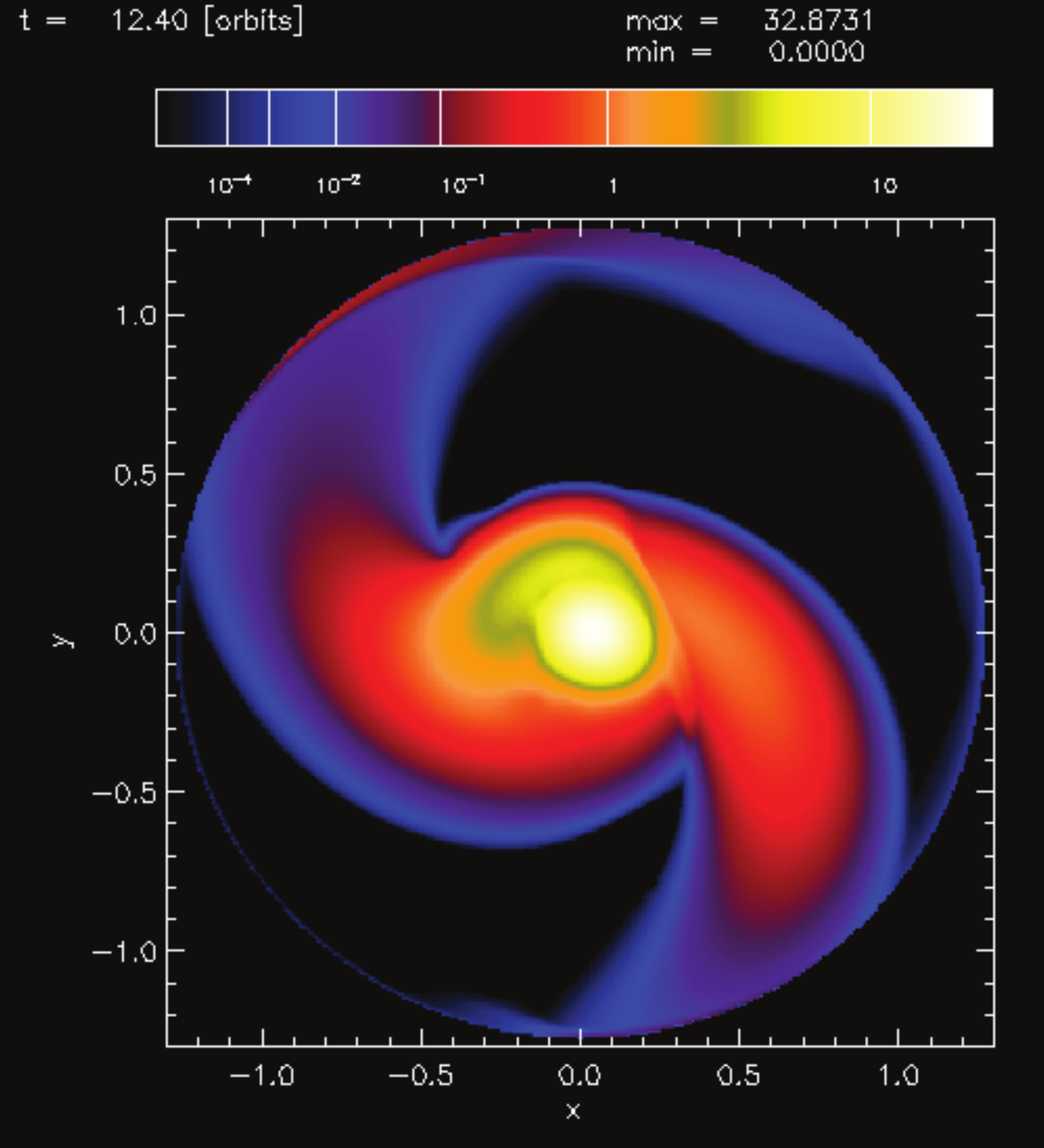} video01 & TI &
                  Binary merges at $12.4 P_{0}$ \\
              & $S1$ & $7.57$ & $+ 4.81$ & \includegraphics[scale=0.03]{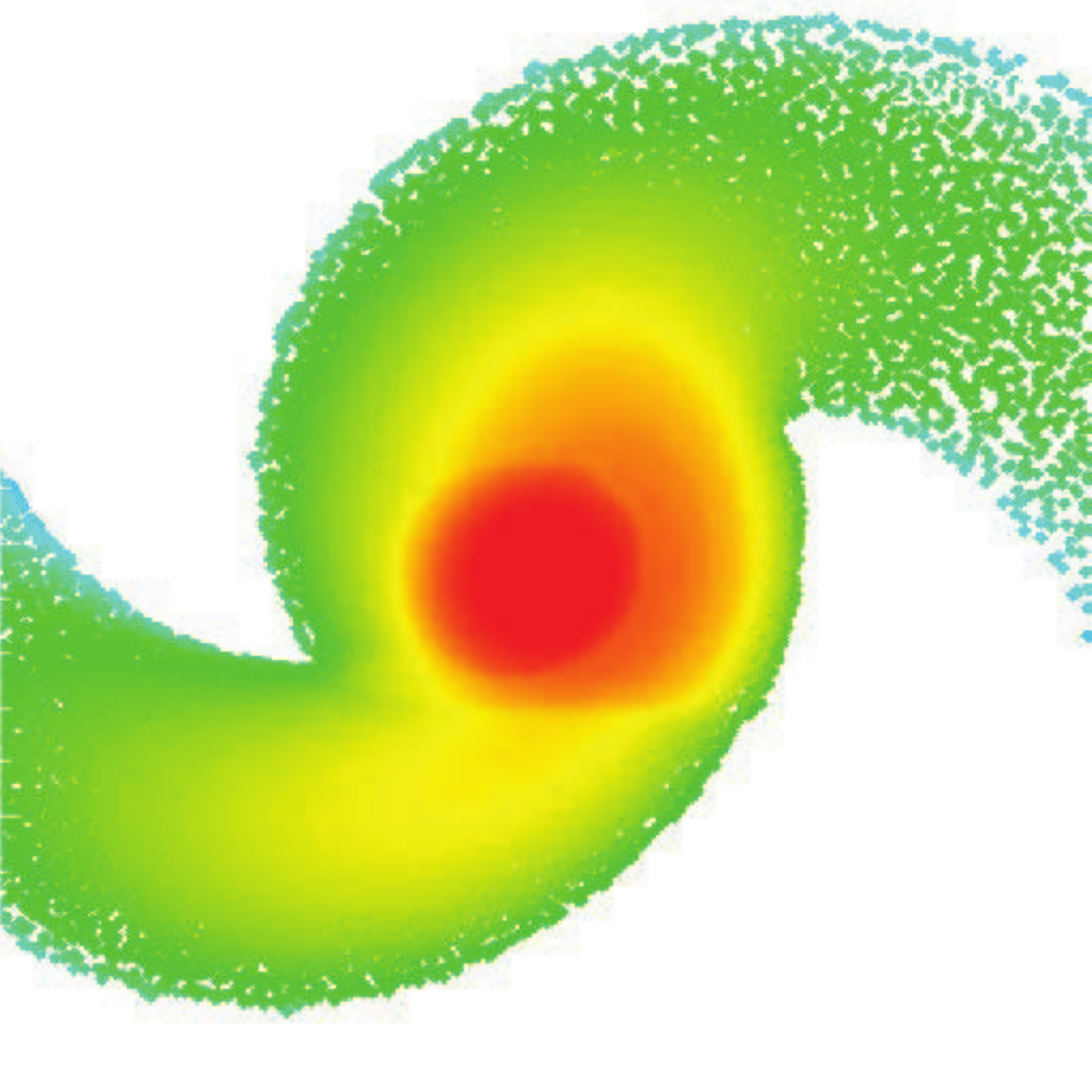} video02 & TI &
                  Binary merges at $7.6 P_{0}$ \\
              & $S2$ & $8.17$ & $+ 4.21$ & \includegraphics[scale=0.03]{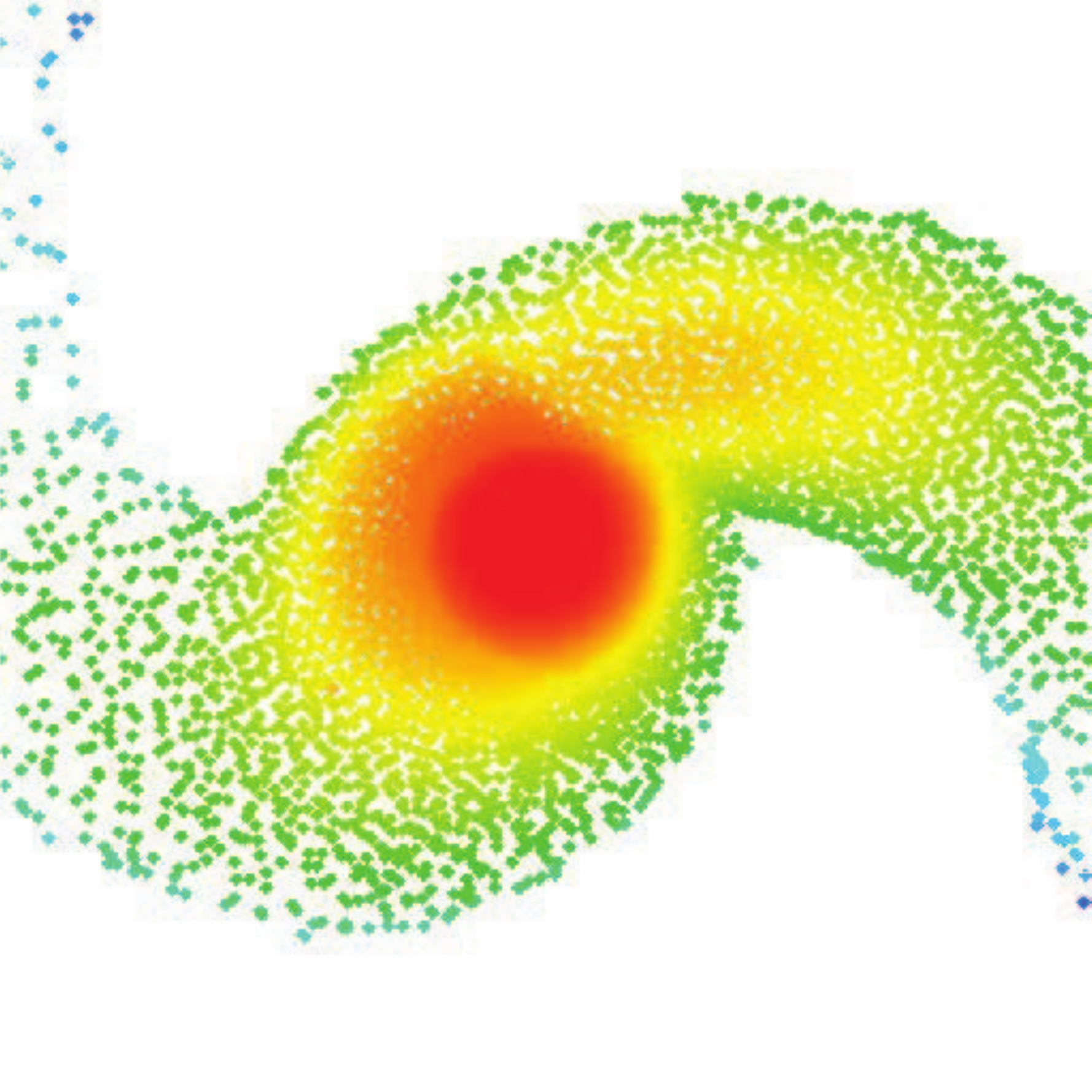} video03 & TI &
                  Binary merges at $8.2 P_{0}$ \\
   Q0.7P & $G1$ & $9.70$ & $0.00$ & \includegraphics[scale=0.03]{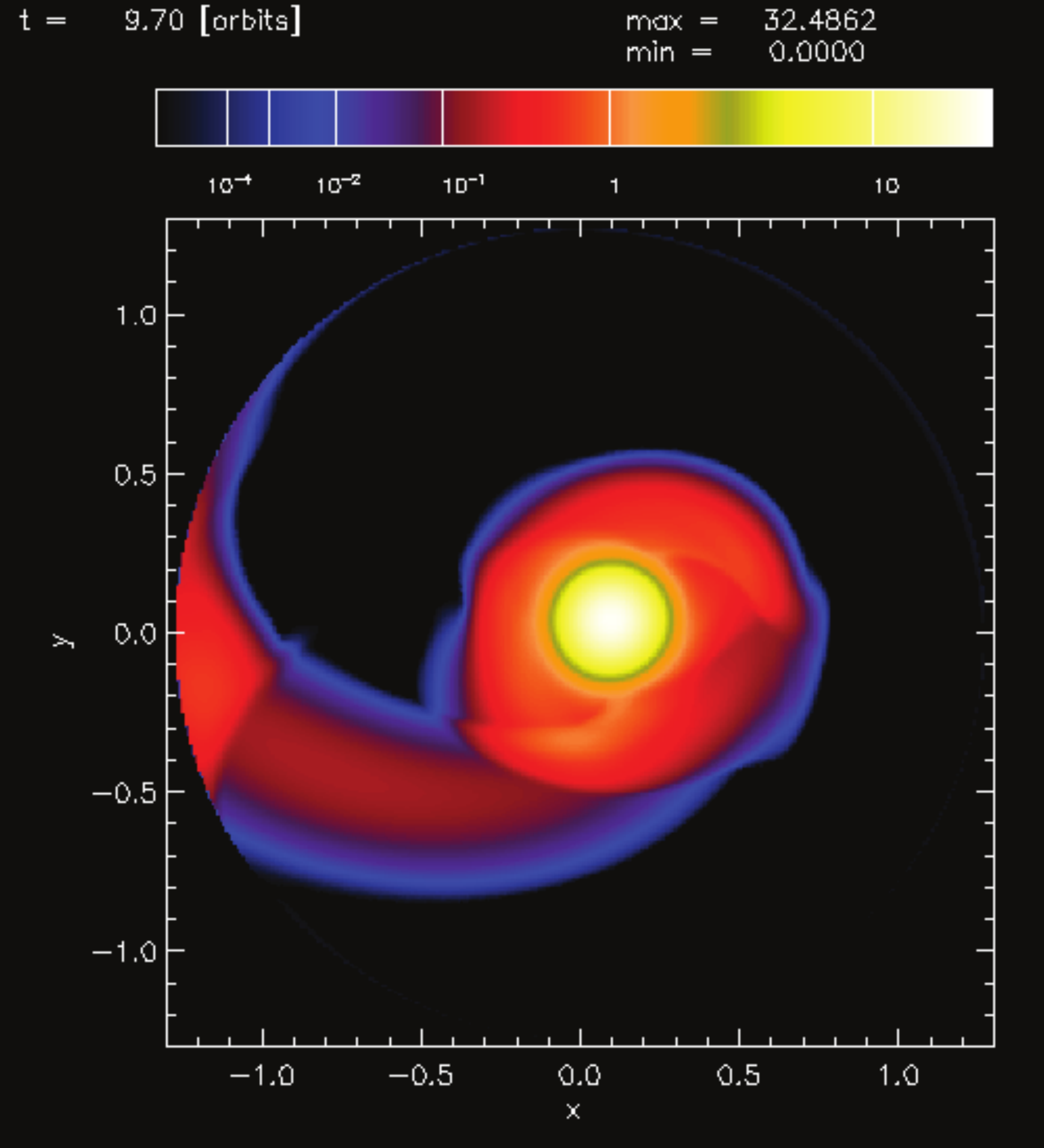} video04 & TI &
                  Donor disrupts at $9.7 P_{0}$ \\
              & $G2$ & $21.03$ & $- 11.41$ & \includegraphics[scale=0.03]{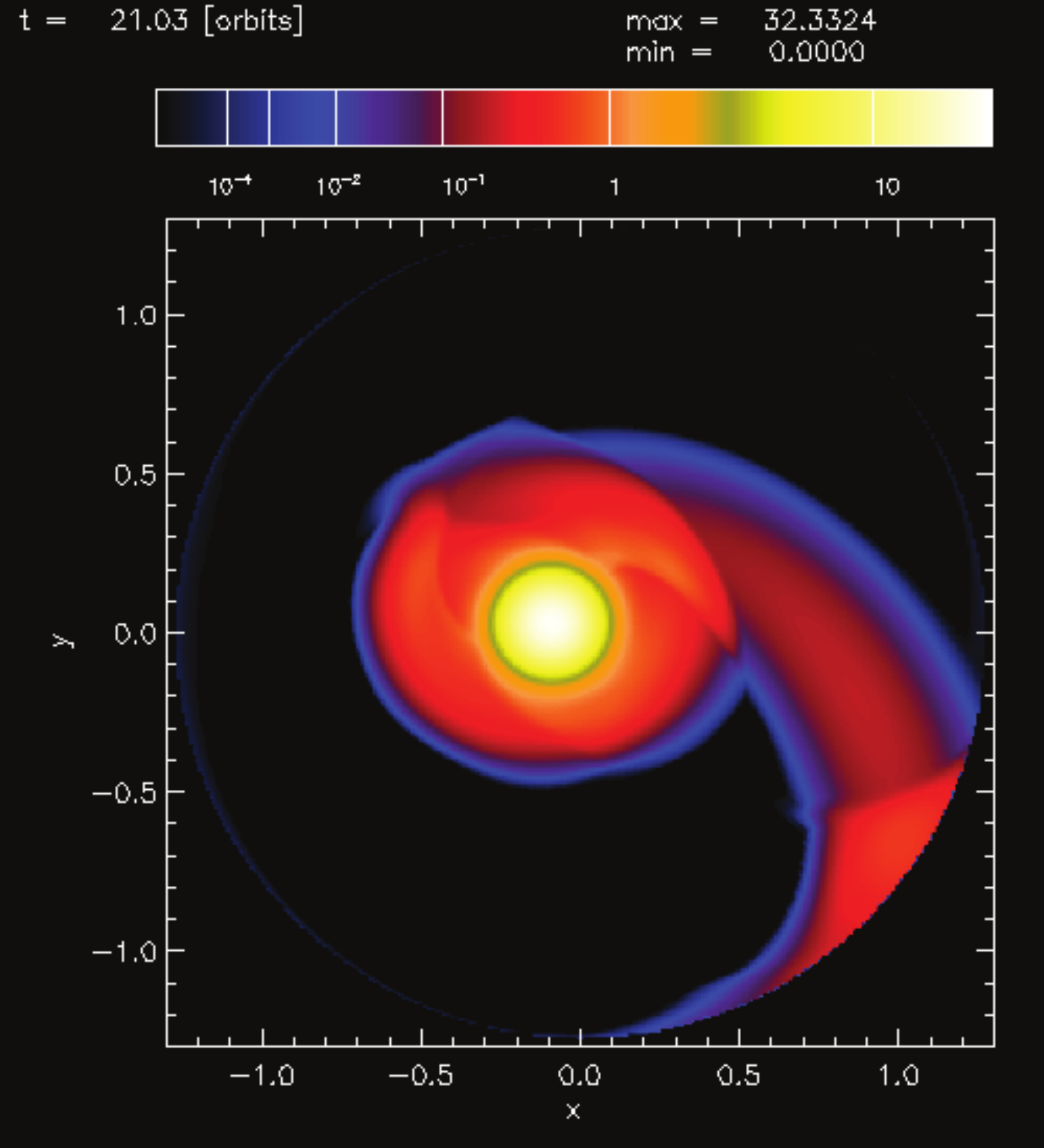} video05 & TI &
                  Donor disrupts at $21.0 P_{0}$ \\
              & $S1$ & $11.51$ & $- 2.11$ & \includegraphics[scale=0.03]{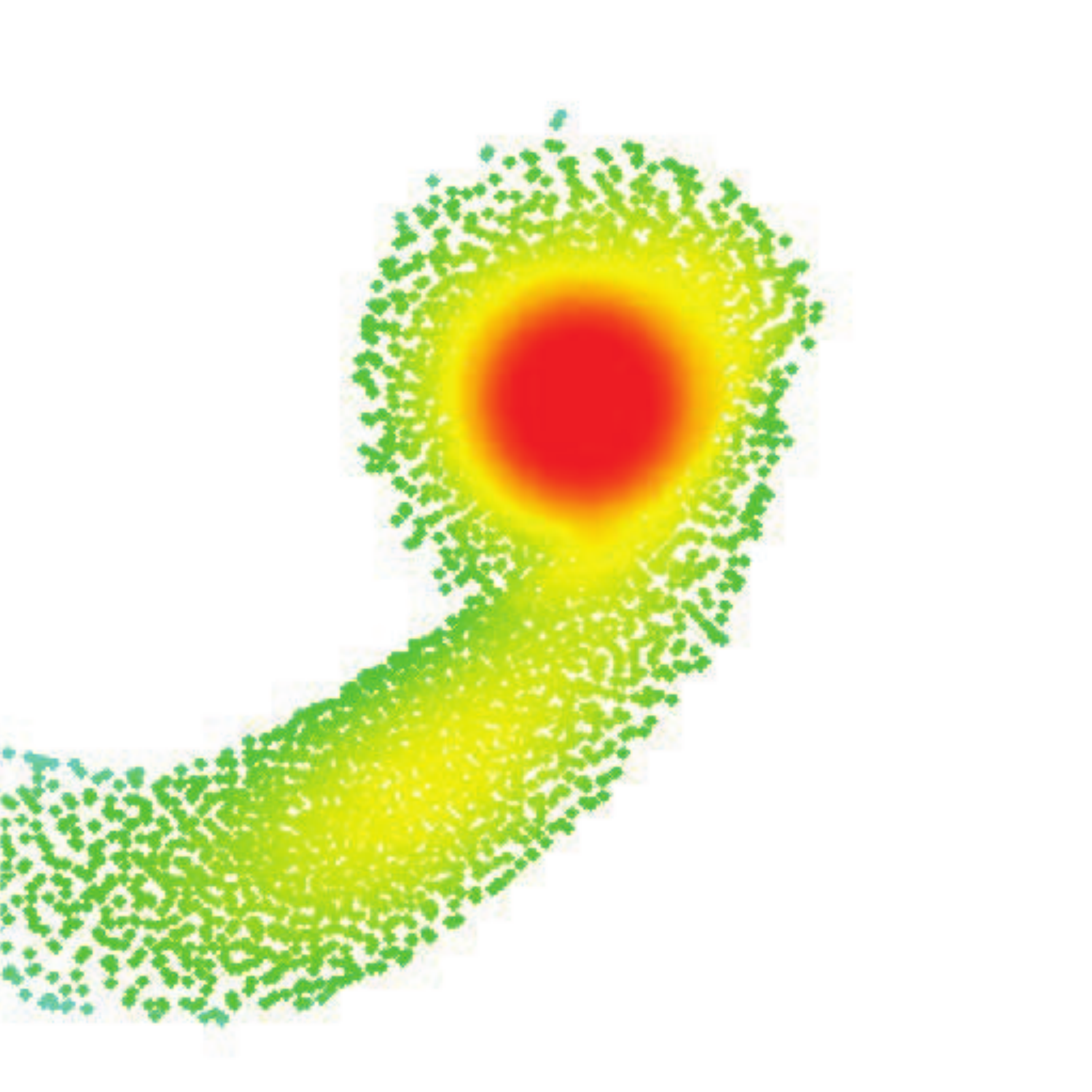} video06 & TI &
                  Donor disrupts $11.5 P_{0}$ \\
   Q0.5P & $G1$ & $\cdots$ & $0.00$ & \includegraphics[scale=0.03]{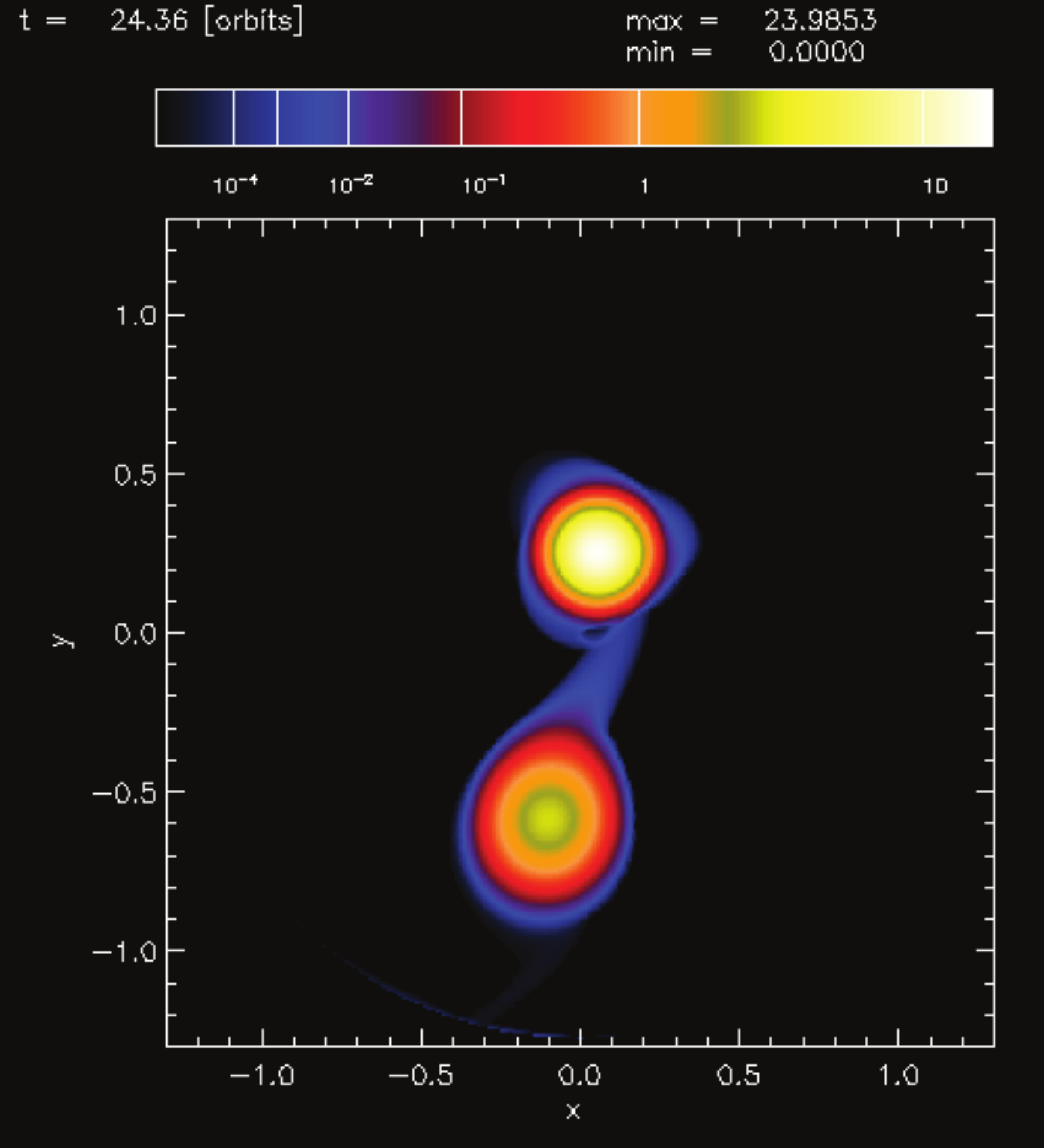} video07 & D &
                  Binary detaches at $34 P_{0}$ \\
              & $S1$ & $\cdots$ & $0.00$ & \includegraphics[scale=0.03]{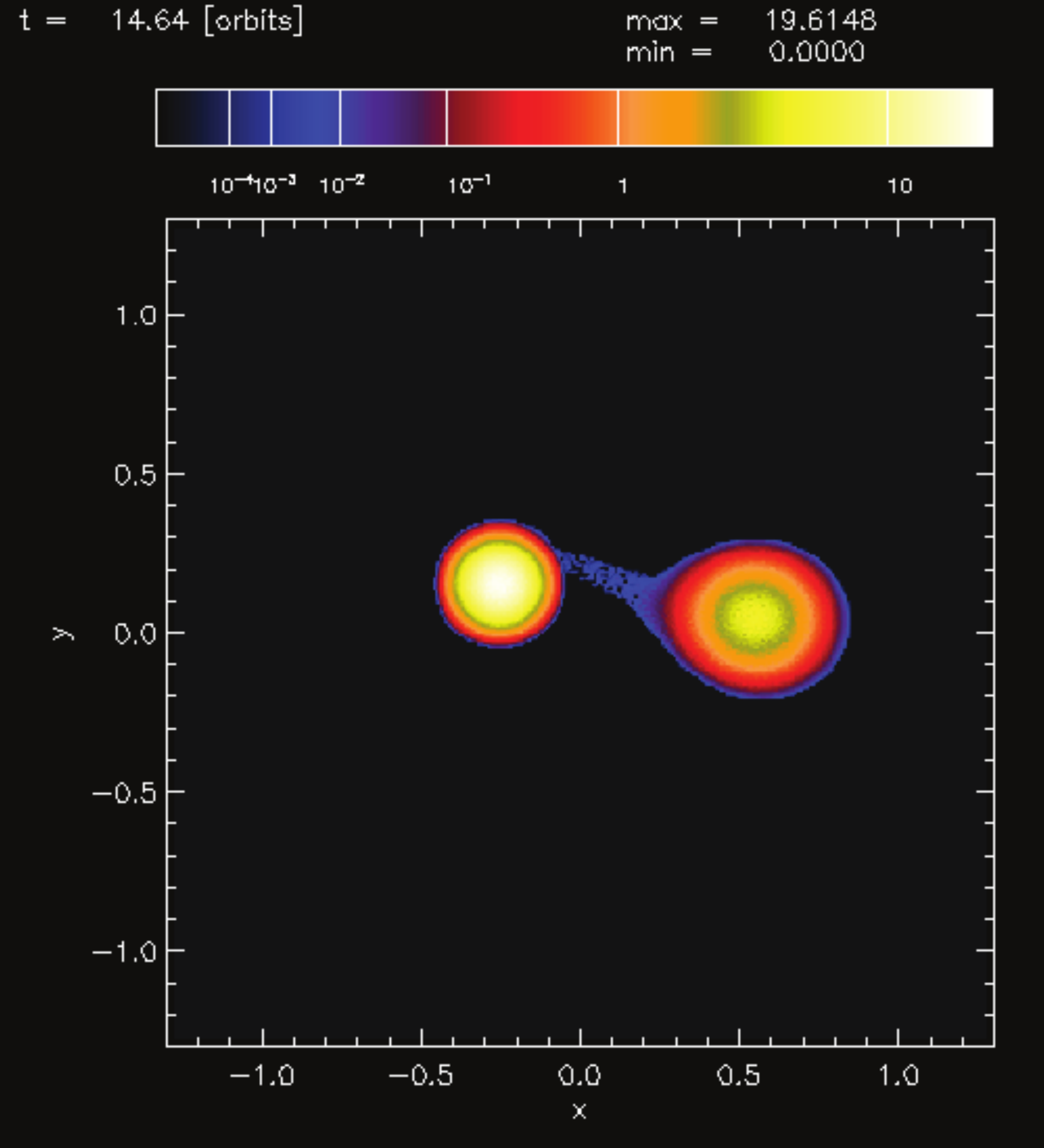} video08  & ? &
                  Undetermined after $14.6 P_{0}$ \\
   Q0.4P & $G1$ & $\cdots$ & $0.00$ & \includegraphics[scale=0.03]{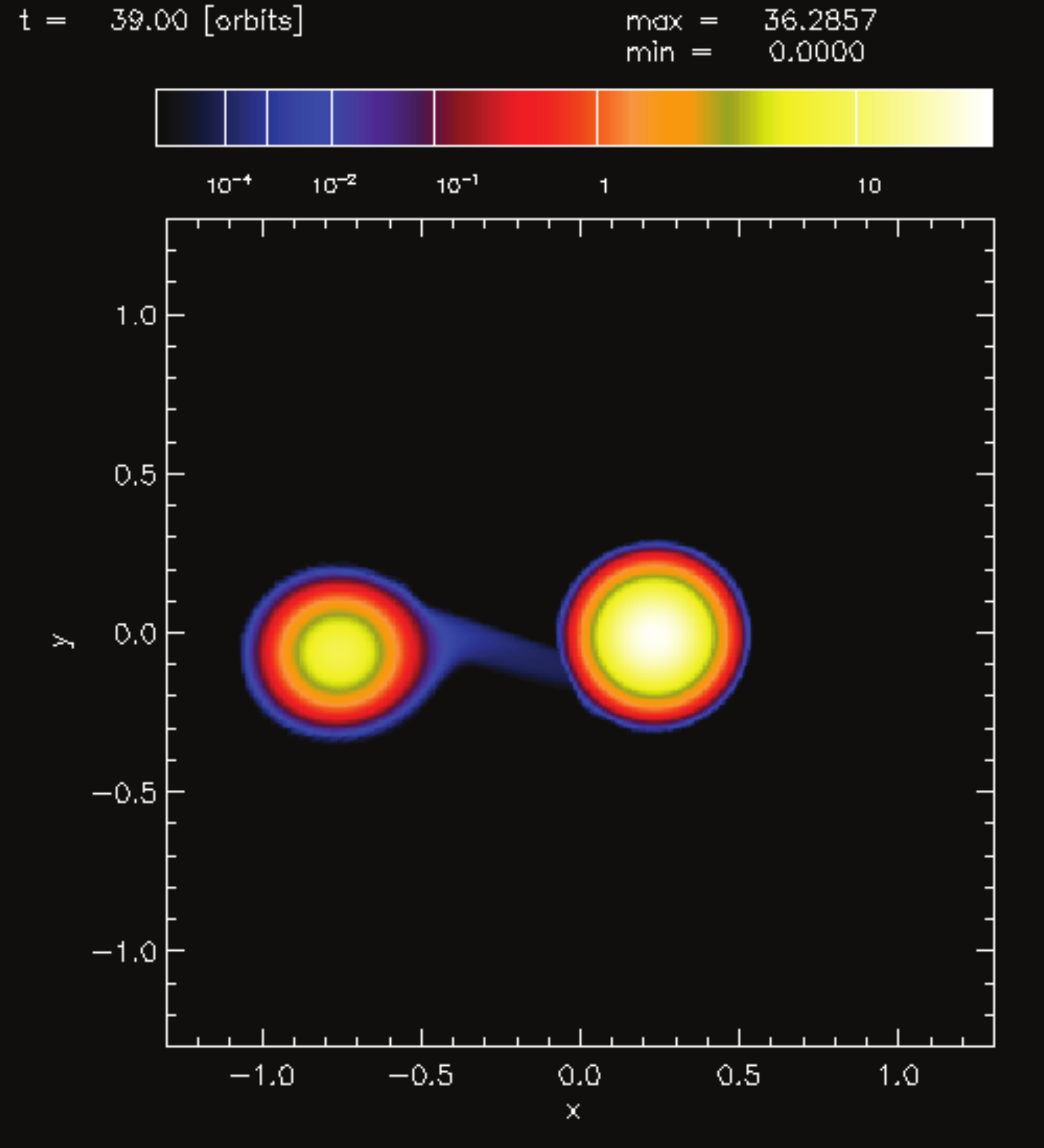} video09 & D &
                  Binary detaches at $40 P_{0}$  \\
              & $G2$ & $\cdots$ & $0.00$ & \includegraphics[scale=0.03]{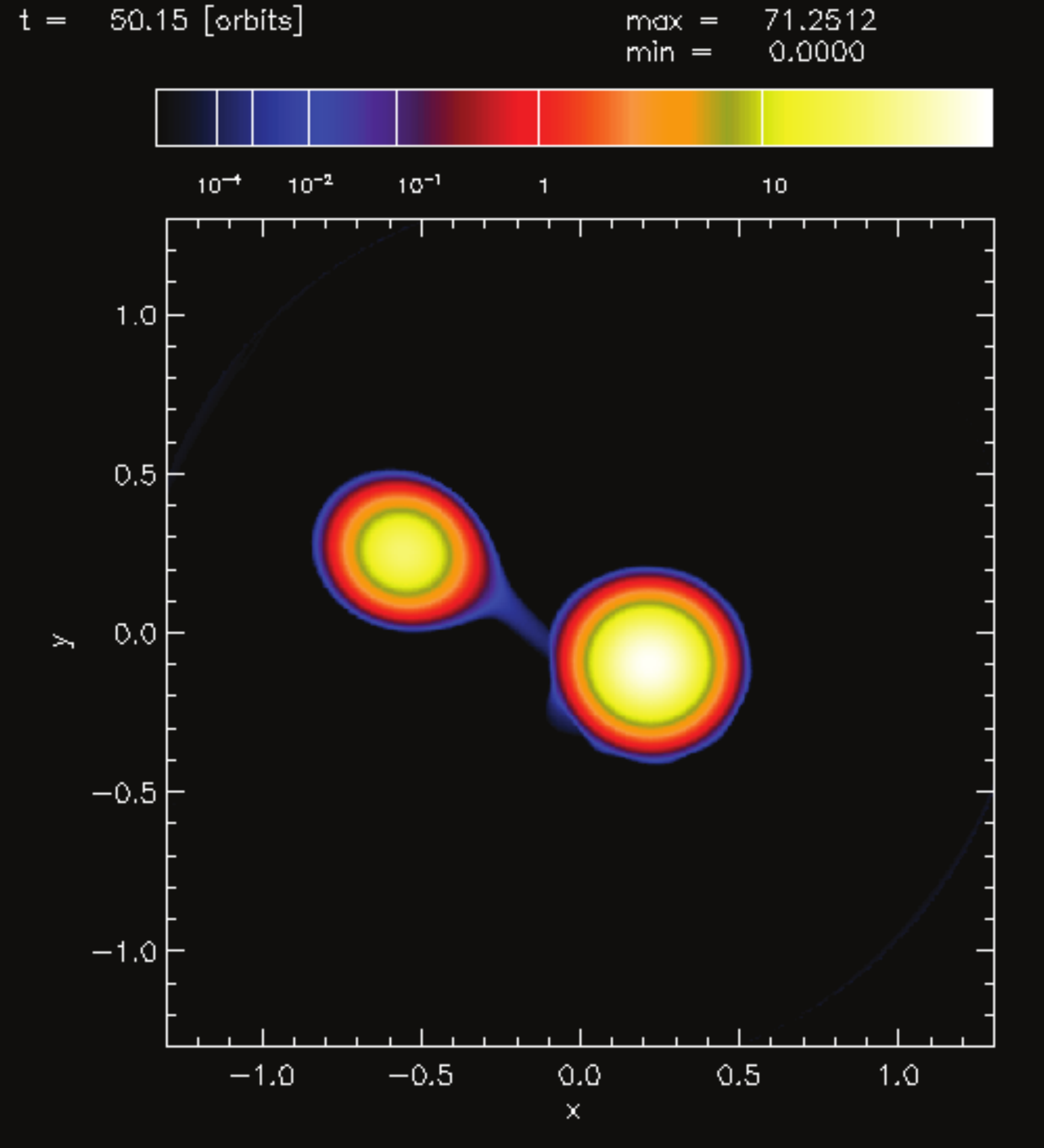} video10 & ?  &
                  Undetermined after $50.2 P_{0}$  \\
              & $G3$ & $\cdots$ & $0.00$ & \includegraphics[scale=0.03]{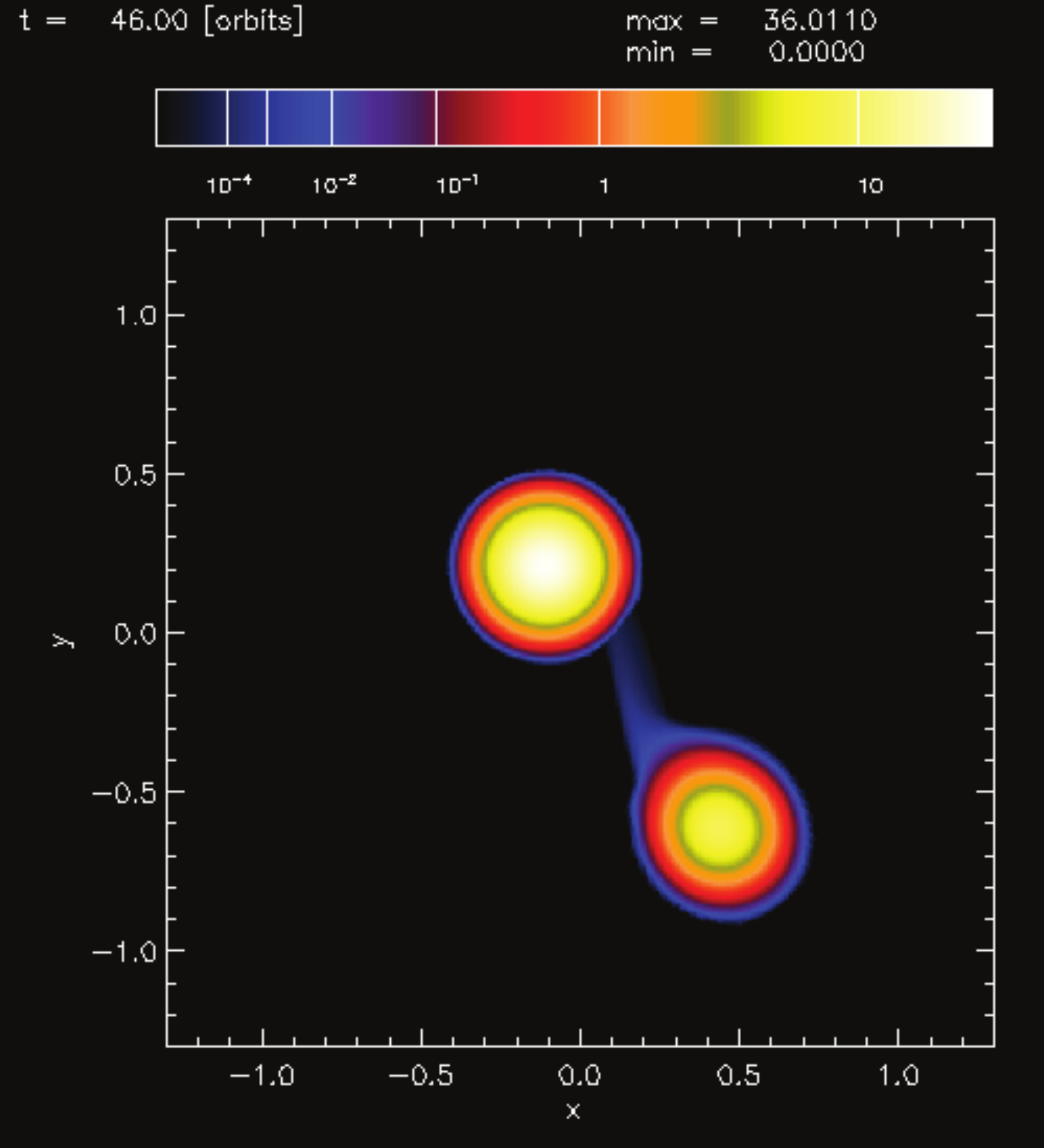} video11 & D &
                   Binary detaches at $46.5 P_{0}$  \\
              & $S1$ & $\cdots$ & $0.00$ & \includegraphics[scale=0.03]{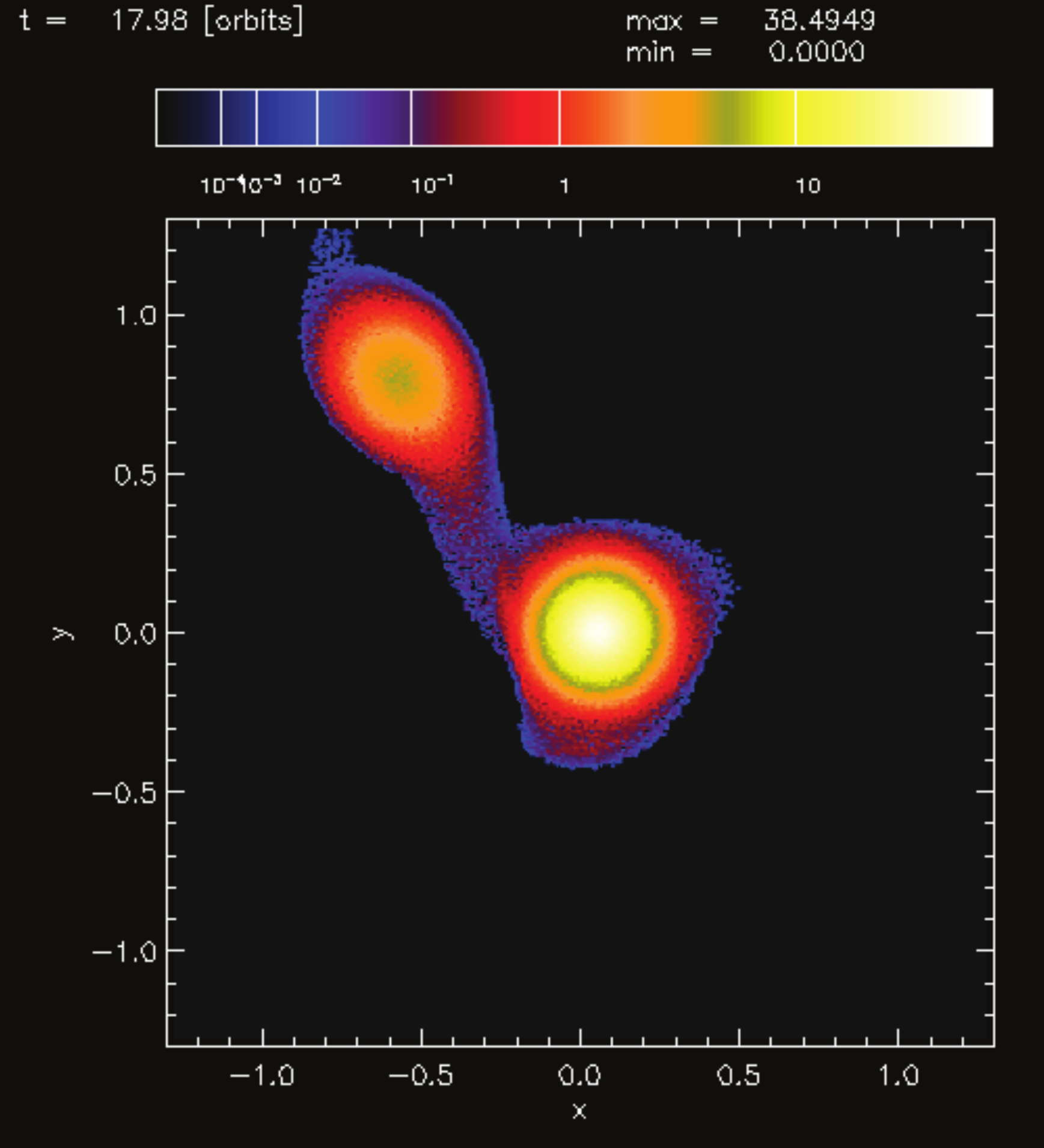} video12 & ? &
                   Undetermined after $18.3 P_{0}$  \\
              & $S2$ & $\cdots$ & $0.00$ & $\cdots$ & ? &
                    (no movie)  \\
   Q1.3I  & $G1$ & $15.54$ & $0.00$ & \includegraphics[scale=0.03]{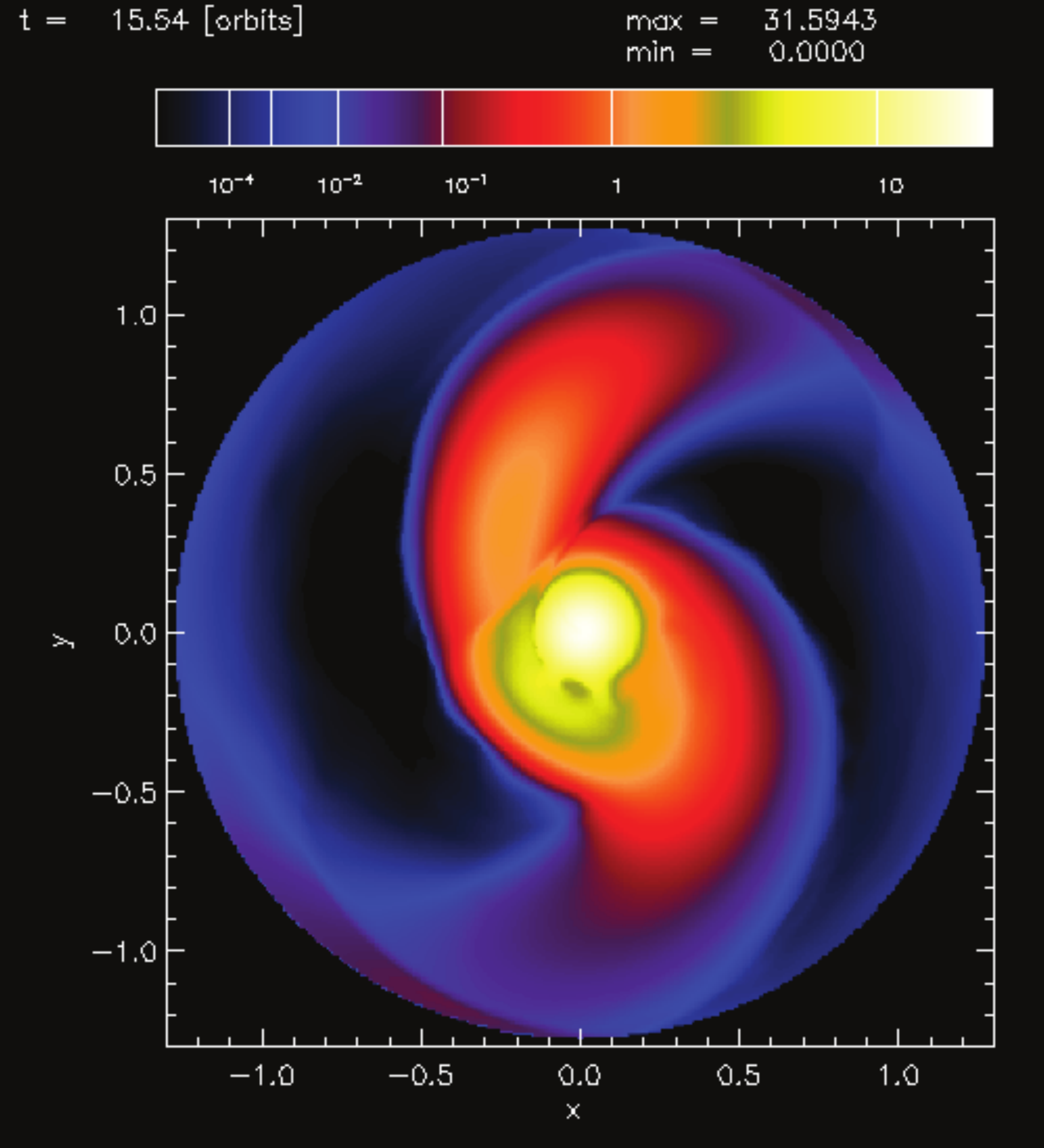} video13 & TI &
                    Binary merges at $15.5 P_{0}$ \\
              & $S1$ & $7.69$ & $+7.85$ & \includegraphics[scale=0.03]{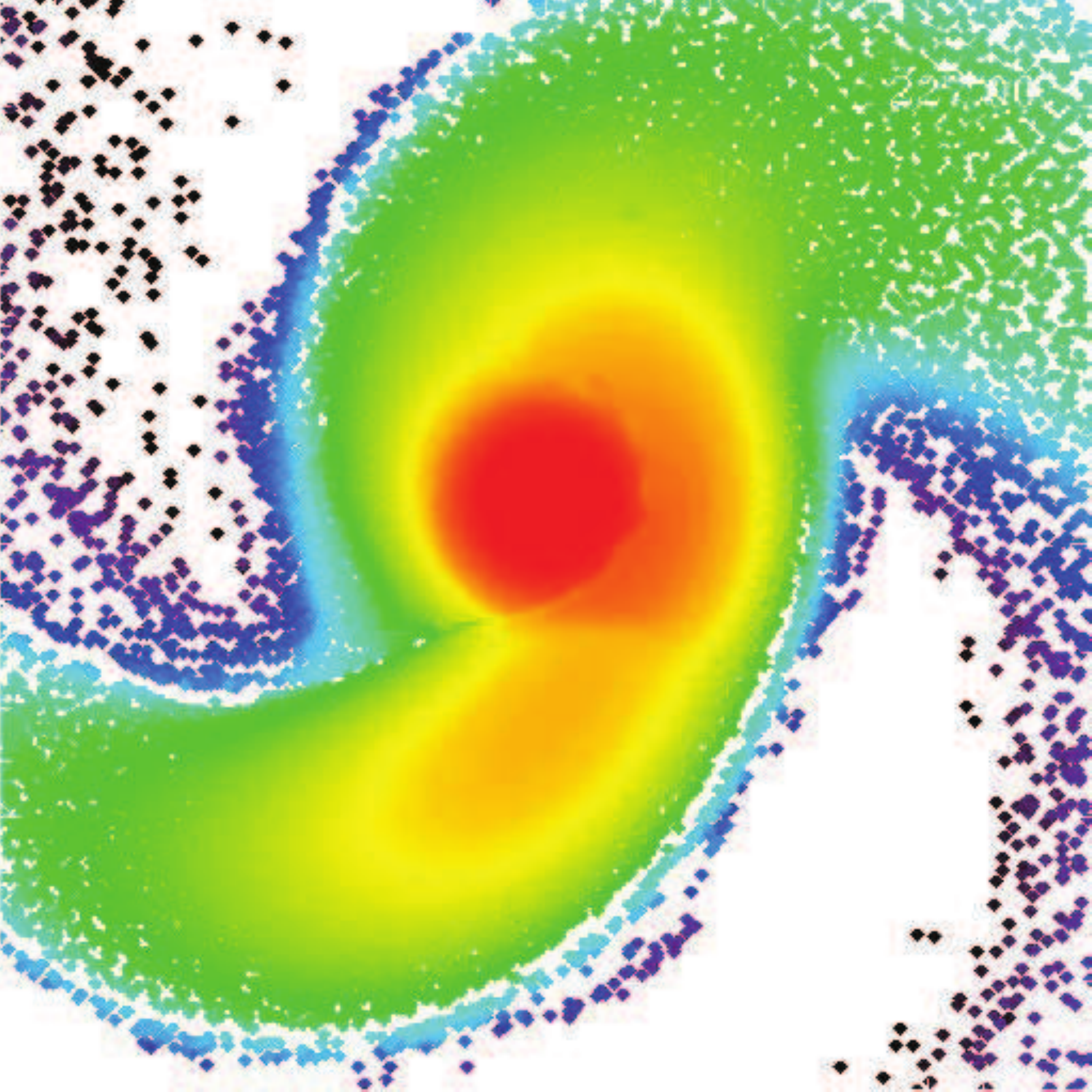} video14 & TI &
                    Binary merges at $7.7 P_{0}$ \\
   Q0.7I  & $G1$ & $12.37$ & $0.00$ & \includegraphics[scale=0.03]{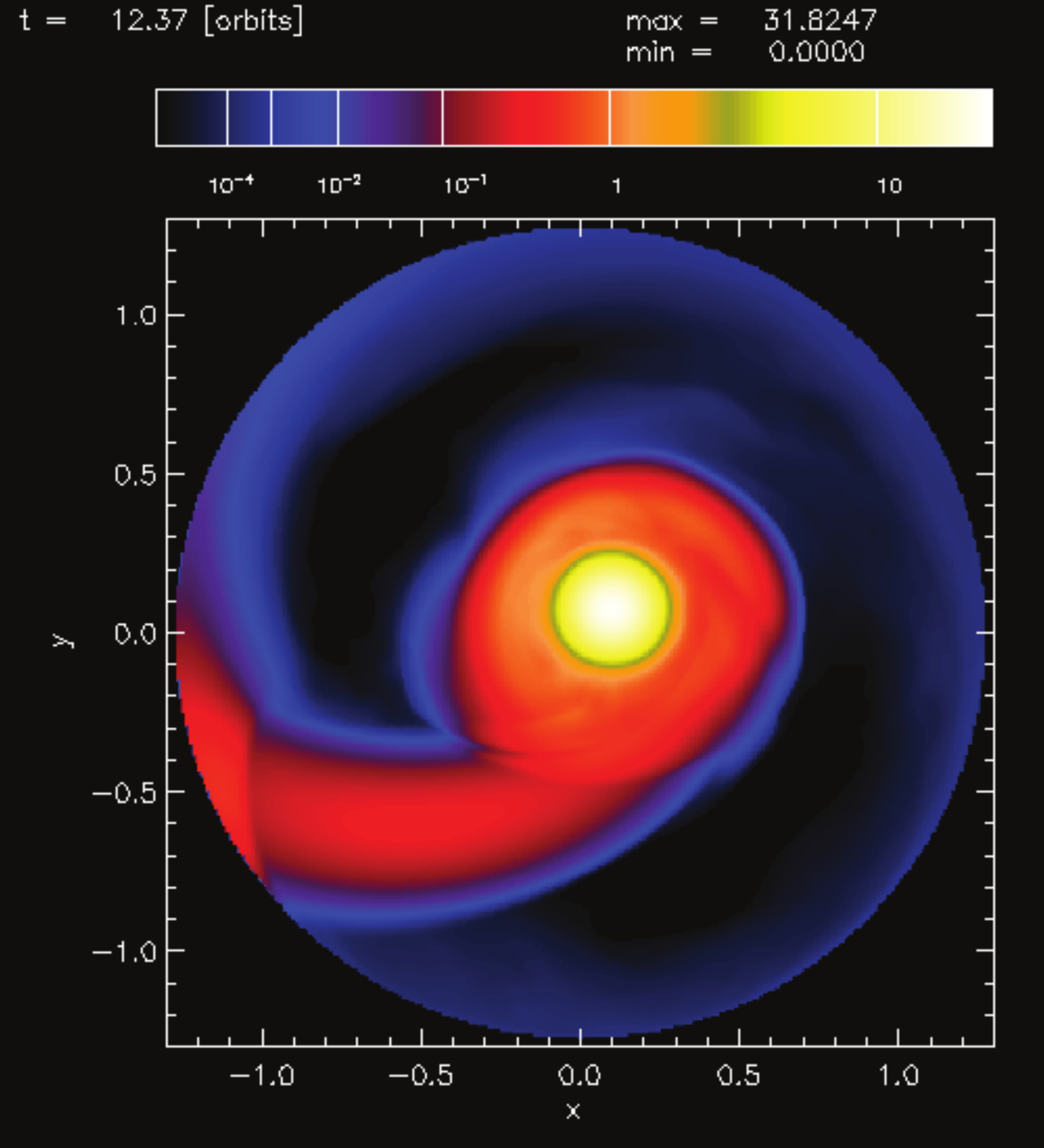} video15 & TI &
                    Donor disrupts at $12.4 P_{0}$  \\
              & $S1$ & $10.04$ & $+2.33$ & \includegraphics[scale=0.03]{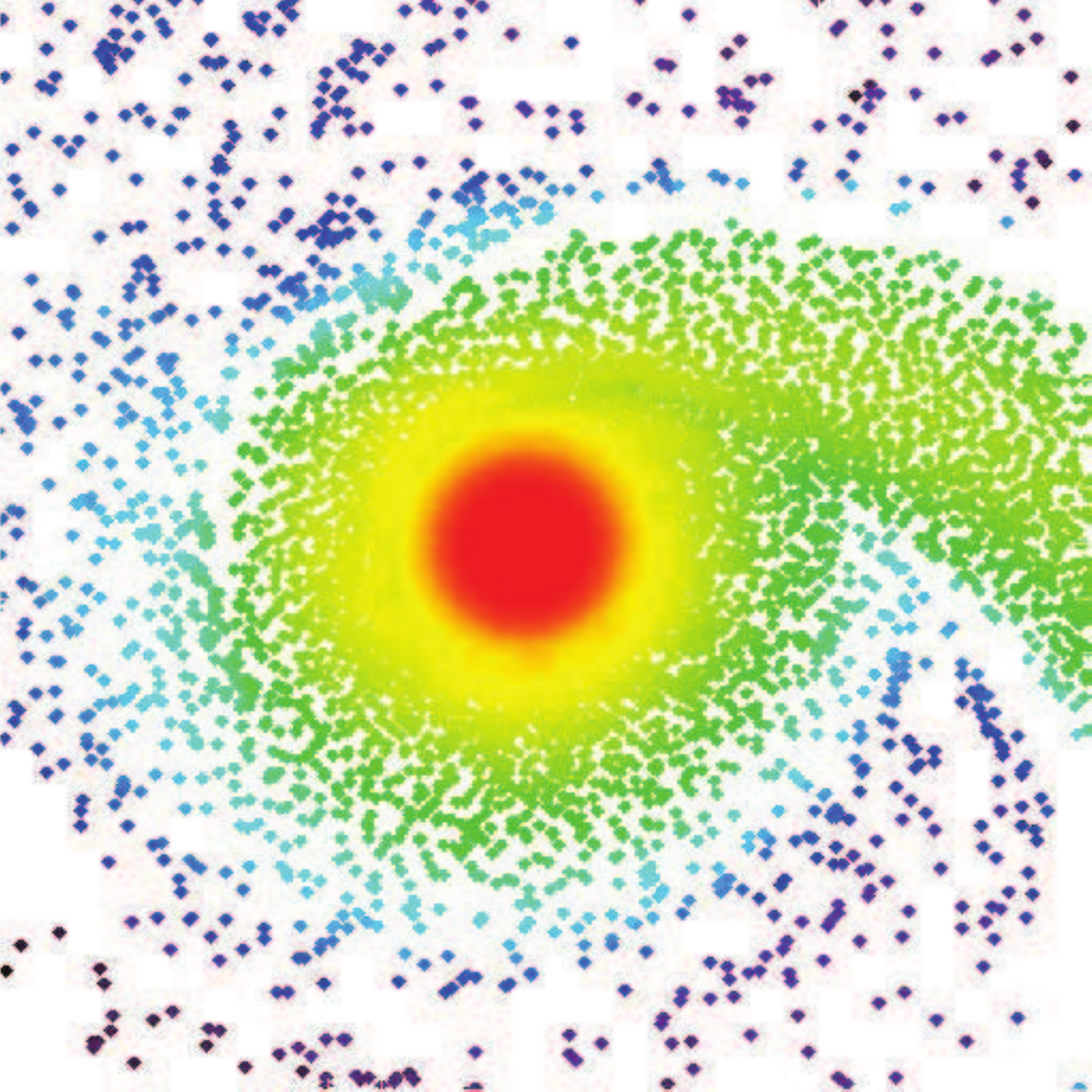} video16 & TI &
                    Donor disrupts at $10.0 P_{0}$  \\
   Q0.5I  & $G1$ & $\cdots$ & $0.00$ & \includegraphics[scale=0.03]{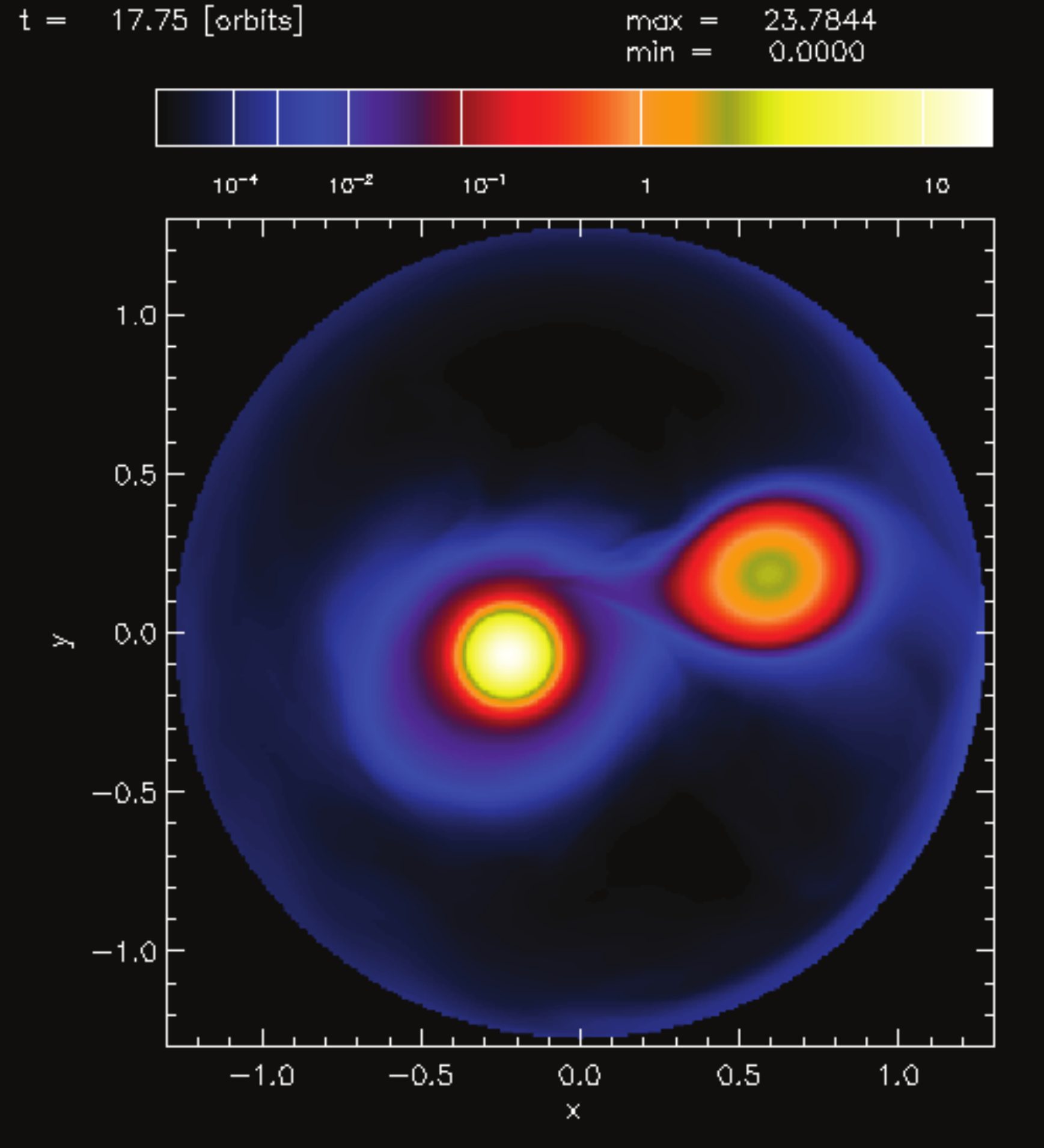} video17 & CE + TI &
                    Donor disrupts at $19.3 P_{0}$ \\
              & $S1$ & $\cdots$ & $0.00$ & \includegraphics[scale=0.03]{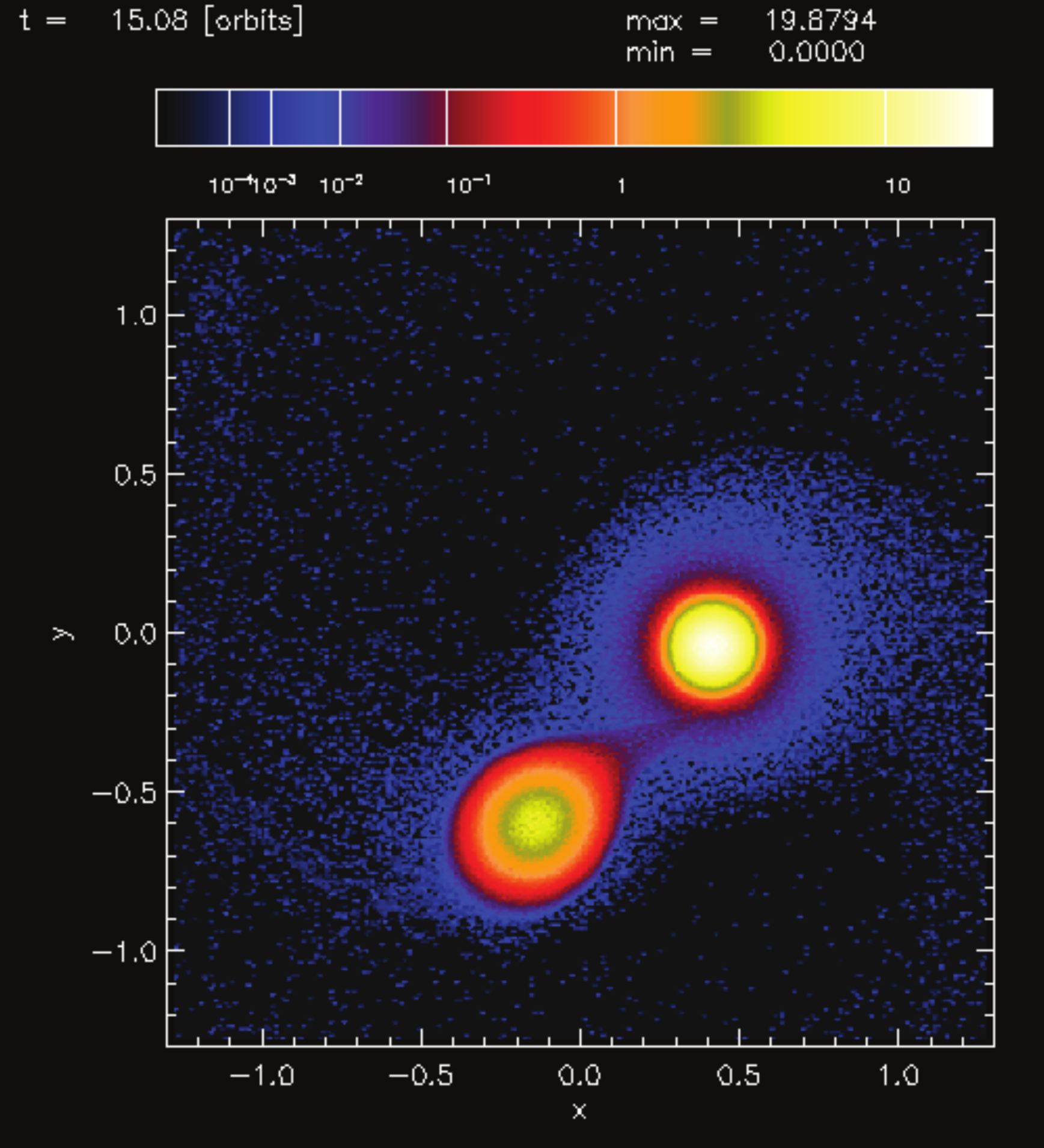} video18 & CE + TD? &
                    Undetermined after $15.0 P_{0}$ \\
   Q0.4I  & $G1$ & $\cdots$ & $0.00$ & \includegraphics[scale=0.03]{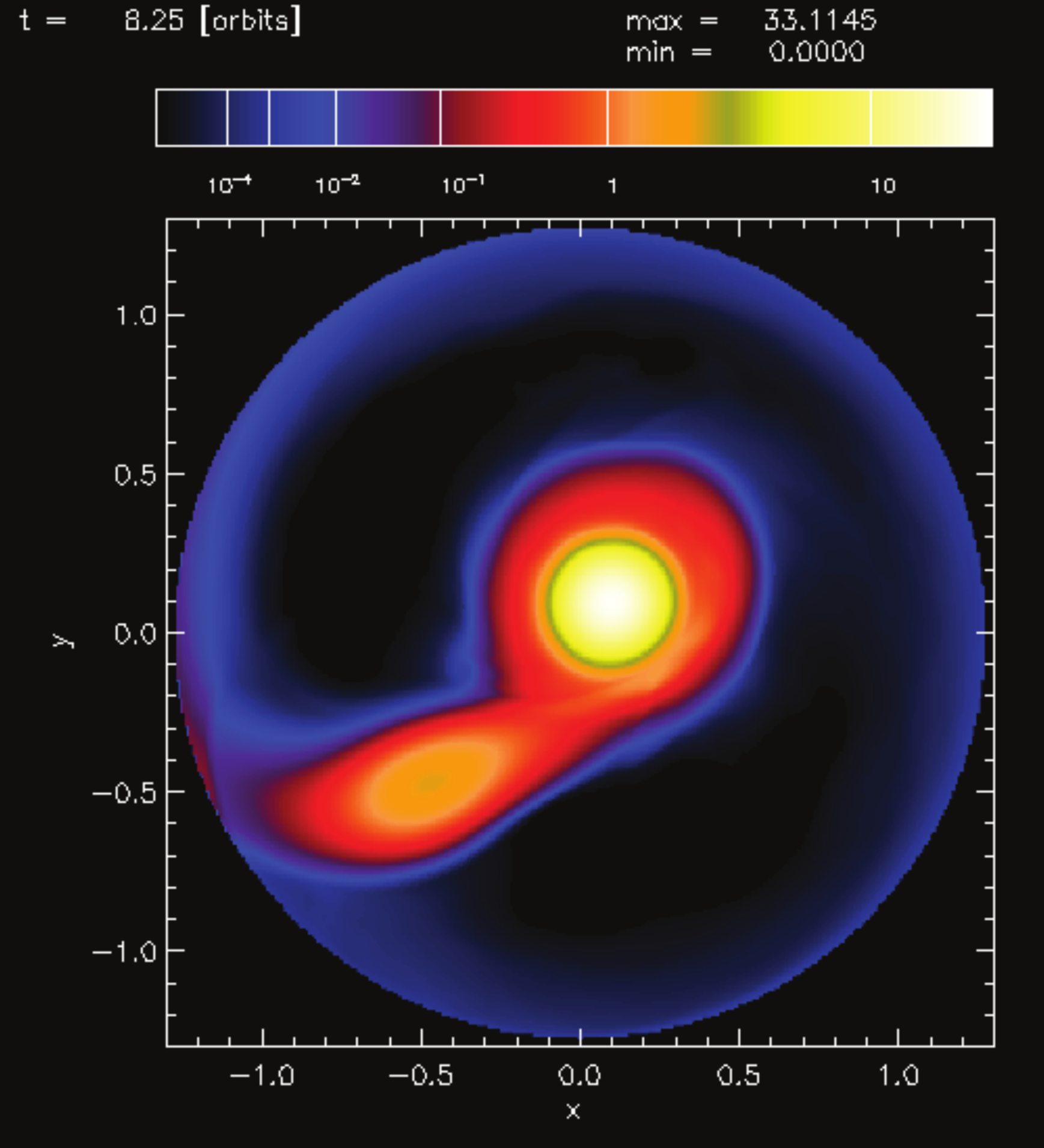} video19 & CE + TI &
                    Donor disrupts at $8.3 P_{0}$\\
              & $G2$ & $\cdots$ & $0.00$ & \includegraphics[scale=0.03]{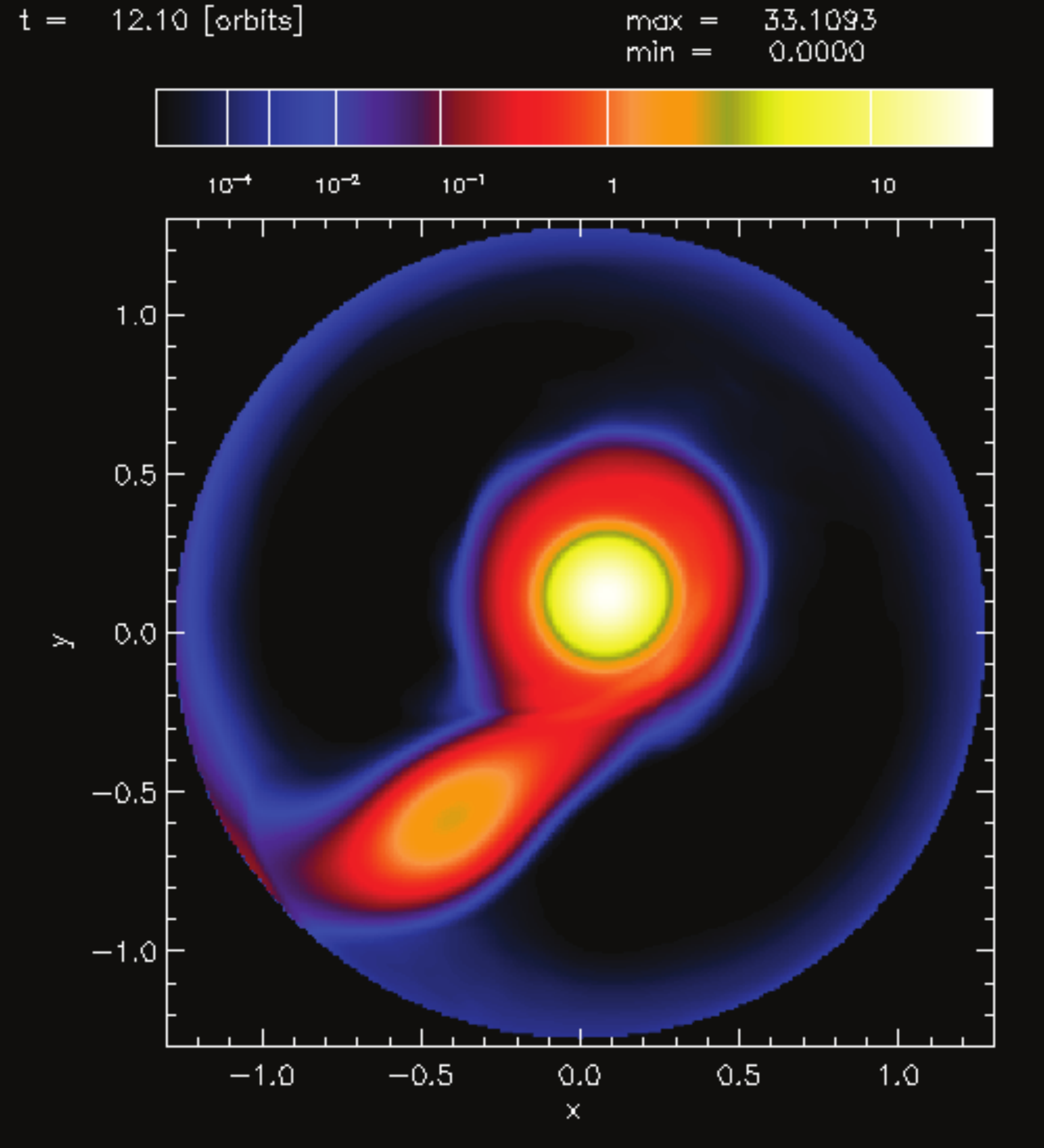} video20 & CE + TI &
                    Donor disrupts at $12.4 P_{0}$ \\
              & $S1$ & $\cdots$ & $0.00$ & \includegraphics[scale=0.03]{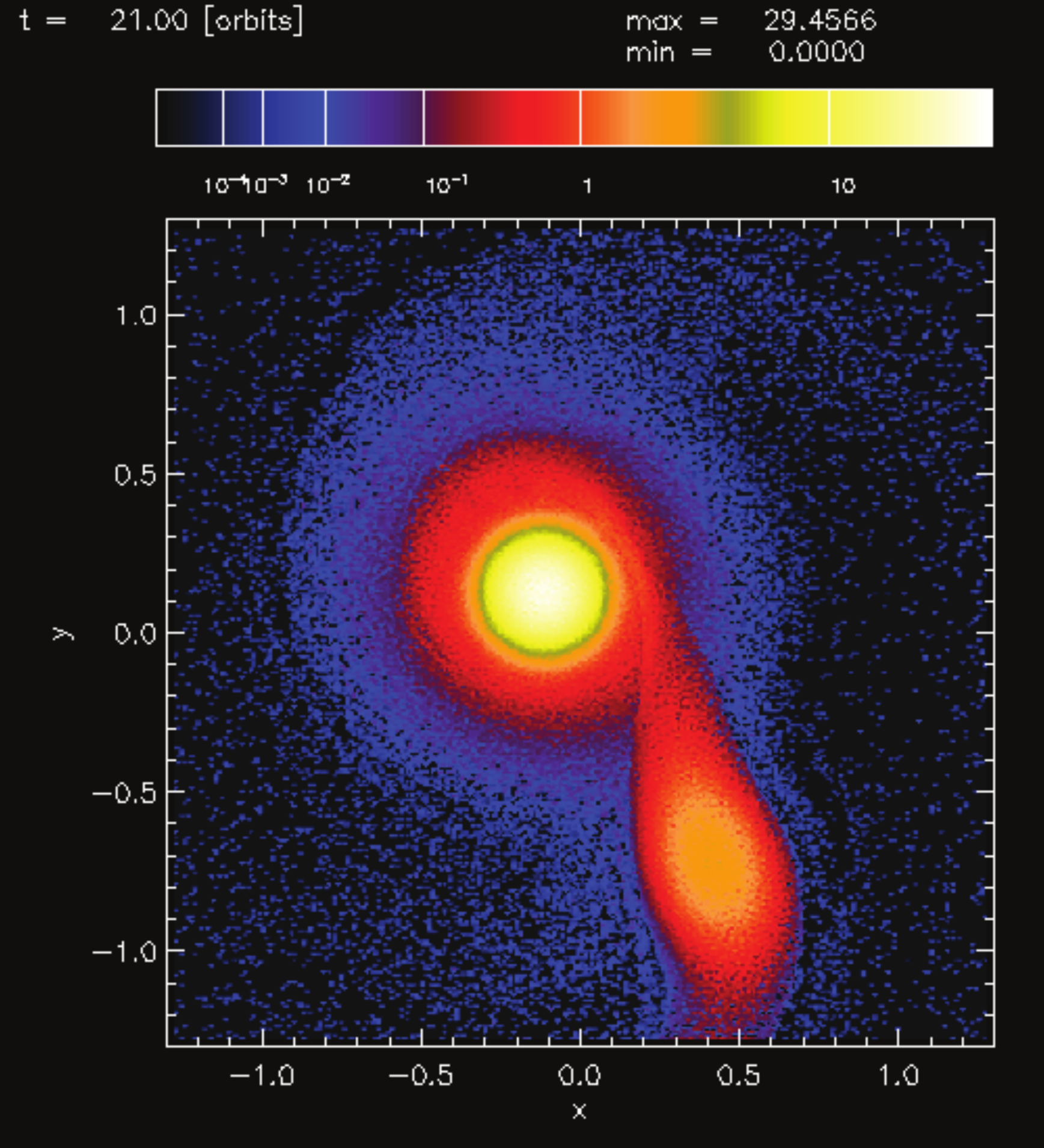} video21 & CE + TI &
                    Donor disrupts at $21.4 P_{0}$ \\
              & $S2$ & $\cdots$ & $0.00$ & $\cdots$ & ? &
                    (no movie) \\
   \enddata
   \tablenotetext{a}{TI = tidal instability; TD = tidal disruption; D = detaches;
                              CE = common envelope}
\end{deluxetable}
\clearpage


\newpage
\clearpage

\newpage
\clearpage

\newpage
\clearpage

\newpage
\setcounter{figure}{3}
\clearpage

\newpage
\begin{figure*}
\centering
\figurenum{4a}
\scalebox{0.8}{\plotone{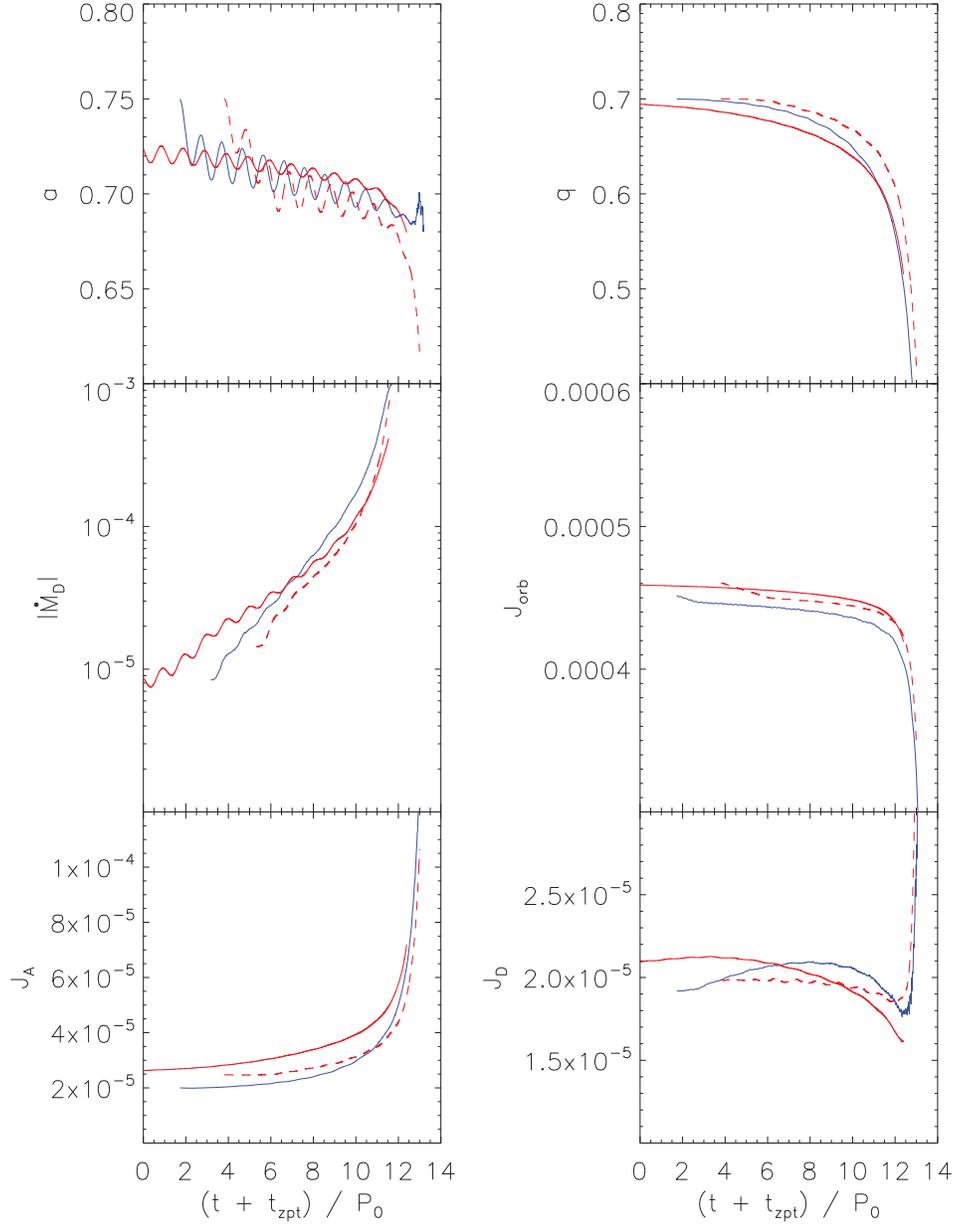}}
\figcaption[Q07Lines]{
Simulation Q0.7P. Curves show the time-dependent behavior of six 
system parameters --- $a$, $q$, $\log|\dot{M}_\mathrm{D}|$,
$J_\mathrm{orb}$, $J_\mathrm{A}$ and $J_\mathrm{D}$ ---
as derived from models Q0.7P$\_G1$ (dashed red curves), 
Q0.7P$\_G2$ (solid red curves), and Q0.7P$\_S1$ (solid blue curves).  
For purposes of presentation, each curve has been terminated at the 
time in the simulation when the system mass ratio has dropped below 
$q = 0.4$ and curves from simulations Q0.7P$\_G2$ and Q0.7P$\_S1$ 
include a zero-point shift in time, $t_\mathrm{zpt}$, as detailed in 
column 4 of Table \ref{tab:Diagnostics}; units used along the vertical 
axis for each plot are defined in Appendix \ref{app:Diagnostics}.} \label{fig:Q07evolutionA}
\end{figure*}
\clearpage

\newpage
\begin{figure*}
\centering
\figurenum{4b}
\scalebox{0.7}{\plotone{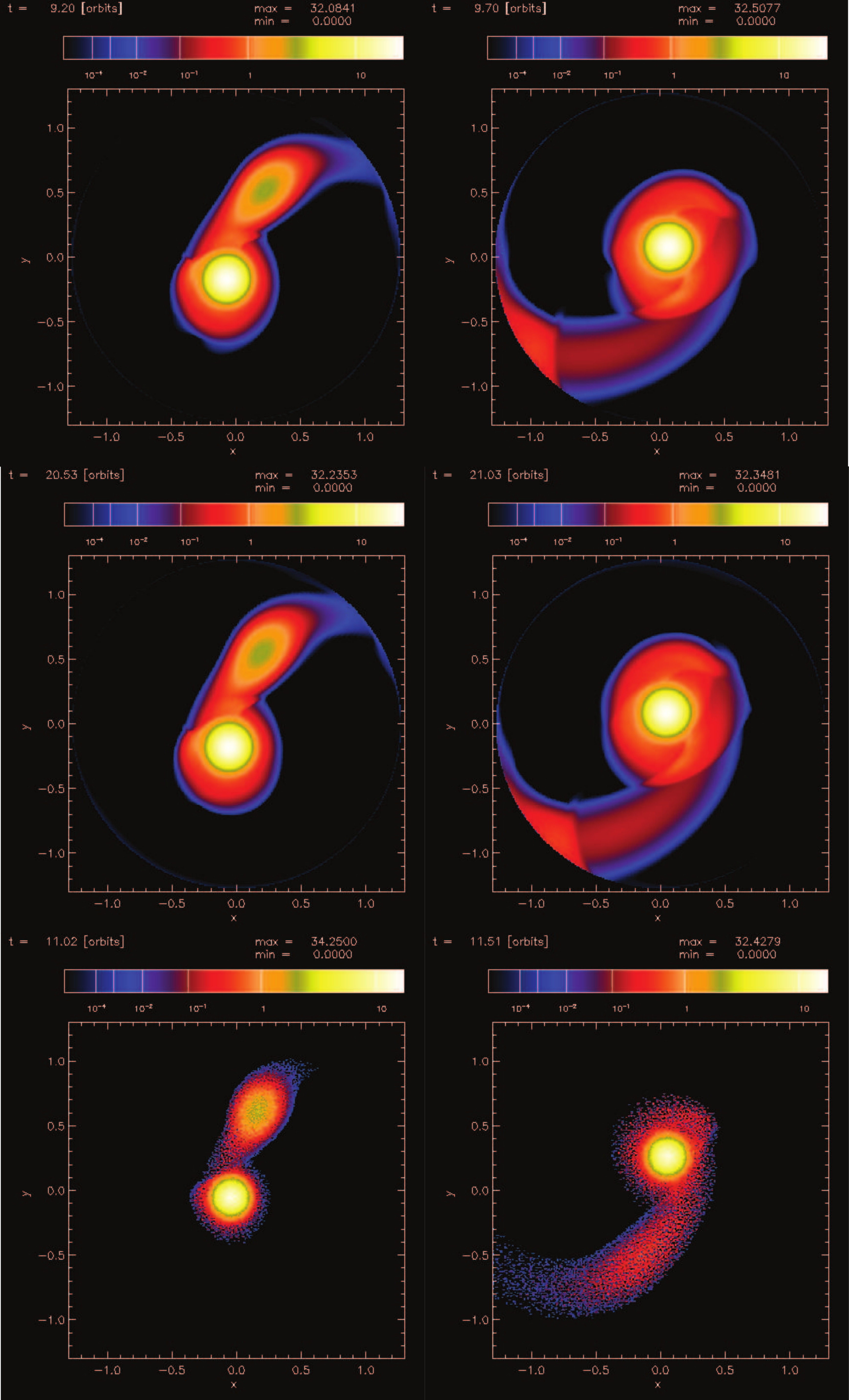}}
\caption[Q07TripleContours]{Simulation Q0.7P.  Images show contours 
of equatorial-plane column densities from  simulation Q0.7P$\_G1$ 
(top), Q0.7P$\_G2$ (middle), and Q0.7P$\_S1$ (bottom) at two
different points in time as tidal disruption of the donor is occurring --- 
specifically, {\it left-most images}: $t_\mathrm{G1} = 12.92 P_0$, 
$t_\mathrm{G2} = 20.53 P_0$, and $t_\mathrm{S1} = 11.02 P_0$;
{\it right-most images}: $t_\mathrm{G1} = 13.42 P_0$, 
$t_\mathrm{G2} = 21.03 P_0$, and $t_\mathrm{S1} = 11.51 P_0$.  
The images displayed for simulations Q$0.7\_G1$ (top),
Q$0.7\-G2$ (middle) and Q$0.7\_S1$ (bottom) 
have been extracted from, respectively, video04, video05 and video06.} 
\label{fig:Q07evolutionB}
\end{figure*}
\setcounter{figure}{4}
\clearpage

\newpage
\begin{sidewaysfigure}[htb!]
\centering
\begin{tabular}{cc}
\textbf{Q0.5} & \textbf{Q0.4} \\
\multicolumn{2}{c}{\scalebox{1.1}{\plottwo{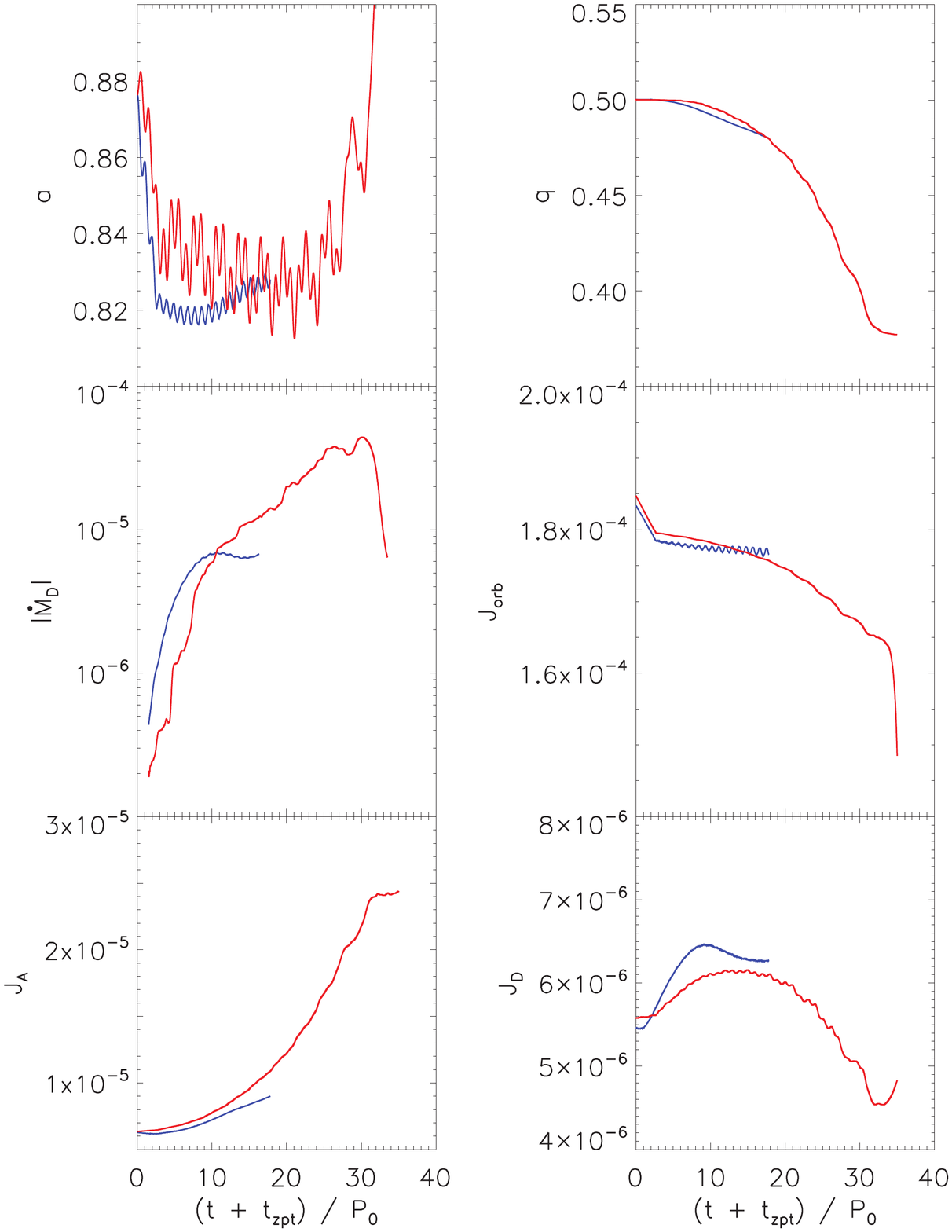}{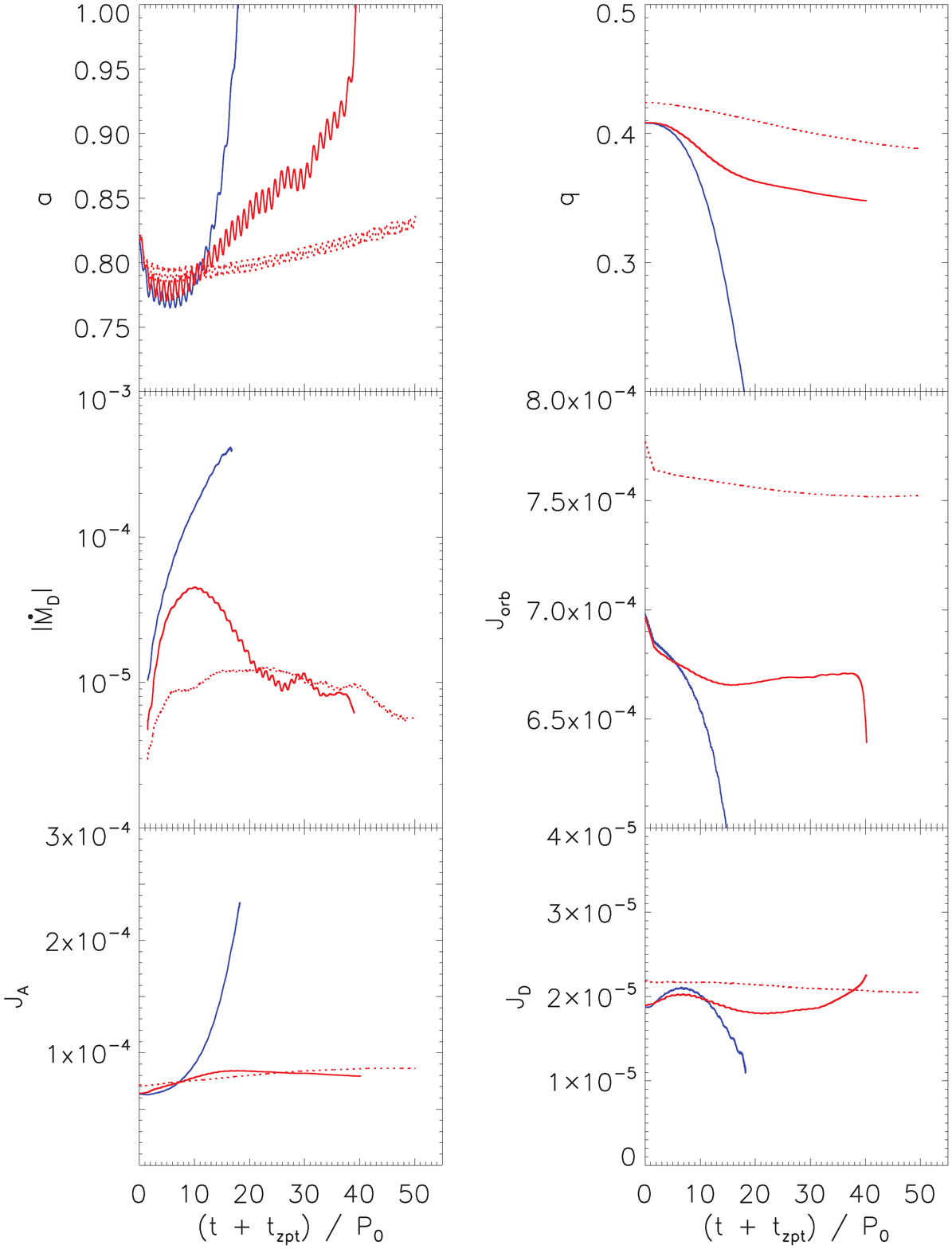}}} \\
\end{tabular}
\caption[COMPOSITE04_V2.jpeg]{Simulations Q0.5P and Q0.4P.
Curves show the time-dependent
behavior of six binary system parameters: $a$, $q$, $\log|\dot{M}_\mathrm{D}|$,
$J_\mathrm{orb}$, $J_\mathrm{A}$ and $J_\mathrm{D}$. The units used
to normalize each quantity are defined in Appendix \ref{app:Diagnostics}.
Left Panel: Results derived from models Q0.5P$\_G1$ (red curves) and
Q0.5P$\_S1$ (blue curves).  Right Panel: Results derived from models
Q0.4P$\_G1$ (red solid curves), Q0.4P$\_G2$ (dashed red curves) 
and Q0.4P$\_S1$ (blue curves).} 
\label{fig:Q05PQ04P}
\end{sidewaysfigure}
\clearpage

\newpage
\begin{figure}[htb!]
\centering
\scalebox{0.6}{\includegraphics{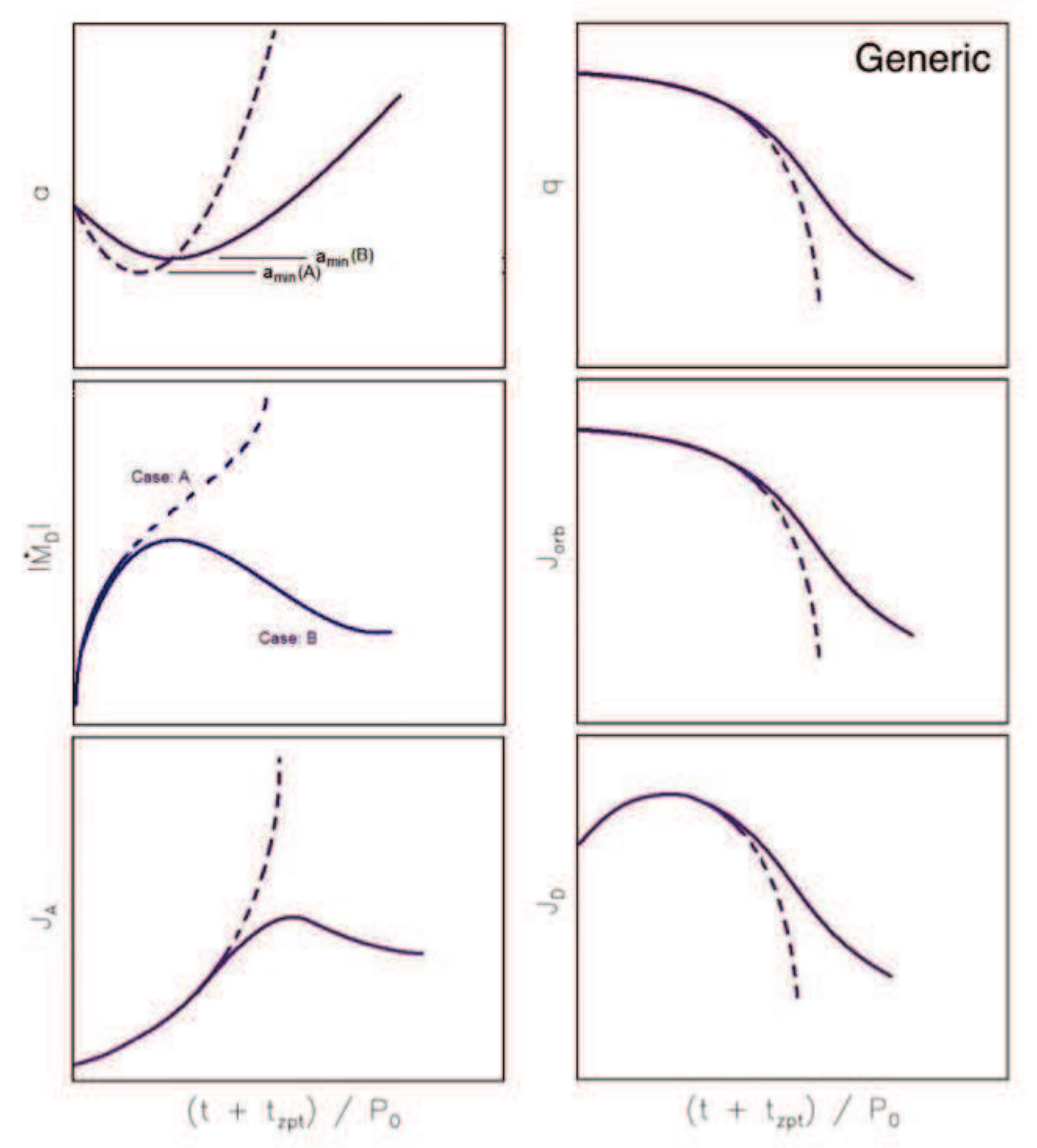}}
\caption[generic_combined7.jpeg]{Accompanying the
\S\ref{subsec:Q0.5PandQ0.4P} discussion,
a schematic illustration is presented of the
generic, time-dependent behavior of six binary system parameters in
``Case A'' (dashed curves) and ``Case B'' (solid curves) evolutions.} \label{fig:Q05PQ04Pgeneric}
\end{figure}
\clearpage

\newpage
\begin{figure}[htb!]
\centering
\scalebox{0.9}{\plotone{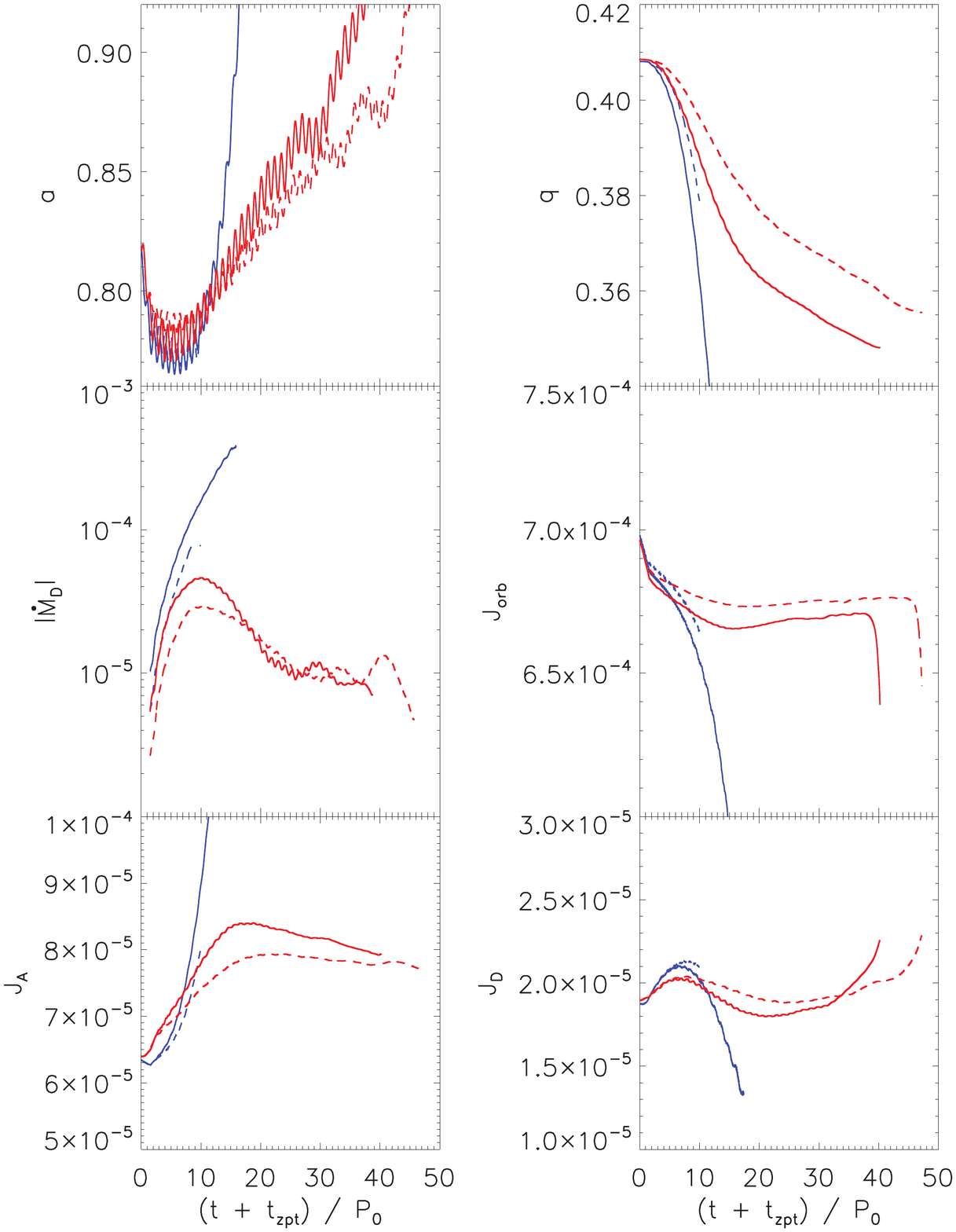}}
\caption[COMPOSITE04_V2.jpeg]{As in Figure \ref{fig:Q05PQ04P}b,
curves show the time-dependent behavior of six binary system parameters
from model simulations Q0.4P.  Blue curves are derived from SPH-code
simulations and red curves are derived from grid-code simulations.  Solid
curves exactly reproduce information provided in the left panel of 
Figure \ref{fig:Q05PQ04P}
from simulations that were driven into contact for $1.6P_0$; for comparison,
dashed curves present results from simulations Q0.4P$\_G3$ and Q0.4P$\_S2$
that were driven into contact for only $1.16P_0$.}
\label{fig:Q04Pmerged}
\end{figure}
\clearpage

\newpage
\begin{figure}[htb!]
\centering
\scalebox{0.8}{\plotone{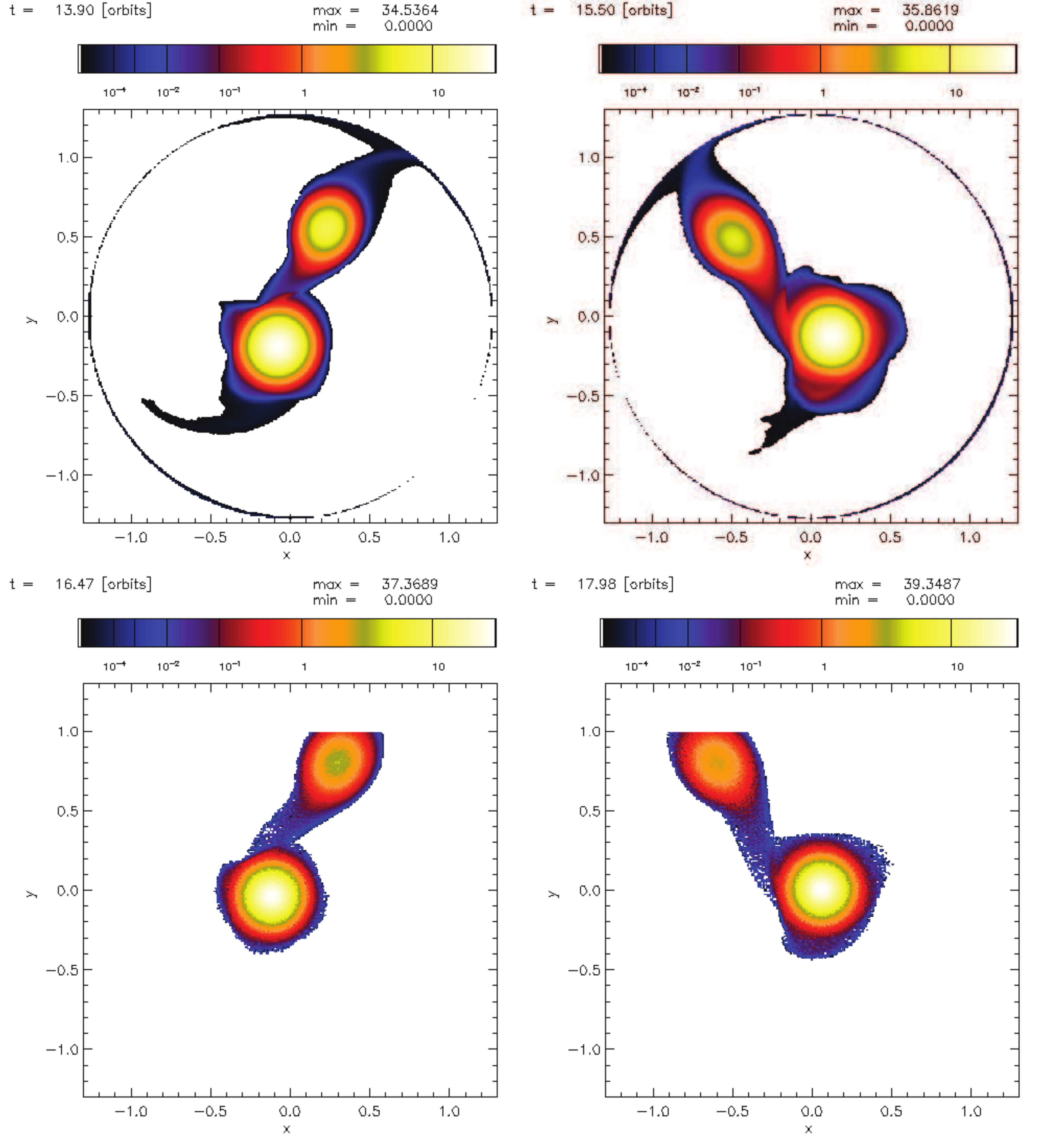}}
\caption[Q04ResonanceComparison.jpeg]{Equatorial-plane surface-density
contours from (bottom) model Q0.4P$\_S1$ and from (top) MFTD07's model 
Q04D at two different times late in their simulations as similar nonlinear-amplitude 
box-like (left) and triangular-shaped (right) resonances are being excited in the
equatorial-belt region of the accretor --- specifically, {\it left-most images}: 
$t_\mathrm{Q04D} = 13.9 P_0$ and $t_\mathrm{S1} = 16.5 P_0$;
{\it right-most images}: $t_\mathrm{Q04D} = 15.5 P_0$ and 
$t_\mathrm{S1} = 18.0 P_0$. The bottom two images have been extracted 
from our video08; a movie showing the development of the structures displayed 
in the top two images has been published in MFTD07.} 
\label{fig:Q04Presonance}
\end{figure}
\clearpage

\newpage
\begin{figure}[htb!]
\centering
\scalebox{0.8}{\plotone{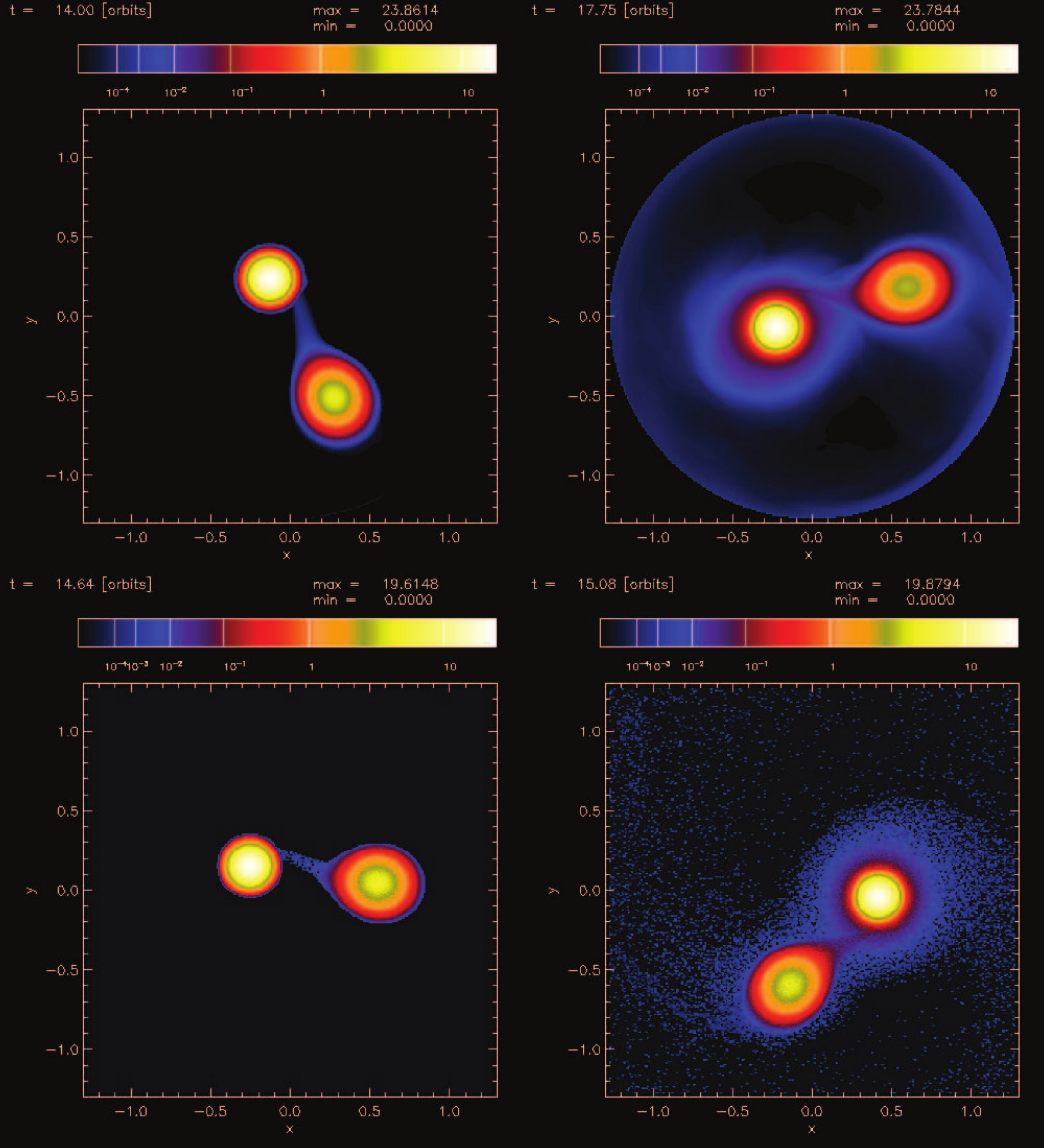}}
\caption[q05_PolyIdealMosaic.jpeg]{The images presented exemplify the
qualitative difference between 
polytropic and ideal gas simulations.  A hot tenuous envelope develops 
around one or both stars due to shock heating in the ideal gas simulations. 
Four selected frames from four simulations starting with initial
model Q0.5 are shown: top row grid results, bottom row SPH results, left column 
polytropic simulations, right column ideal gas simulations. 
Since $t_{\rm zpt}$ is undefined in this case, the four frames were 
selected by eye to represent a similar stage in their time evolution.} 
\label{fig:Q05PolyIdeal}
\end{figure}
\clearpage

\newpage
\begin{figure}[htb!]
\centering
\figurenum{10a}
\scalebox{0.9}{\plotone{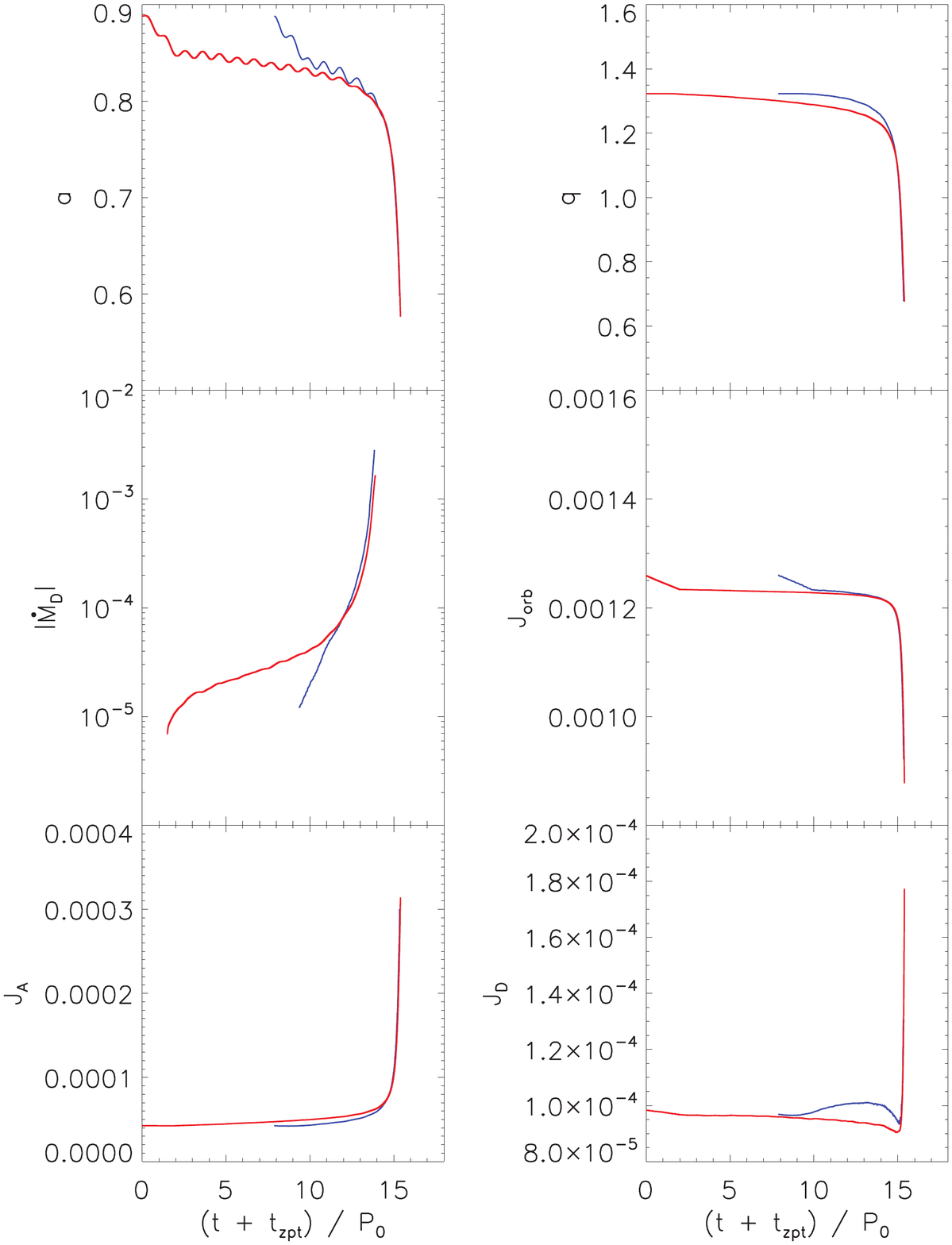}}
\figcaption[Q13ideal_Lines.jpeg]{Simulation Q1.3I. Curves
show the time-dependent behavior of six system parameters 
as derived from models Q1.3I$\_G1$ (solid red curves) and
Q1.3I$\_S1$ (solid blue curves). As detailed in Table 
\ref{tab:Diagnostics}, the SPH-code simulation includes a 
zero-point shift in time of $t_\mathrm{zpt}=7.85$.}
\label{fig:Q13idealA}
\end{figure}
\clearpage

\newpage
\begin{figure}[htb!]
\centering
\figurenum{10b}
\scalebox{0.8}{\plotone{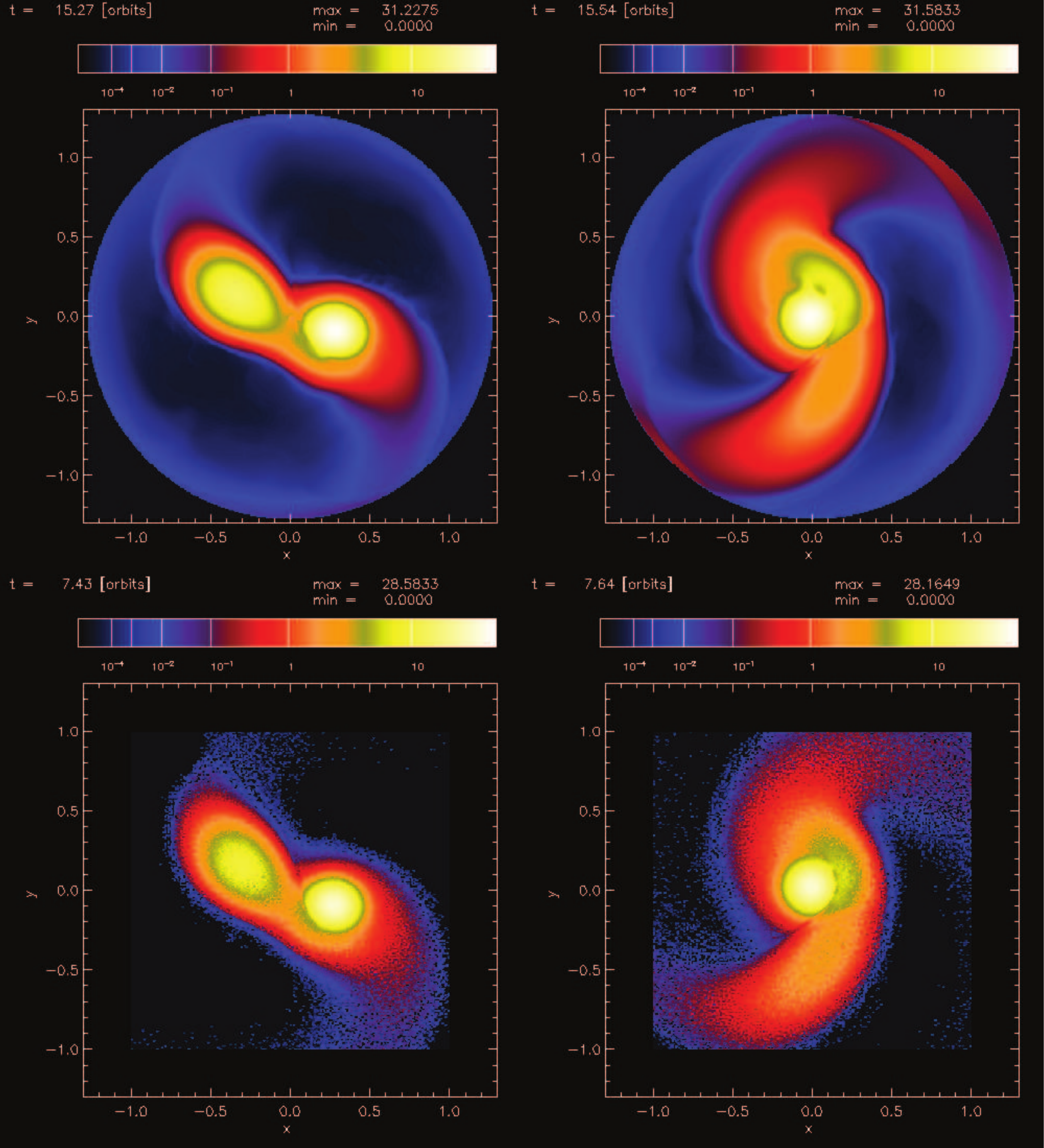}}
\figcaption[Q13ideal_Contours.jpeg]{Simulation Q1.3I.
Images show contours of equatorial-plane column densities 
from model Q1.3I$\_G1$ (top) and model Q1.3I$\_S1$
(bottom) at two different points in time as the binary merger 
is occurring --- specifically, {\it left-most images}: 
$t_{G1} = 15.27P_0$ and $t_{S1} = 7.43P_0$;
{\it right-most images}: $t_{G1} = 15.54P_0$ and 
$t_{S1} = 7.64P_0$.}
\label{fig:Q13idealB}
\end{figure}
\setcounter{figure}{10}
\clearpage

\newpage
\begin{figure}[htb!]
\centering
\figurenum{11a}
\scalebox{0.9}{\plotone{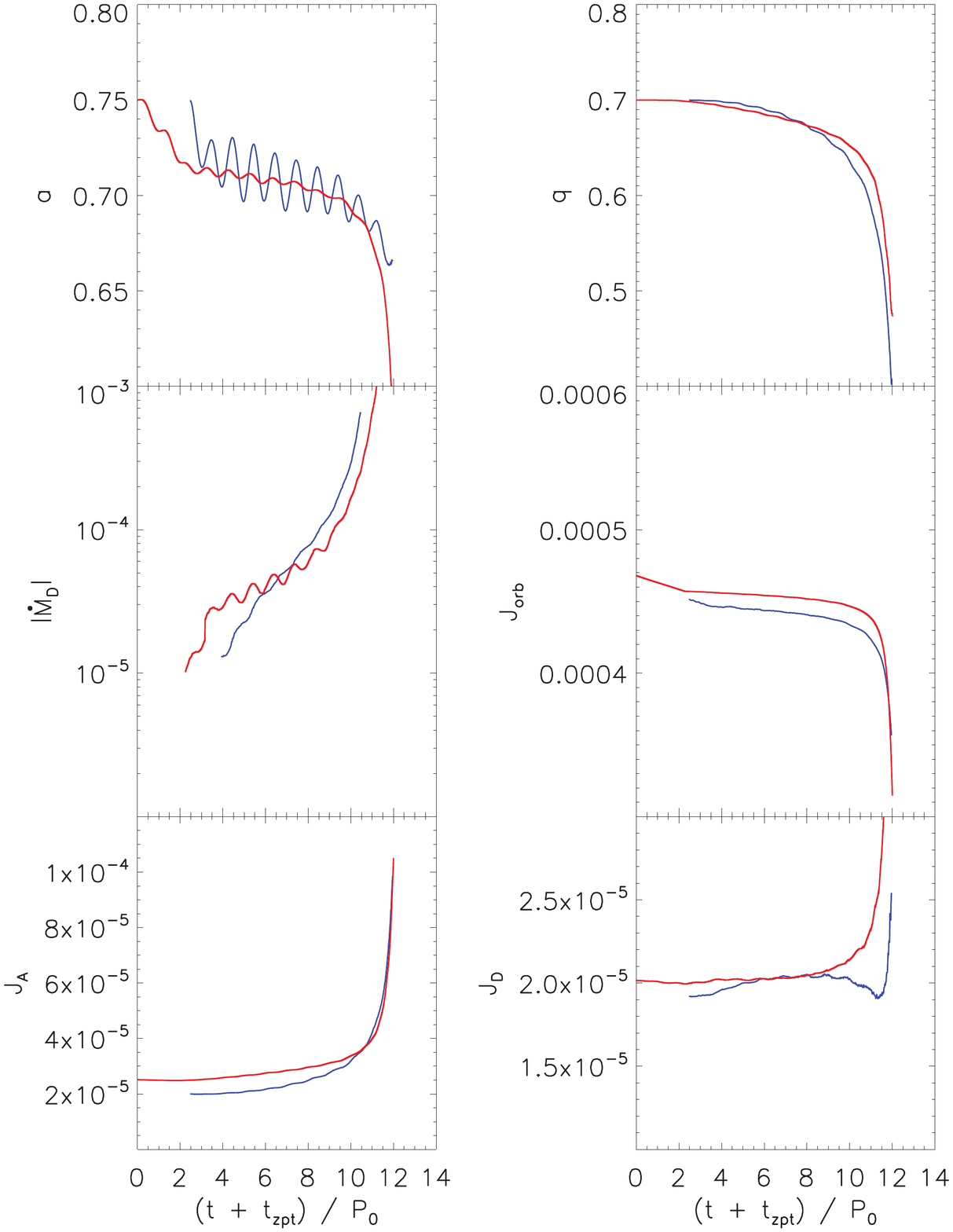}}
\figcaption[Q07ideal_Lines.jpeg]{Simulation Q0.7I. Curves
show the time-dependent behavior of six system parameters as
derived from models Q0.7I$\_G1$ (solid red curves) and
Q0.7I$\_S1$ (solid blue curves). As detailed in Table \ref{tab:Diagnostics},
the SPH-code simulation includes a zero-point shift in time of 
$t_\mathrm{zpt}= - 8.7$.}
\label{fig:Q07idealA}
\end{figure}
\clearpage

\newpage
\begin{figure}[htb!]
\centering
\figurenum{11b}
\scalebox{0.8}{\plotone{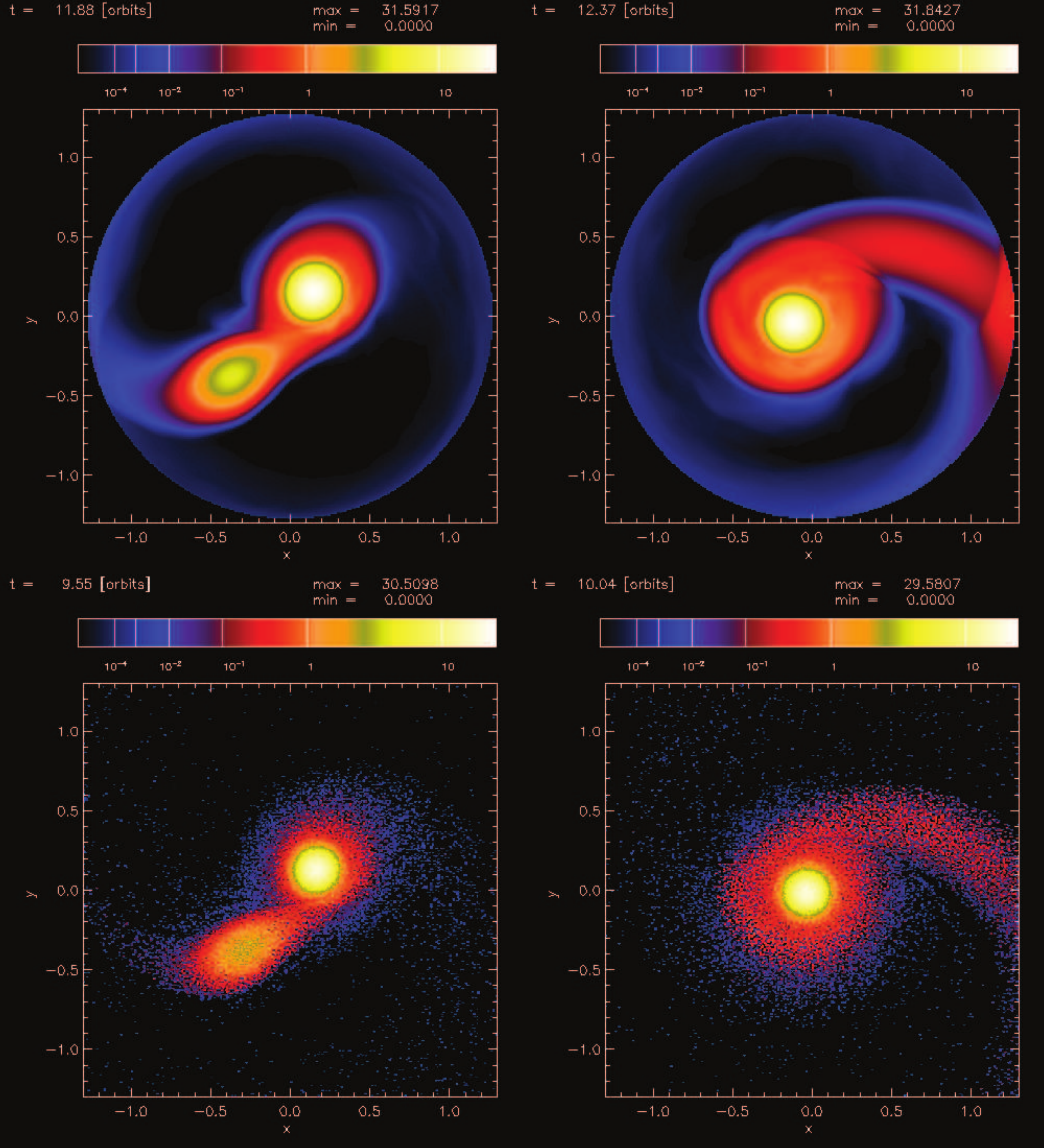}}
 \figcaption[Q07_ideal_composite.jpeg]{Simulation Q0.7I.
Images show contours of equatorial-plane column densities from 
model Q0.7I$\_G1$ (top) and model Q0.7I$\_S1$ (bottom) at two 
different points in time as the tidal destruction of the donor star is occurring --- specifically,
{\it left-most images}: $t_{G1} =  11.88 P_0$ and $t_{S1} = 9.55 P_0$;
{\it right-most images}: $t_{G1} = 12.37 P_0$ and $t_{S1} = 10.04P_0$.}
\label{fig:Q07idealB}
\end{figure}
\setcounter{figure}{11}
\clearpage

\newpage
\begin{sidewaysfigure}[htb!]
\centering
\begin{tabular}{cc}
\textbf{Q0.5} & \textbf{Q0.4} \\
\multicolumn{2}{c}{\scalebox{1.1}{\plottwo{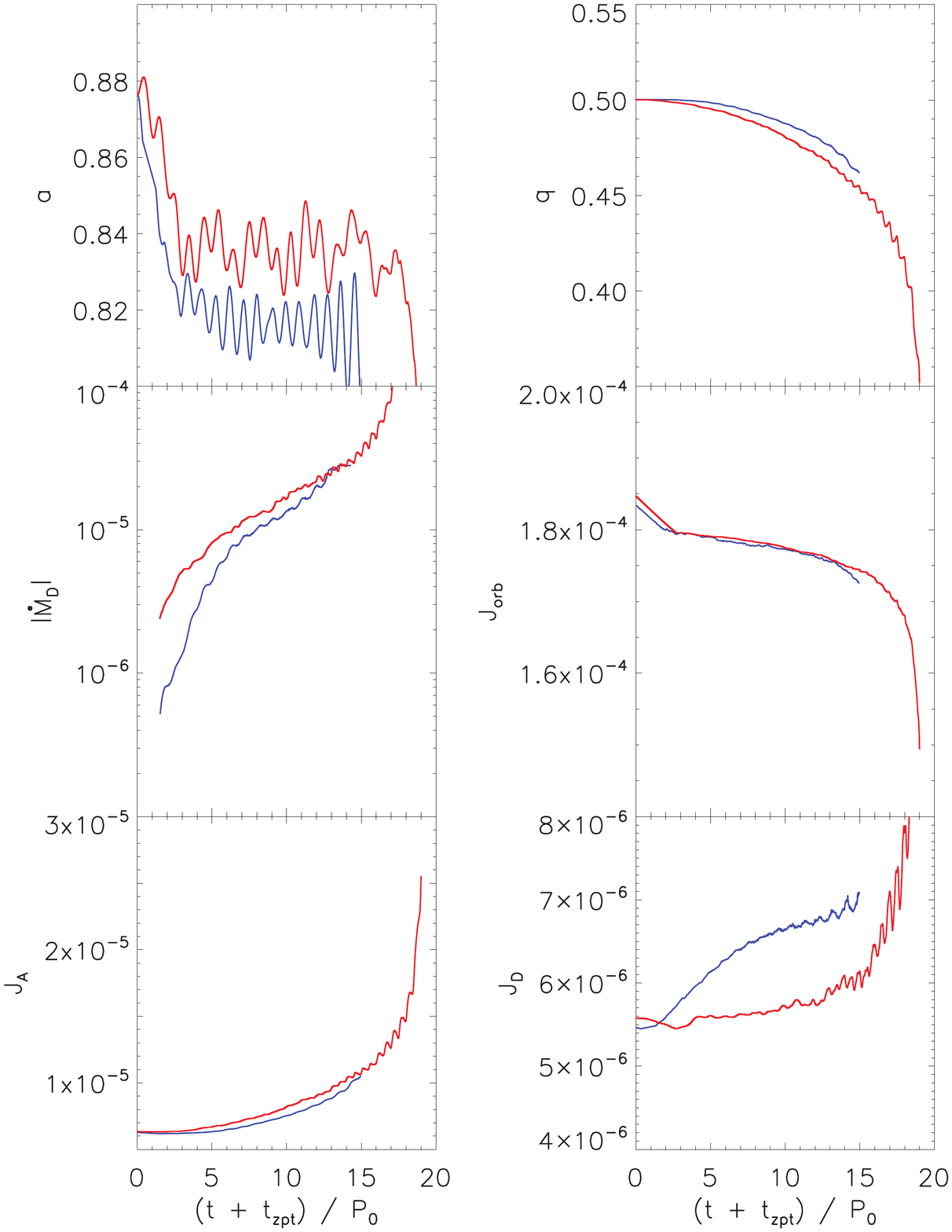}{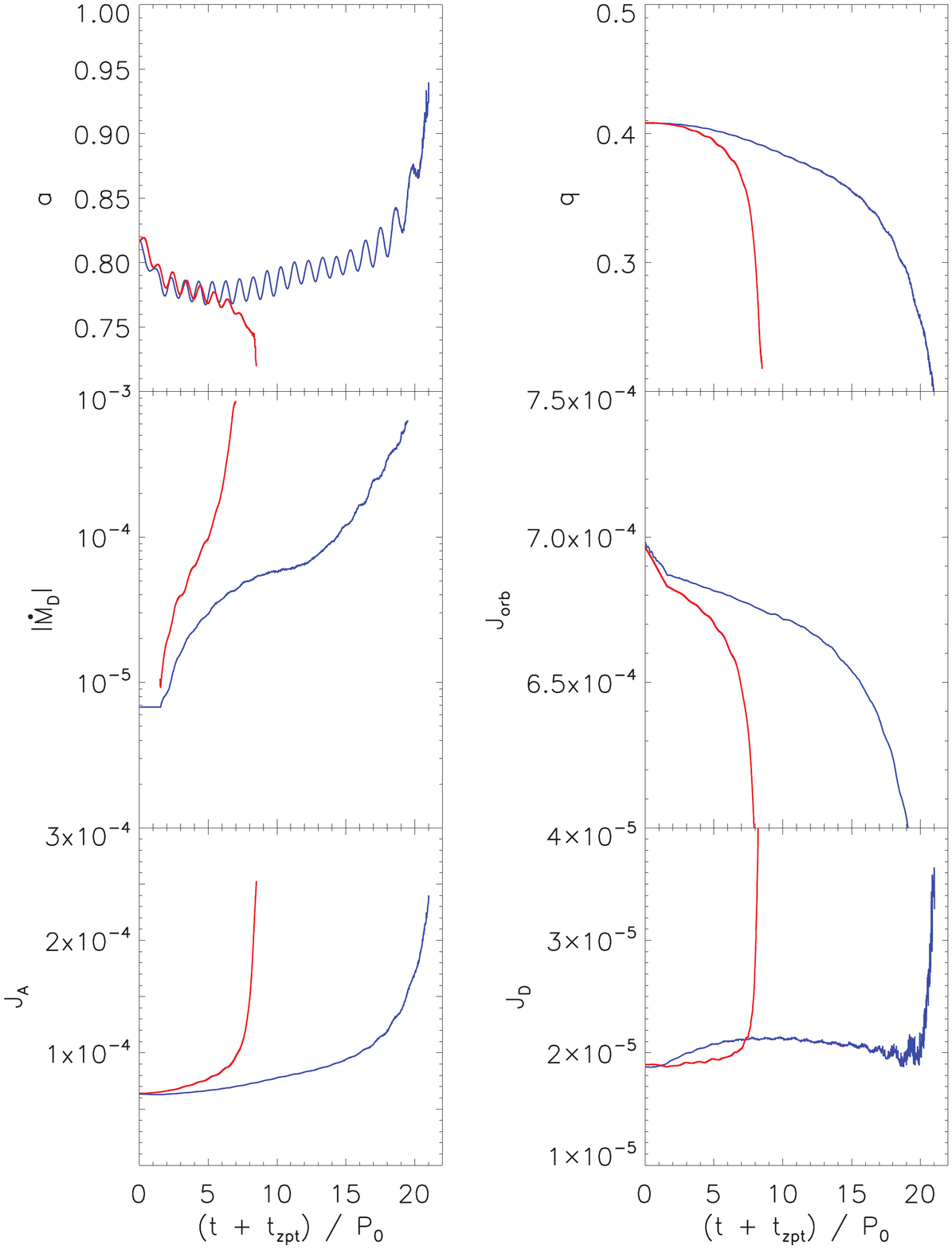}}} \\
\end{tabular}
\caption[Q05ideal_compositeA.jpeg]{Simulations Q0.5I and Q0.4I.
Same as Figure \ref{fig:Q05PQ04P} except for ideal-gas equation of state.
Left Panel: Results derived from models Q0.5I$\_G1$ (red curves) and
Q0.5I$\_S1$ (blue curves).  Right Panel: Results derived from models
Q0.4I$\_G1$ (red curves) and Q0.4I$\_S1$ (blue curves).}
\label{fig:Q05ideal}
\end{sidewaysfigure}
\clearpage

\newpage
\begin{figure}[htb!]
\centering
\scalebox{0.9}{\plotone{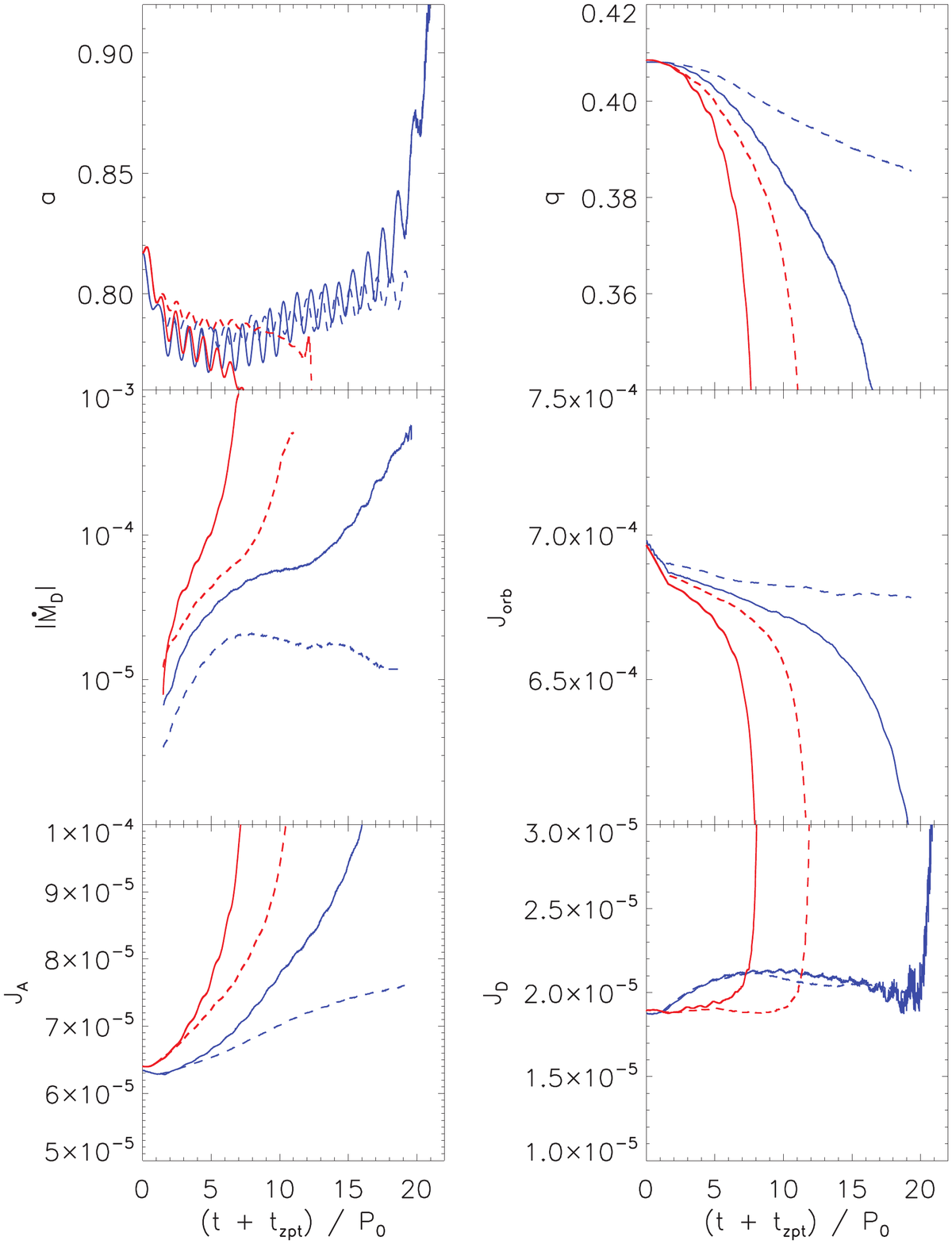}}
\caption[Q04idealMergedPanels.jpeg]{As in Figure 
 \ref{fig:Q04Pmerged},
curves show the time-dependent behavior of six binary system parameters
from model simulations Q0.4I.  Blue curves are derived from SPH-code
simulations and red curves are derived from grid-code simulations.  Solid
curves exactly reproduce information provided in the right panel of Figure 
\ref{fig:Q05ideal}
from simulations that were driven into contact for $1.6P_0$; for comparison,
dashed curves present results from simulations Q0.4I$\_G2$ and Q0.4I$\_S2$
that were driven into contact for only $1.16P_0$.} 
\label{fig:Q04Imerged}
\end{figure}
\clearpage

\begin{sidewaysfigure}[htb!]
\centering
\scalebox{0.9}{\plotone{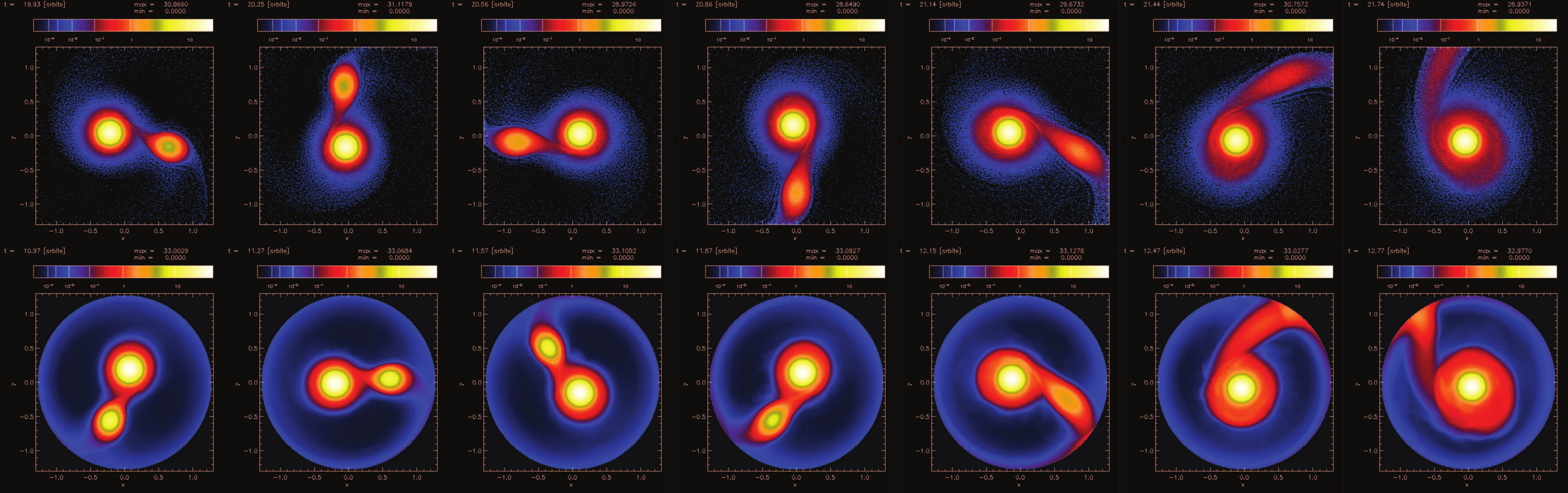}}
\caption[Q04I_montageBoth02small.jpeg]{
Seven images extracted from video19 and video21
that illustrate 
(top row) simulation Q0.4I$\_S1$ over the time period $19.93 \le 
t/P_0 \le 21.74$ and (bottom row) simulation Q0.4I$\_G1$ over the 
time period $10.97 \le t/P_0 \le 12.77$, during the episode of tidal 
disruption of the donor.  In both cases, adjacent images are separated
in time by approximately $0.3P_0$ and the image sequence covers a 
total of $1.8 P_0$.}
\label{fig:Q04idealComparison}
\end{sidewaysfigure}
\clearpage

\end{document}